\documentclass[letter,11pt]{article}
 \pdfoutput=1
 \usepackage{jheppub}
 
 \usepackage{graphicx}
 \usepackage{hyperref}
 \usepackage[utf8]{inputenc}
 \usepackage{amssymb}
\usepackage{multirow}
\usepackage{soul}
 \usepackage{ulem}
 
 \usepackage{booktabs}
\usepackage{mathrsfs}
\usepackage{epsfig}
\usepackage{graphicx}
\usepackage{subfigure}
\usepackage{dcolumn}
\usepackage{bm}
\usepackage{amsmath}
\usepackage{slashed}
\usepackage{multirow}
\usepackage{color}
\usepackage{dcolumn}
\usepackage{bm}
\usepackage{color}
\usepackage[dvipsnames]{xcolor}

\allowdisplaybreaks

\newcommand{\lsim}{{\;\raise0.3ex\hbox{$<$\kern-0.75em\raise-1.1ex\hbox{$\sim$}}\;}}
\newcommand{\gsim}{{\;\raise0.3ex\hbox{$>$\kern-0.75em\raise-1.1ex\hbox{$\sim$}}\;}}
\newcommand{\beq}{\begin{equation}}
\newcommand{\eeq}{\end{equation}}
\newcommand{\bea}{\begin{eqnarray}}
\newcommand{\eea}{\end{eqnarray}}

\def\baa{\begin{array}}
\def\eaa{\end{array}}

\mathchardef\minus="002D

\newcommand{\sm}{\mathrm{SM}}

\def\figureautorefname~#1\null{Fig.\,#1\null}
\def\tableautorefname~#1\null{Table\,#1\null}

\def\equationautorefname~#1\null{Eq.\,(#1)\null}

\newcommand{\tanb}{\tan \beta}
\def\cosba{\cos(\beta-\alpha)}
\def\gev{\textrm{~GeV}}
\newcommand{\lambvs}{\sqrt{\lambda v^2}}

\preprint{ADP-20-31/T1141}

\title{Strong first order electroweak phase transition in 2HDM confronting future Z \& Higgs factories}
\author[*]{Wei Su}
\author[*]{, ~Anthony G. Williams}
\author[\dagger]{, ~Mengchao Zhang}

\affiliation[\dagger]{Department of Physics and Siyuan Laboratory, Jinan University, Guangzhou 510632, P.R. China} 

\affiliation[*]{ARC Centre of Excellence for Dark Matter Particle Physics, Department of Physics, University of Adelaide, South Australia 5005, Australia}

\abstract{The electroweak phase transition can be made first order by extending the Standard Model (SM) Higgs sector with extra scalars. The same new physics can explain the matter-antimatter asymmetry of the universe by supplying an extra source of CP violation and sphaleron processes 
. In this paper we study the existence of a strong first order electroweak phase transition (SFOEWPT) in the type-I and type-II two Higgs doublet models (2HDM). We focus on how the SFOEWPT requirement constraints the spectrum of non-SM Higgs. Through a parameter space scan, we find that SFOEWPT suggests upper limits on the masses of heavy Higgs $m_{A/H/H^\pm}$, which is less than 1 TeV. High temperature expansion and Higgs vacuum uplifting is used for an analytical understanding of our results.
We also study the probe ability on SFOEWPT from Higgs and $Z$-pole precision measurements at one-loop level at future Higgs \& $Z$ factories. And together with theoretical constraints, sizeable loop corrections require $m_A \approx m_{H^\pm} > m_H$ to meet SFOEWPT condition in Type-II 2HDM.
}

\makeatletter
\def\@fpheader{\relax}
\makeatother

\date{2020.08.06}

\begin{document} 
\maketitle
\flushbottom

\section{Introduction}
The discovery of the Higgs boson in 2012 completes the Standard Model (SM)~\cite{Aad:2012tfa,Chatrchyan:2012xdj}, yet there remain observations that cannot be explained by it. 
One of the most famous puzzles is the baryon asymmetry of the universe (BAU), which sees the visible matter in our universe being dominated by baryons whilst the amount of anti-baryons is negligible. 
Particle physics models that can successfully explain the BAU need to satisfy the three Sakharov conditions~\cite{Sakharov:1967dj}. 
The SM was once considered as a candidate model~\cite{Kuzmin:1985mm,Shaposhnikov:1986jp,Shaposhnikov:1987tw}, since baryon number conservation can be broken by an electroweak sphaleron process~\cite{Manton:1983nd,Klinkhamer:1984di,Kuzmin:1985mm}, the CKM matrix provides CP violation, and the electroweak phase transition can induce a departure from equilibrium if the Higgs boson is light enough. 
However, such an electroweak baryogenesis (EWBG) mechanism in the SM framework turns out to fail, since the CP violation present in the CKM matrix is too small~\cite{Huet:1994jb} and the measured Higgs mass is too heavy to trigger a strong first order electroweak phase transition (SFOEWPT)~\cite{Kajantie:1996mn,Csikor:1998eu}. 
Thus for a successful baryogenesis, new physics beyond the SM (BSM) is required to supply a new source of CP violation and a strong out-of-equilibrium process~\cite{Trodden:1998ym,Konstandin:2013caa}. 
In this work we focus on the latter issue. 

In order to obtain a SFOEWPT, generally we need to extend the scalar sector of the SM. 
Additional parameters in the scalar sector help to change the shape of Higgs potential whilst leaving the Higgs vacuum expectation value (VEV) and the mass of Higgs same. 
Simple SM extensions include the addition of a $SU(2)$ singlet 
~\cite{Carena:2018vpt,Cline:2012hg,Cline:2017qpe,Carena:2018cjh,Cline:2009sn,Profumo:2014opa,Curtin:2014jma,Huang:2015bta,Kotwal:2016tex,Vaskonen:2016yiu,Beniwal:2017eik,Kurup:2017dzf,Chiang:2017nmu,Alves:2018jsw,Li:2019tfd,Bell:2019mbn,Grzadkowski:2018nbc,Huang:2018aja}, 
an extra doublet~\cite{Bochkarev:1990fx,McLerran:1990zh,Bochkarev:1990gb,Turok:1990zg,Cohen:1991iu,Turok:1991uc,Nelson:1991ab,Funakubo:1993jg,Davies:1994id,Funakubo:1995kw,Funakubo:1996iw,Cline:1995dg,Fuyuto:2015jha,Chiang:2016vgf,Dorsch:2013wja,Dorsch:2014qja,Cline:1996mga,Fromme:2006cm,Cline:2011mm,Dorsch:2016nrg,Basler:2016obg,Haarr:2016qzq,Fuyuto:2017ewj,Dorsch:2017nza,Cherchiglia:2017gko,Basler:2017uxn,Andersen:2017ika,Bernon:2017jgv,Gorda:2018hvi,Basler:2018cwe,Wang:2018hnw,Kainulainen:2019kyp,Wang:2019pet,Borah:2012pu,Cline:2013bln,Fuyuto:2015ida,Modak:2018csw,Chao:2015uoa}, 
an extra triplet~\cite{Inoue:2015pza,Niemi:2018asa,Chala:2018opy,Zhou:2018zli}, 
or extra higher dimensional operators~\cite{Ellis:2019flb,Cao:2017oez,Huang:2015izx,Huang:2016odd,Balazs:2016yvi,deVries:2017ncy}.
Consideration of the hierarchy problem leads to further solutions such as embedding the Higgs boson in a 
Composite Higgs model~\cite{Cline:2008hr,Bian:2019kmg,Xie:2020bkl} or 
a supersymmetric model~\cite{Cline:1997vk,Menon:2004wv,Carena:2011jy,Bi:2015qva,Demidov:2016wcv,Huang:2014ifa,Cheung:2012pg,Balazs:2013cia,Huber:2006wf,Bian:2017wfv,Kozaczuk:2014kva,Katz:2015uja,Akula:2017yfr,Lee:2004we,Balazs:2004ae,Liebler:2015ddv} to obtain a SFOEWPT.
For a SFOEWPT in other models see~\cite{Kobakhidze:2015xlz,Ramsey-Musolf:2017tgh,YaserAyazi:2019caf,Mohamadnejad:2019vzg}.
In this work we study the existence of a SFOEWPT in the type-I and type-II 2HDMs~\cite{Lee:1973iz,Branco:2011iw}.
These models are attractive to study because the number of new parameters is relatively small.
In addition, both Higgs doublets in 2HDM models are charged under $SU(2)\times U(1)$ and couple to SM fermions. 
This gives a greater range of observations that can probe the models relative to models that are extended by SM singlet scalars.

It is well known that, compared with other baryogenesis mechanisms, e.g. leptogenesis~\cite{Fukugita:1986hr} or the Affleck–Dine mechanism~\cite{Affleck:1984fy}, EWBG can be detected at the electroweak scale and part of the parameter space can be covered by current or expected collider experiments. 
In 2HDMs, in addition to the SM-like Higgs boson $h$, there are three non-SM Higgs bosons, $H$/$A$/$H^{\pm}$.
$H$/$A$/$H^{\pm}$ couple to $h$ and help to build an energy barrier between the symmetric phase and the $SU(2)\times U(1)$ broken phase when the temperature of the universe is around the electroweak scale. 
Then the phase transition, which is tunneling through the energy barrier, can be first order and strong enough. 
In our study, we will show that, in order for this strong first order phase transition to occur, the masses of $H$, $A$ and $H^{\pm}$ bosons should all be smaller than about 1 TeV, and generally there needs to be a relatively large mass splitting between the heavy Higgs bosons $H$, $A$ and $H^{\pm}$.

$H$, $A$, and $H^{\pm}$ bosons with a mass lighter than 1 TeV can be directly produced at the current LHC or future hadron colliders like the HE-LHC~\cite{Abada:2019ono} or the SPPC~\cite{CEPCStudyGroup:2018ghi,CEPCPhysics-DetectorStudyGroup:2019wir}. Channels like $A/H \to t\bar{t}/b\bar{b}/\tau\bar{\tau}$, $H^{\pm}\to t\bar{b}$, or $A\to HZ$~\cite{Li:2020hao,Kling:2018xud,Chen:2017dwb,Craig:2016ygr} can be used for detection or exclusion.  
Besides, through mixing and loop effects, the non-SM Higgs bosons also change the predicted value of the oblique parameters $S$, $T$ and $U$, and reduce the Higgs couplings $\kappa_i=g_hii^{\rm 2HDM}/g_hii^{\rm SM}$ relative to the SM expectation. 
Future $e^+e^-$ colliders like the ILC~\cite{Bambade:2019fyw}, FCC-ee~\cite{Abada:2019lih,Abada:2019zxq} and CEPC~\cite{CEPCStudyGroup:2018ghi,CEPCPhysics-DetectorStudyGroup:2019wir} will copiously produce $Z$ and Higgs bosons, and thus those observables (especially the $hZZ$ coupling) can be measured with unprecedented precision.
In this work, we perform a global fit to obtain the parameter space of 2HDMs that simultaneously satisfies a SFOEWPT and the expected measurement precision at future $Z$ and Higgs factories.

The structure of this paper is as follows. In Section 2 we briefly introduce our 2HDM models and calculation methods. In Section 3 we list all relevant measurements that can be used to constrain the parameter space of the type-I and type-II 2HDMs. 
Section 4 starts with an analytic analysis which helps readers to understand the features of the electroweak phase transition in 2HDMs. 
Then we study three simplified typical cases, and present the most general scan result. We conclude this work in Section 5.


\section{The electroweak phase transition in 2HDMs}

\subsection{Two Higgs Doublet Models}
2HDMs without a $\mathbb{Z}_2$ symmetry generally induce dangerous flavour-violating couplings at tree level.
In this work we therefore consider 2HDMs with a soft  $\mathbb{Z}_2$ symmetry breaking. The tree-level scalar potential for a 2HDM can be written as:
\begin{eqnarray}
\nonumber V^\text{0}(\Phi_1,\Phi_2) &=& m^2_{11}\Phi^{\dagger}_1 \Phi_1 + m^2_{22}\Phi^{\dagger}_2 \Phi_2 - m^2_{12} \left( \Phi^{\dagger}_1 \Phi_2 + h.c. \right) + \frac{\lambda_1}{2} \left( \Phi^{\dagger}_1 \Phi_1 \right)^2 + \frac{\lambda_2}{2} \left( \Phi^{\dagger}_2 \Phi_2 \right)^2 \\
& & + \lambda_3 \left( \Phi^{\dagger}_1 \Phi_1 \right) \left( \Phi^{\dagger}_2 \Phi_2 \right) + \lambda_4 \left( \Phi^{\dagger}_1 \Phi_2 \right) \left( \Phi^{\dagger}_2 \Phi_1 \right) + \frac{\lambda_5}{2} \left[ \left( \Phi^{\dagger}_1 \Phi_2 \right)^2 + h.c. \right] .
\label{potential}
\end{eqnarray}
We consider a CP-conserving case, in which all mass parameters $m^2_{ij}$ and quartic couplings $\lambda_i$ are real.
After electroweak symmetry breaking (EWSB), the two $SU(2)_L$ Higgs doublets $\Phi_i$ obtain VEVs $v_i$, and they can be expanded in the component real scalar fields:
\begin{eqnarray}
\Phi_1 =  \left(\begin{array}{c} \phi^+_1 \\ \frac{1}{\sqrt{2}} ( v_1 + h_1 + ia_1 )  \end{array}\right) \quad , \quad \Phi_2 = \left(\begin{array}{c} \phi^+_2 \\ \frac{1}{\sqrt{2}} ( v_2 + h_2 + ia_2 )  \end{array}\right) . 
\end{eqnarray}
with $v_1^2 + v_2^2  \equiv v^2 \approx (246\text{~GeV})^2$  .
We further define the ratio of VEVs as $\tan\beta \equiv v_2/v_1$.

Two of the three $m^2_{ij}$ can be replaced by other parameters by imposing conditions that result from minimising the Higgs potential 
\begin{eqnarray}
m^2_{11} &=& m^2_{12} \frac{v_2}{v_1} - \frac{v^2_1}{2}\lambda_1 - \frac{v^2_2}{2} (\lambda_3 + \lambda_4 + \lambda_5) \label{eq:m11}\\
m^2_{22} &=& m^2_{12} \frac{v_1}{v_2} - \frac{v^2_2}{2}\lambda_2 - \frac{v^2_1}{2} (\lambda_3 + \lambda_4 + \lambda_5) \label{eq:m22}. 
\end{eqnarray}
Thus the squared mass matrices of the CP-even, CP-odd, and charged Higgs are:
\begin{eqnarray} \label{mass1}
 \mathcal{M}^2_{\text{even}} &=& \left(\begin{array}{cc} m^2_{12}\tan\beta + \lambda_1 v^2_1 \ \ & -m^2_{12} + v_1 v_2 \lambda_{345} \\-m^2_{12} + v_1 v_2 \lambda_{345} \ \ & m^2_{12}/\tan\beta + \lambda_2 v^2_2\end{array}\right) ,  \\\label{mass2}
\mathcal{M}^2_{\text{odd}} &=& \left( m^2_{12} - v_1 v_2 \lambda_5   \right) \left(\begin{array}{cc}\tan\beta & -1 \\ -1 & 1/\tan\beta \end{array}\right) ,  \\\label{mass3}
\mathcal{M}^2_{\text{charged}} &=& \left(  m^2_{12} - \frac{1}{2} v_1 v_2 \left( \lambda_4 + \lambda_5 \right) \right) \left(\begin{array}{cc} \tan\beta & -1 \\ -1 & 1/\tan\beta \end{array}\right).
\end{eqnarray}
Here $\lambda_{345} \equiv \lambda_3 + \lambda_4 + \lambda_5$. 
After diagonalization, the mass eigenstates are related to the original fields by the rotation matrices:
\begin{eqnarray}
\left(\begin{array}{c}H \\ h \end{array}\right) =  \left(\begin{array}{cc}\cos\alpha & \sin\alpha \\ -\sin\alpha & \cos\alpha \end{array}\right)  \left(\begin{array}{c} h_1 \\ h_2 \end{array}\right) , \\
\left(\begin{array}{c}G^0 \\ A \end{array}\right) =  \left(\begin{array}{cc}\cos\beta & \sin\beta \\ -\sin\beta & \cos\beta \end{array}\right)  \left(\begin{array}{c} a_1 \\ a_2 \end{array}\right) , \\
\left(\begin{array}{c}G^{\pm} \\ H^{\pm} \end{array}\right) =  \left(\begin{array}{cc}\cos\beta & \sin\beta \\ -\sin\beta & \cos\beta \end{array}\right)  \left(\begin{array}{c} \phi^{\pm}_1 \\ \phi^{\pm}_2 \end{array}\right) 
\end{eqnarray}
We choose our input parameters to be: 
\begin{eqnarray}
\cos(\beta - \alpha) \ , \ \tan\beta \ , \ m^2_{12} \ ,  \ m_H \ , \ m_A \ , \ m_{H^{\pm}} .
\label{input} 
\end{eqnarray}
The mass of the SM-like Higgs boson $m_h$ is fixed to the current central measured value 125.09 GeV~\cite{Lacey:2283082}. 
Then the $\lambda_i$ can be re-expressed in terms of these input parameters. Considering the theoretical constraints, including vacuum stability, perturbativity, and unitarity, we introduce 
\begin{equation}
    \lambda v^2 \equiv m_H^2-\frac{m^2_{12}}{\sin\beta  \cos\beta},
\end{equation}
following the notation in~\cite{Gu:2017ckc}. Under the assumption of degenerate heavy Higgs masses $m_H=m_A=m_{H^\pm}$, there is no theoretical restriction on the $\tanb$ range when $\lambvs=0$.


Type-I and Type-II 2HDMs have different $\mathbb{Z}_2$ parity assignments, and thus the couplings between scalar and other particles have a different dependence on $\tan\beta$ and the mixing angle $\alpha$. 
The main difference between the Type I and Type II models is the dependence of the couplings $Af\bar{f}$ and $Hf\bar{f}$ on the value of $\tan\beta$.
Couplings between $A/H$ and down-type fermions are suppressed by $\frac{1}{\tan\beta}$ in the Type I model, but are enhanced by $\tan\beta$ in the Type II model. 
Thus the Type II model is generally more constrained by experiments than the Type I model when $\tan\beta$ is large.

Here we need to emphasize that in the 2HDM we can set the mass of the non-SM Higgs bosons $A/H/H^{\pm}$ to an arbitrarily high scale. This is because of the presence of $m^2_{11, 12, 22}$ , with $m^2_{12}$ breaking $\mathbb{Z}_2$ symmetry in Eq.~\ref{potential}.
As can be seen from Eq.~\ref{mass1} to Eq.~\ref{mass3} 
, the squared masses of $A/H/H^{\pm}$ arise from two types of contribution. 
One of them involves terms of the form $\lambda_i v_j v_k$, which are bounded by perturbative unitarity and thus cannot be too large. 
Upper-limits on these terms are roughly given by $4\pi v^2 \approx \ (870 \text{ GeV})^2$. 
Another part of Higgs mass squares come from $m^2_{12}$ ($m^2_{11/22}$ are transformed through~\autoref{eq:m11} and \autoref{eq:m22}),
and these terms can in principle be set to any value without violating theoretical requirements. 
This makes the search for evidence of 2HDMs an endless game: you can never completely falsify a New Physics model containing hypothetical particles which have no upper limits on their mass.

However, in the following part of this work we will show that the requirement of a SFOEWPT imposes upper limits on the masses of the $A/H/H^{\pm}$ bosons, making it possible to fully verify or falsify the idea of EWBG in 2HDMs in the near future.

\subsection{Thermal effective potential}
To study the phase transition in the early universe, we need to study the dependence of the free energy density on the order parameter.
In our case, the free energy density is the thermal effective potential, and the order parameter is the homogeneous scalar VEV~\cite{Laine:2016hma}. 
The thermal effective potential $V(\phi_1,\phi_2,T)$ at temperature $T$ is composed of four parts:
\begin{eqnarray}
V(\phi_1,\phi_2,T) = V^\text{0}(\phi_1,\phi_2) + V^\text{CW}(\phi_1,\phi_2) + V^\text{CT}(\phi_1,\phi_2) + V^\text{T}(\phi_1,\phi_2,T) .
\label{eq:TEP}
\end{eqnarray}
Here $V^\text{0}$ is the tree-level potential of our model, $V^\text{CW}$ is one-loop Coleman-Weinberg potential, $V^\text{CT}$ is the counter term, and $V^\text{T}$ is the thermal correction. 

The tree-level potential $V^\text{0}(\phi_1,\phi_2)$ is obtained by replacing the field operators $\Phi_1(x)$ and $\Phi_2(x)$ in $V^\text{0}(\Phi_1,\Phi_2)$ with the homogeneous field values $\frac{1}{\sqrt{2}}(0,\phi_1)^T$ and $\frac{1}{\sqrt{2}}(0,\phi_2)^T$:
\begin{eqnarray}
V^{0}(\phi_1,\phi_2) = \frac{1}{2}m^2_{11}\phi_1^2 +  \frac{1}{2}m^2_{22}\phi_2^2  - m^2_{12}\phi_1\phi_2  + \frac{1}{8}\lambda_1 \phi_1^4 + \frac{1}{8}\lambda_2 \phi_2^4 + \frac{1}{4} \lambda_{345} \phi_1^2 \phi_2^2 .
\end{eqnarray}

The one-loop Coleman-Weinberg potential $V^\text{CW}(\phi_1,\phi_2)$ is given in the $\overline{\text{MS}}$ renormalization scheme by~\cite{Coleman:1973jx}:  
\begin{eqnarray}
V^\text{CW}(\phi_1,\phi_2) &=& \frac{1}{64 \pi^2} \sum_i n_i m_i^4(\phi_1,\phi_2) \left[ \ln\frac{m_i^2(\phi_1,\phi_2)}{\mu^2} - c_i \right], 
\end{eqnarray}
with the index $i$ running over all massive particles. $n_i$ is the degrees of freedom  of particle $i$ multiplied by $(-1)^{2s}$ ($s$ is the spin of particle $i$), which is -12, -4, 6, 3, 2, 1, 2 and 1 for quarks, leptons, $W^{\pm}$, $Z$, $H^{\pm}$, $G^0$, $G^{\pm}$, and neutral scalars, respectively.
$c_i$ is $\frac{5}{6}$ for gauge bosons, and $\frac{3}{2}$ for other particles. $m_i^2(\phi_1,\phi_2)$ is the mass square of particle $i$ with $v_1$ and $v_2$ in its expression being replaced by scalar field value $\phi_1$ and $\phi_2$.
The renormalization scale $\mu$ is set to the zero temperature VEV $v$.

In Eq.~\ref{input} we choose the scalar masses, mixing angle, and VEV ratio as our input parameters. 
These parameters are considered as physical parameters. 
It means that the VEVs are determined by the position of the minimum of the scalar potential, and squared masses are given by the second order partial derivatives of the scalar potential with respect to the scalar fields at the position of the minimum.  
Adding the Coleman-Weinberg correction will shift both the position of the minimum and the second order partial derivatives of the tree-level potential. 

Thus, in order to offset the modification, counter terms $V^\text{CT}(\Phi_1,\Phi_2)$ need to be added to the Lagrangian. 
For a CP-conserving 2HDM, $V^\text{CT}(\Phi_1,\Phi_2)$ can be expressed as~\cite{Basler:2016obg}:
\begin{eqnarray}
\nonumber V^\text{CT}(\Phi_1,\Phi_2) &=& \delta m^2_{11}\Phi^{\dagger}_1 \Phi_1 +\delta m^2_{22}\Phi^{\dagger}_2 \Phi_2 -\delta m^2_{12} \left( \Phi^{\dagger}_1 \Phi_2 + h.c. \right) + \frac{\delta\lambda_1}{2} \left( \Phi^{\dagger}_1 \Phi_1 \right)^2 + \frac{\delta\lambda_2}{2} \left( \Phi^{\dagger}_2 \Phi_2 \right)^2 \\\nonumber
& & + \delta\lambda_3 \left( \Phi^{\dagger}_1 \Phi_1 \right) \left( \Phi^{\dagger}_2 \Phi_2 \right) + \delta\lambda_4 \left( \Phi^{\dagger}_1 \Phi_2 \right) \left( \Phi^{\dagger}_2 \Phi_1 \right) + \frac{\delta\lambda_5}{2} \left[ \left( \Phi^{\dagger}_1 \Phi_2 \right)^2 + h.c. \right] \\
& & + \delta t_1  \phi_1  + \delta t_2 \phi_2  .
\end{eqnarray}
Coefficients of counter terms, those $\delta$s, need to be fixed by ``on-shell'' conditions:
\begin{eqnarray}
\nonumber  \partial_{\psi_i} \left( V^\text{CT}(\Phi_1,\Phi_2) + V^\text{CW}(\Phi_1,\Phi_2) \right) &=& 0 \\
 \partial_{\psi_i}\partial_{\psi_j} \left( V^\text{CT}(\Phi_1,\Phi_2) + V^\text{CW}(\Phi_1,\Phi_2) \right) &=& 0,
\end{eqnarray}
with $\psi_i$ denoting all of the component scalar fields of $\Phi_1$ and $\Phi_2$.
These conditions are evaluated at the minimum of the scalar potential at zero temperature, where $\Phi_1 =  \frac{1}{\sqrt{2}}(0,v_1)^T$ and $\Phi_2 =  \frac{1}{\sqrt{2}}(0,v_2)^T$\footnote{Second order derivatives of $V^\text{CW}$ suffer from an infrared divergence originating from the massless Goldstone boson when $T = 0$. This problem can be solved by introducing an IR cut-off mass~\cite{Cline:2011mm}.}.
After adding these counter terms, our input parameters can be treated as physical parameters which are directly connected to observables.  

The thermal correction with ring resummation included is~\cite{Quiros:1999jp,Arnold:1992rz}:
\begin{eqnarray}
 V^{T}(\phi_1,\phi_2,T) &=& \frac{T^4}{2 \pi^2} \sum_i n_i J_B\left(\frac{m^2_i(\phi_1,\phi_2)}{T^2}\right) + \frac{T^4}{2 \pi^2} \sum_j n_j J_F\left(\frac{m^2_j(\phi_1,\phi_2)}{T^2}\right) \\\nonumber
& &-\frac{T^4}{12 \pi} \sum_k n_k \left[ \left( \frac{\tilde{m}^2_k(\phi_1,\phi_2,T)}{T^2} \right)^{3/2} - \left( \frac{m^2_k(\phi_1,\phi_2)}{T^2} \right)^{3/2} \right] .
\label{VT}
\end{eqnarray}
Here, the index $i$ denotes all gauge bosons and scalars, $j$ denotes leptons and quarks, and $k$ denotes scalars and the longitudinal component of gauge bosons. 
The functions $J_{B,F}$ are two integrals which come from the scalar and fermion thermal corrections respectively:
\begin{eqnarray}
J_B(x) &=& \int^{\infty}_0 dk \ k^2 \ln \left[ 1-\exp(-\sqrt{k^2+x}) \right] , \\
J_F(x) &=& \int^{\infty}_0 dk \ k^2 \ln \left[ 1+\exp(-\sqrt{k^2+x}) \right] .
\end{eqnarray}
The second line in~\ref{VT} comes from ring resummation, which is used to avoid the infrared divergence that occurs when the scalar mass is much smaller than the temperature.
$\tilde{m}^2_k(\phi_1,\phi_2,T)$ is the thermal Debye mass, an expression for which can be found in the literature~\cite{Arnold:1992rz,Basler:2016obg}.

\subsection{Numerical analysis method}
An electroweak phase transition is considered to be strong enough only if the net baryon number generated around the bubble wall is not significantly washed out by the sphaleron process inside the bubble.  
This condition can be converted to the requirement on the value of ``wash out'' parameter~\cite{Moore:1998swa}: 
\begin{eqnarray}
\xi_c \equiv \frac{v_c}{T_c} > 0.9
\end{eqnarray}
Here $T_c$ is the critical temperature where a second minimum of $V(\phi_1,\phi_2,T)$ that breaks $SU(2)\times U(1)$ appears, and $v_c \equiv \sqrt{v_1^2(T_c)+v_2^2(T_c)}$ reflects the scale of electroweak symmetry breaking.
Here $v_1(T_c)$ and $v_2(T_c)$ are the scalar field values which minimize $V(\phi_1,\phi_2,T_c)$.

The calculation of $\xi_c$ suffers from theoretical uncertainties. 
The first problem is that the $\xi_c$ induced by $V(\phi_1,\phi_2,T)$ is not gauge independent by itself~\cite{Nielsen:1975fs,DiLuzio:2014bua,Patel:2011th}. 
Missing higher-order quantum corrections also induce a theoretical uncertainty~\cite{Laine:2017hdk}.
For a concrete model, one can use lattice simulations to obtain a reliable value of $\xi_c$~\cite{Kainulainen:2019kyp}, but such a non-perturbative calculation is very computationally expensive. 
Being aware of the theoretical uncertainty in the calculation of $\xi_c$, in this work we relax the criterion of a SFOEWPT to $\xi_c \equiv \frac{v_c}{T_c} > 0.9$.
On the other hand, for a first order phase transition to really happen in the universe, the bubble nucleation rate should be larger than the Hubble expansion rate at the nucleation temperature~\cite{McLerran:1990zh,Dine:1991ck}.
This requirement can be considered as a further constraint on the 2HDM parameter space. For a conservative estimate, in this work we will not consider a requirement on the bubble nucleation rate.

Analytically, $T_c$ and $v_c$ can be obtained by solving the following equations:
\begin{eqnarray}
& & V(0,0,T_c) = V(v_1(T_c),v_2(T_c),T_c) , \\
& & \frac{\partial}{\partial \phi_1}V(\phi_i,\phi_2,T_c)\Big|_{\phi_1=v_1(T_c),\phi_2=v_2(T_c)}=0  \ (i=1,2) , \\
& & \frac{\partial}{\partial \phi_1}V(\phi_i,\phi_2,T_c)\Big|_{\phi_1=0,\phi_2=0}=0  \ (i=1,2) .
\end{eqnarray}
To make $(0,0)$ and $(v_1(T_c),v_2(T_c))$ as local minimum points of $V(\phi_i,\phi_2,T_c)$, Hessian matrix of $V(\phi_i,\phi_2,T_c)$ at $(0,0)$ and $(v_1(T_c),v_2(T_c))$ also need to be positive definite. 
However, due to the complicated form of $V(\phi_1,\phi_2,T)$, solving these equations analytically is quite difficult. 
Instead, one can search for the critical temperature using a numerical method. 
There are already public packages which can be used for numerical thermal phase transition analysis, such as  $\texttt{CosmoTransitions}$~\cite{Wainwright:2011kj}, $\texttt{BSMPT}$~\cite{Basler:2018cwe}, and $\texttt{PhaseTracer}$~\cite{Athron:2020sbe}.
We choose $\texttt{BSMPT}$ for our numerical analysis, since the 2HDM has been implemented in $\texttt{BSMPT}$ as a benchmark model, and $\texttt{BSMPT}$ is written in $\texttt{C++}$ which helps to save numerical calculation time.
In $\texttt{BSMPT}$, the search for $T_c$ is started from a high temperature (the default value is 300 GeV), where the minimum position of $V(\phi_1,\phi_2,T)$ is $(0,0)$. 
Then $\texttt{BSMPT}$ traces the minimum position of $V(\phi_1,\phi_2,T)$ with decreasing temperature. 
If $\texttt{BSMPT}$ detects a minimum position jumping $(0,0) \Rightarrow (v_1(T'),v_2(T'))$ at a certain temperature $T'$, the search stops and the output $T'$ is the desired critical temperature $T_c$. 

The full thermal phase transition history of the 2HDM could be complicated~\cite{Bernon:2017jgv}. Multiple phase transition processes are possible.  
For baryogenesis, however, only the phase transition that transfers $(0,0) \Rightarrow (v_1(T),v_2(T))$ is relevant. 
This is because a successful baryogenesis requires the sphaleron rate to be very fast outside the bubble wall, i.e. $\Gamma_{\text{Sph}} \sim (\alpha_W T)^4$. 
While in the electroweak symmetry breaking phase, the sphaleron rate will be strongly suppressed as $\Gamma_{\text{Sph}} \propto \exp{(-E_{Sph}(T_c)/T_c)}$. Here the sphaleron energy $E_{Sph}(T_c) \sim \text{10TeV}\times\frac{v_c}{v}$.  
Thus another phase transition $(v_1(T''),v_2(T'')) \Rightarrow (w_1(T''),w_2(T''))$ has nothing to do with baryogenesis, because the sphaleron rate outside the bubble will be too low to generate baryon number.
We will therefore not take this kind of phase transition into account in this work.

\section{Current and expected bounds}

2HDMs are constrained by various theoretical considerations and experimental measurements, such as vacuum stability,  perturbativity and unitarity, as well as heavy flavor observations~\cite{Han:2015yys},  electroweak precision measurements, and LHC Higgs measurements and non-SM Higgs searches ~\cite{Kling:2020hmi}.  We briefly summarize below the constraints we adopt in the following sections.

\subsection{Theoretical constraints}


\begin{itemize} 
\item \textbf{Vacuum stability}

In order to make the vacuum stable, the scalar potential should be bounded from below~\cite{Deshpande:1977rw,Sher:1988mj,Nie:1998yn,Kanemura:1999xf}:
\begin{eqnarray}
\lambda_1 > 0 \ , \ \lambda_2 > 0 \ , \ \lambda_3 > -\sqrt{\lambda_1 \lambda_2} \ , \ \lambda_3 + \lambda_4 - |\lambda_5| > -\sqrt{\lambda_1 \lambda_2} 
\end{eqnarray}

\item \textbf{Perturbativity and unitarity}

We adopt a general perturbativity condition of $|\lambda_i|\leq  4\pi$, and for the unitarity bound~\cite{Huffel:1980sk,Maalampi:1991fb,Kanemura:1993hm,Akeroyd:2000wc,Ginzburg:2005dt}:
\begin{eqnarray}
& &\left|3(\lambda_1 + \lambda_2) \pm \sqrt{9(\lambda_1 - \lambda_2)^2 + 4(2\lambda_3 + \lambda_4)^2 }\right| < 16\pi \ , \ \\
& &\left|(\lambda_1 + \lambda_2) \pm \sqrt{(\lambda_1 - \lambda_2)^2 + 4 \lambda_4^2 }   \right| < 16\pi \ , \  \\
& &\left|(\lambda_1 + \lambda_2) \pm \sqrt{(\lambda_1 - \lambda_2)^2 + 4 \lambda_5^2 }   \right| < 16\pi \ , \ \\
& &\left|\lambda_3 + 2\lambda_4 \pm 3\lambda_5 \right| < 8\pi \ , \ \left| \lambda_3 \pm \lambda_4 \right| < 8\pi \ , \ \left| \lambda_3 \pm \lambda_5 \right| < 8\pi  
\end{eqnarray}
\end{itemize}

\begin{figure*}[t]
\centering
\includegraphics[width=0.325\textwidth]{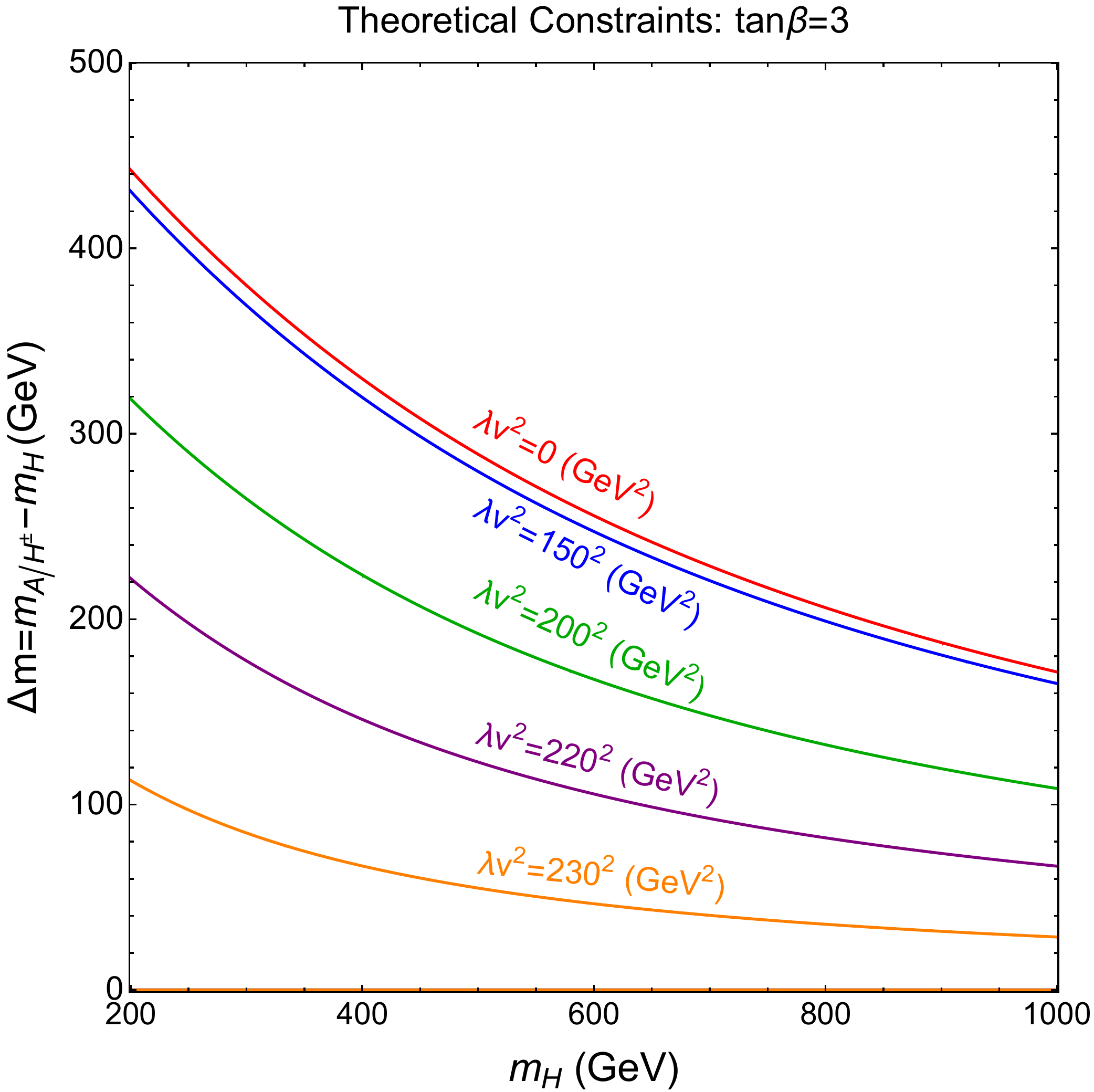}
\includegraphics[width=0.31\textwidth]{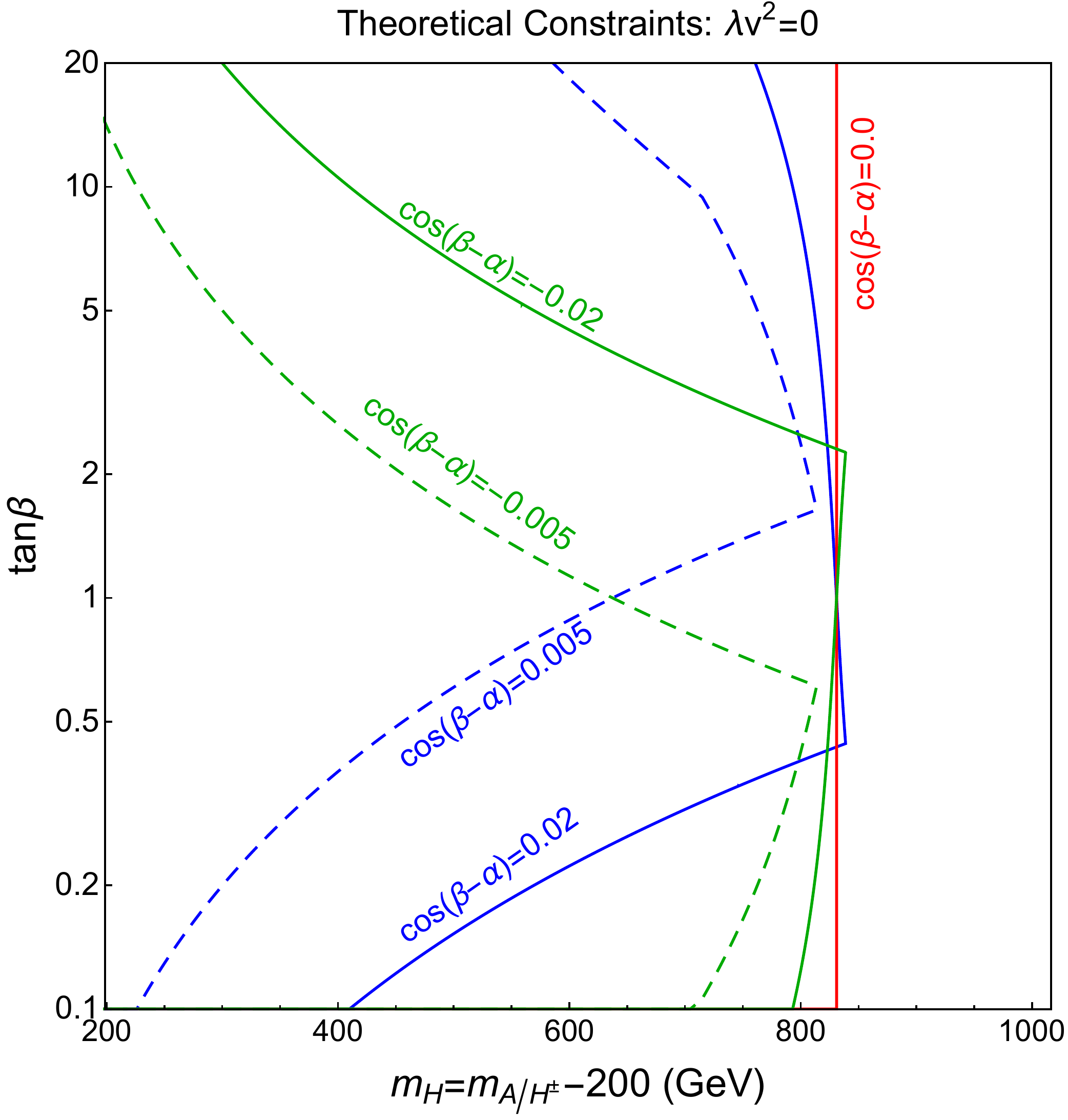}
\includegraphics[width=0.315\textwidth]{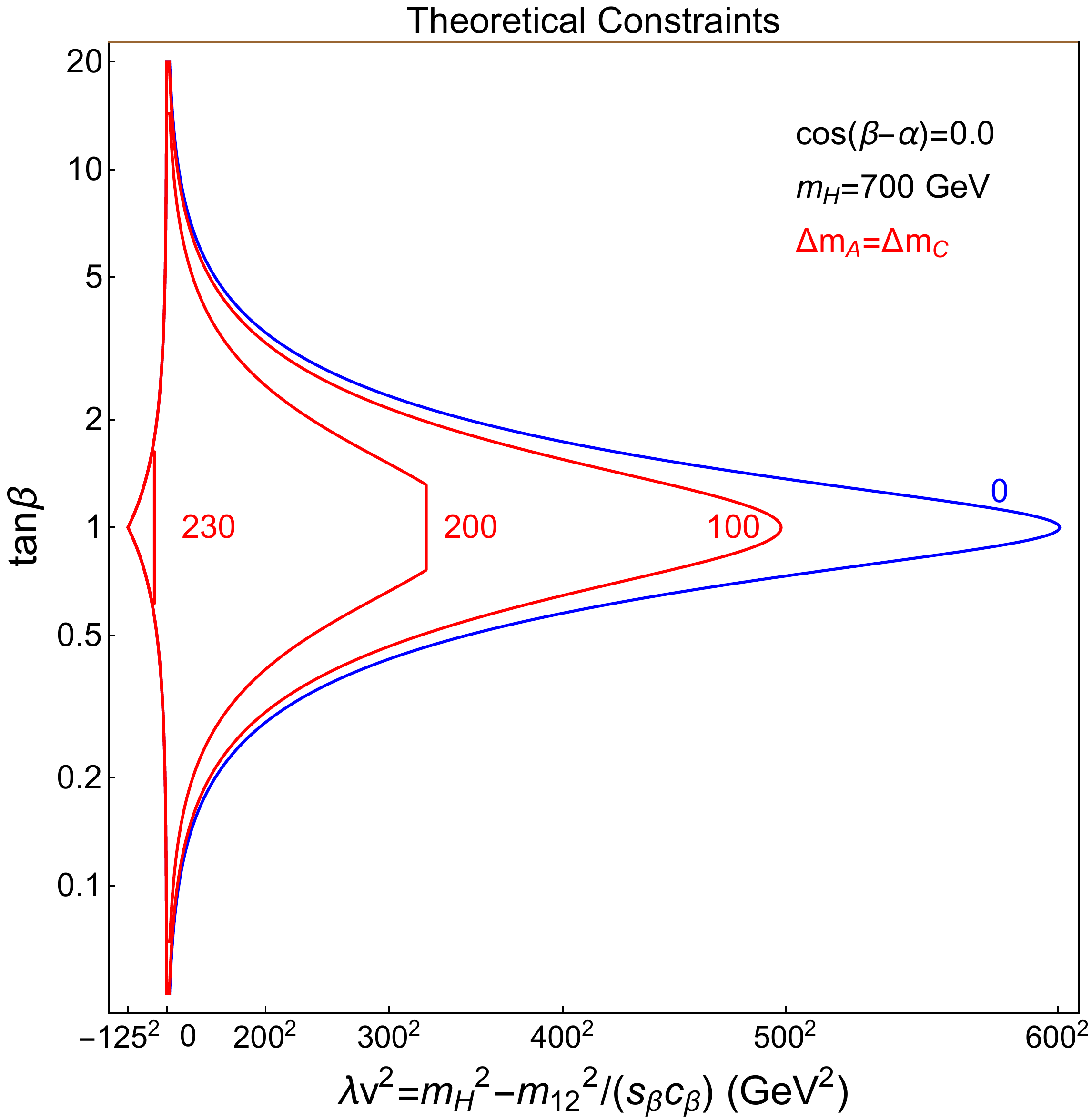}
\caption{
The impact of theoretical constraints in the $m_H - \Delta m$ (left), $m_H - \tanb$ (middle), $\lambda v^2 - \tanb$ (right) planes. In the left panel, the allowed region is under the lines with $\tanb=3, \cosba=0$ . In the middle panel, $\lambvs=0$ GeV, and the allowed region is to the left of the corresponding lines. In the right panel, $m_H$ is fixed at 700 GeV, and the allowed region is inside of the boundary line. See text for full details.
}
\label{fig:theory} 
\end{figure*}

To provide some general insights into the impact of these theoretical constraints, we show in~\autoref{fig:theory} the allowed regions in the $m_H - \Delta m$ (left), $m_H - \tanb$ (middle), and $\lambda v^2 - \tanb$ (right) planes, for various fixed values of the other parameters. In the left panel, we take $\tanb=3, \cosba=0$, fixing $m_A=m_{H^\pm}$. Here $\lambvs= 0, 150, 300, 220, 230 \gev$ are represented by the red, blue, green, purple, and orange lines, and the region under the lines is allowed by the theoretical constraints. Generally, a larger heavy Higgs mass $m_H$ corresponds to a smaller allowed mass splitting $\Delta m$ for any specific $\lambvs$. The allowed  $\Delta m$ also gets smaller when $\lambvs$ gets larger, and here there is no region left for $\lambvs > 232 \gev$.

In the middle panel with $\lambvs=0$ GeV, we explore the effect of the parameter $\cosba$. Here, based on the allowed $|\cosba|$ at the current LHC Run-II \cite{Su:2019ibd}, we take $\cosba=\pm 0.005$ (dashed lines), and $\cosba=\pm 0.02$ (solid lines) and show the allowed region, which is to the left of the corresponding lines. We fix the mass splitting $\Delta m=m_{A/H^\pm}-m_H = 200 \gev$. Under $\cosba=0$, $m_H<820 \gev$ is allowed, independently of $\tanb$. If $\cosba\ne 0$, such as the 0.005 region shown by the dashed lines, the allowed regions are reduced. As discussed in \cite{Gu:2017ckc}, the allowed regions for opposite-sign $\cosba$ are symmetric around the line $\tanb=1$.

In the right panel, $m_H$ is fixed at 700 GeV, and $\Delta m=m_{A/H^\pm}-m_H =$ 0, 100,  200 and 230 are shown. The allowed region is inside of the boundary line. Larger $\Delta m$ leads to a smaller allowed $\lambda v^2$ range, and $\Delta m>230 \gev$ is no longer allowed. For $\lambvs=0$, there is no restriction on $\tanb$.
\subsection{Direct searches at LHC Run-II}

\begin{figure*}[t]
\centering
\includegraphics[width=0.49\textwidth]{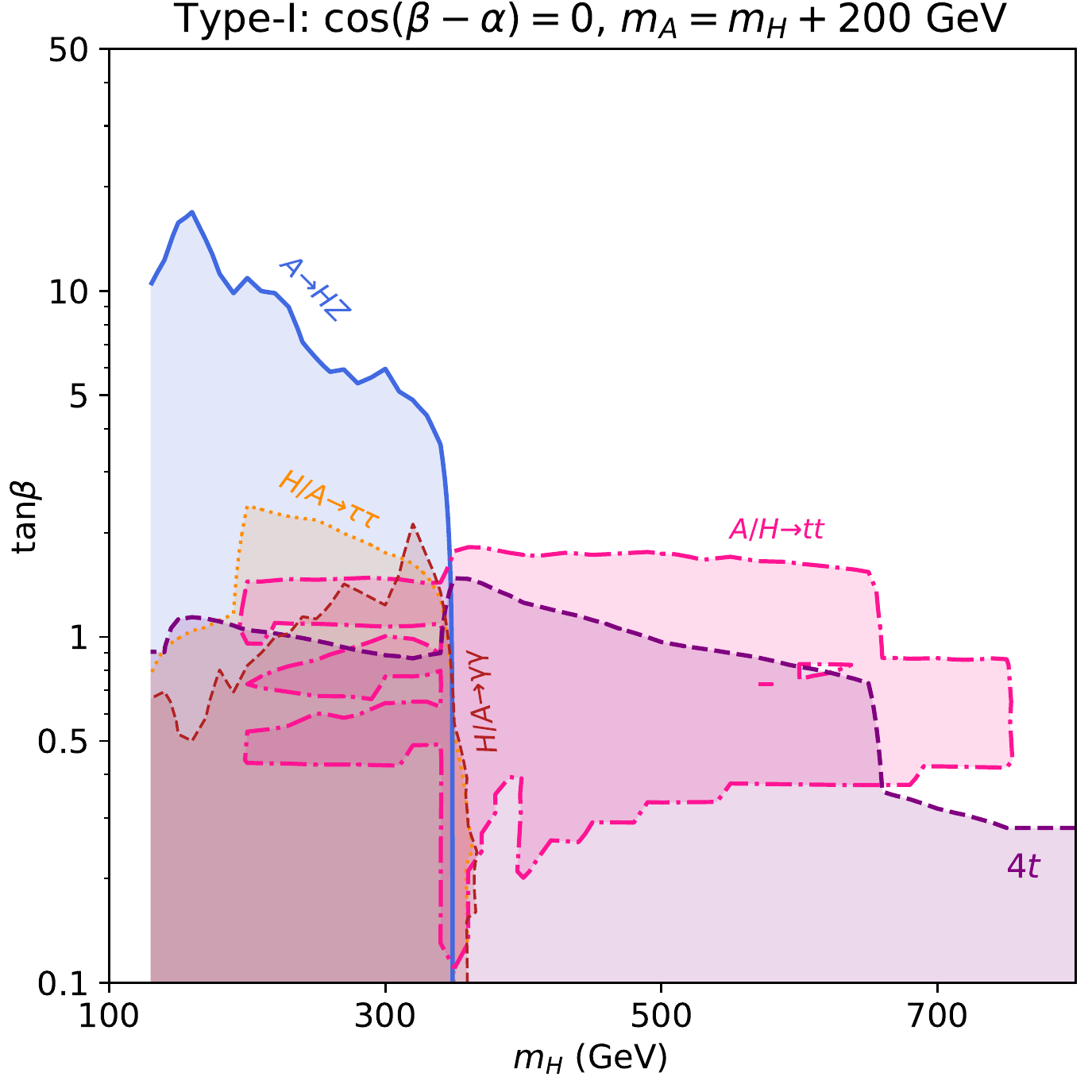}
\includegraphics[width=0.49\textwidth]{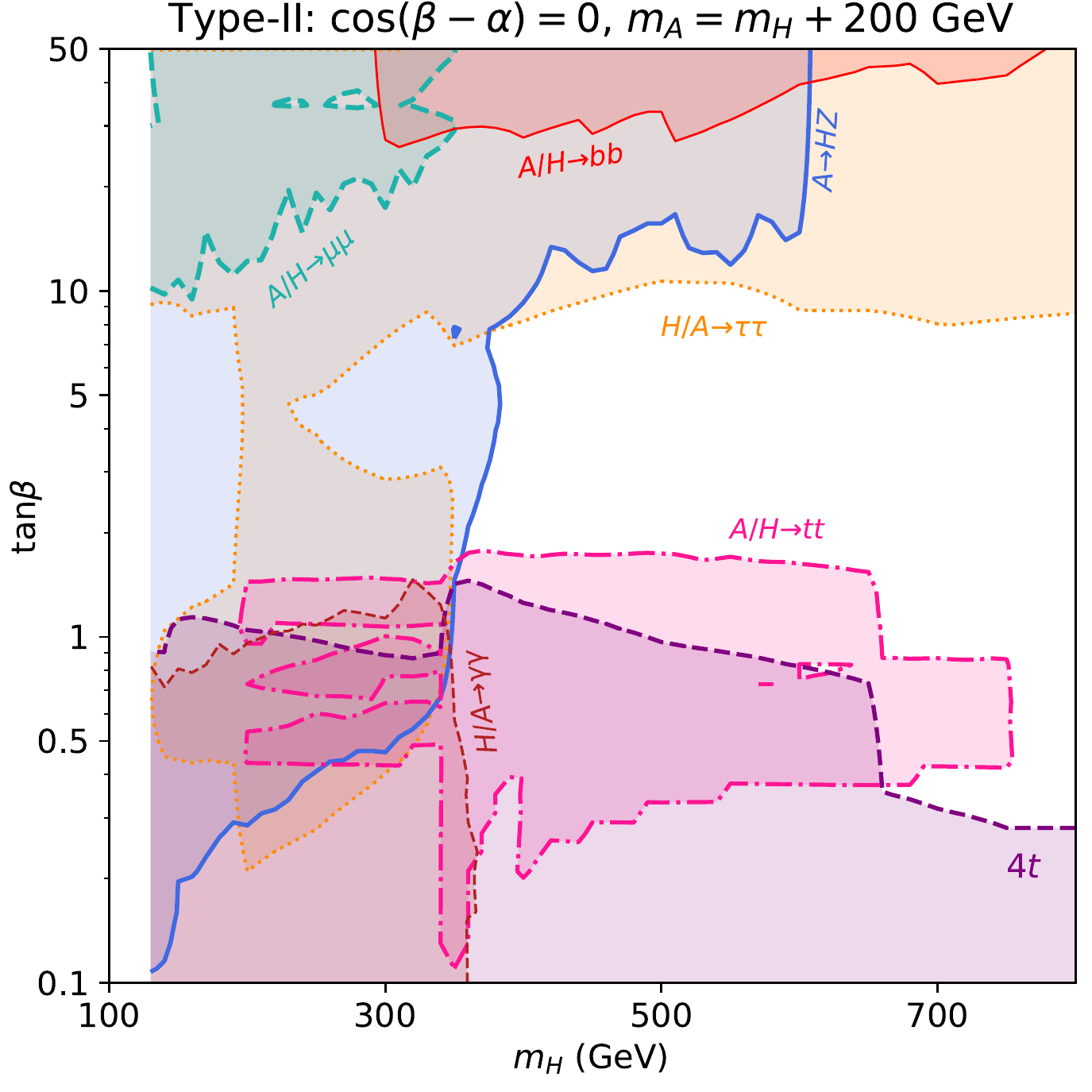}
\caption{
Direct search constraints for the heavy Higgs mass spectrum $m_A=m_{H^\pm}=m_H+200$ GeV. We show the 95\% C.L. exclusion region in the $m_{H} - \tanb$ plane for the Type-I 2HDM (left) and Type-II 2HDM (right) with $\cosba=0$. The various heavy Higgs decays include i) the conventional search results on $H/A \to bb$ (red), $H/A \to \tau\tau$ (dotted orange), $H/A \to \mu\mu$ (dot-dashed cyan), $H/A \to \gamma\gamma$ (dashed brown), $H/A \to tt$ (dot-dashed magenta) and $4t$ production (dashed purple),  and ii) exotic decay channels $A\to HZ$ (blue). Other considered decays such as $A\to Zh, H \to hh$ are not relevant since their couplings are proportional to $\cosba$.
%
}
\label{fig:LHC} 
\end{figure*}

We take into account the latest heavy Higgs searches at LHC Run-II, including $A/H \to \mu\mu$~\cite{Sirunyan:2019tkw,Aaboud:2019sgt}, $A/H \to bb$ ~\cite{Sirunyan:2018taj,Aad:2019zwb}, $A/H \to \tau\tau$~\cite{Sirunyan:2018zut,CMS:2019hvr,Aad:2020zxo}, $A/H \to \gamma\gamma$~\cite{Sirunyan:2018aui, Sirunyan:2018wnk,Aad:2014ioa,Aaboud:2017yyg, ATLAS:2018xad}, $A/H \to tt$~\cite{Sirunyan:2019wph}, $H \to ZZ$~\cite{Sirunyan:2018qlb,Aaboud:2017rel}, $H\to WW$~\cite{Sirunyan:2019pqw,Aaboud:2017gsl}, $A \to hZ \to bb\ell\ell$~\cite{Khachatryan:2015lba,Sirunyan:2019xls,Aad:2015wra,Aaboud:2017cxo}, $A \to hZ \to \tau\tau\ell\ell$~\cite{Khachatryan:2015tha,Sirunyan:2019xjg,Aad:2015wra}, $H \to hh $~\cite{Sirunyan:2017tqo,Sirunyan:2018two,Aad:2015xja,Aad:2019uzh}, and $A/H\!\to\! HZ\!/\!AZ$ \cite{Aaboud:2018eoy,Sirunyan:2019wrn}. 
To investigate the impact on the 2HDM parameter space of the published null results in these searches, we take the cross section times branching fraction limits, $\sigma \times \text{BR}$, from the LHC studies and reinterpret them for our 2HDM model points using the {\tt SusHi} package \cite{Liebler:2016ceh}  to calculate the production cross-section at NNLO level, and  the {\tt 2HDMC}~\cite{Eriksson:2009ws} code for Higgs decay branching fractions at tree level. 

As a first example, taking the benchmark point $\cosba=0$,  $m_{12}^2=m_H^2 \cos\beta \sin\beta$ and $m_A=m_{H^+}=m_H+200 ~\gev$, we show the current collider limits in the $m_{H} - \tanb$ plane in \autoref{fig:LHC}, for both the Type I and Type II models. The various channels include $H/A \to bb$ (red), $H/A \to \tau\tau$ (dotted orange), $H/A \to \mu\mu$ (dot-dashed cyan), $H/A \to \gamma\gamma$ (dashed brown), $H/A \to tt$ (dot-dashed magenta) and $4t$ production (dashed purple),  as well as the exotic decay channel $A\to HZ$ (blue). Other decays such as $A\to Zh$ and $H \to hh$ will only contribute if $\cosba$ deviates from zero at tree level~\cite{Su:2019dsf}.

For the Type-I model (left panel of~\autoref{fig:LHC}), the exotic decay $A\to HZ$ channel covers $m_H<2m_t, \tanb<5$ totally, and can reach to $\tanb=10$. Top quarks searches, $4t+A/H\to tt$, cover $m_H<800~\gev$ for $\tanb<0.3$, and $m_H<650~\gev$ for $\tanb<1.1$ 
$A/H \to \tau\tau, \gamma\gamma$ then exclude the region $m<350~\gev, \tanb<1$ 
Generally because of the $\cot\beta$-enhanced Yukawa coupling in Type-I model, only the small $\tanb$ region can be explored\cite{Kling:2020hmi}. 
In the Type-II 2HDM (right panel), the top quark and $H/A \to \gamma\gamma$ constraints are similar to those for the Type-I model, while the fermionic decays $A/H \to bb, \tau\tau, \mu\mu$ could exclude $m_H$ to 800 GeV when $\tanb>10$ generally. 
Since the $Hbb$, and $H\tau\tau$ couplings are $\tanb$-enhanced, the $A\to HZ$ decay contributes a lot at medium and large $\tanb$ regions, $\tanb>0.5, m_H<2 m_t$ and $\tanb>15, m_H<600~\gev$ 
Thus $m_H<2m_t$ is totally excluded by all channels together in Type-II model, and only $1.5<\tanb<10$ is allowed for $m_H<650 \gev$, which is important for our later study of the electroweak phase transition. 

\subsection{Higgs and $Z$ pole precision measurements}
\label{sec:cons_indrect}
The SM has been tested with high precision via observables measured at the $Z$-pole from LEP-I \cite{ALEPH:2005ab} and the LHC \cite{Haller:2018nnx}. 
Future lepton colliders will further improve the precision of measurements in the Higgs sector, and we therefore include hypothetical future lepton collider results in our study. In Ref.~\cite{Chen:2019pkq}, it was shown that the precision reached by several future $e^+e^-$ machines, including the CEPC program with an integrated luminosity of 5.6 ab$^{-1}$~\cite{CEPCStudyGroup:2018ghi,CEPCPhysics-DetectorStudyGroup:2019wir}, the FCC-ee program with 5 ab$^{-1}$ of integrated luminosity~\cite{Abada:2019lih,Abada:2019zxq},  and the ILC with various center-of-mass energies~\cite{Bambade:2019fyw}, is similar. 
Thus, following the approach adopted in Ref.~\cite{Chen:2018shg,Chen:2019pkq}, we will explore the CEPC proposals in detail.

In our analyses, we take the $S,T,U$ data at $95\%$ Confidence Level (C.L.) from Table 2 of Ref.~\cite{Chen:2019pkq}, and the Higgs precision measurements from Table 3 in the same reference. We use a $\chi^2$  profile-likelihood  fit,
\begin{equation}
\chi^2=\sum_{ij}^Z ( X_i - \hat X_i) (\sigma^2)_{ij}^{-1} ( X_j - \hat X_j )+\sum_i^H \frac{(\mu_i^{\rm{2HDM}}-\mu_i^{\rm{obs}})^2}{\sigma_{\mu_i}^2}\,,
\end{equation}
with $ X_i=(\Delta S\,, \Delta T\,, \Delta U)_{\rm 2HDM}$ being the 2HDM predicted values, and  $\hat{X}_i=(\Delta S\,, \Delta T\,, \Delta U)$ being the  current best-fit central value for current measurements, and 0 for future measurements at the first term for Z sector.  The $\sigma_{ij}$ are the error matrix, $\sigma_{ij}^2\equiv \sigma_i \rho_{ij} \sigma_j$ with $\sigma_i$ and correlation matrix $\rho_{ij} $ given in~\cite{Chen:2019pkq}.
For the second term about Higgs sector, Higgs precision measurements are used to perform global fit with $\mu_i^{\rm{2HDM}}=(\sigma_i\times\textrm{Br}_i)^{\rm{2HDM}}/(\sigma_i\times\textrm{Br}_i)^{\rm{SM}}$ is the signal strength for various Higgs search channels, $\sigma_{\mu_i}$ is the estimated error for each process. The studies~\cite{Chen:2018shg,Chen:2019pkq} show that one-loop level electroweak corrections  to SM Higgs couplings have probe ability to heavy Higgs with Higgs precision measurements, and thus our study of 


For future colliders, the various $\mu_i^{\rm{obs}}$ are set to be unity in the current analyses, assuming no deviations from the SM observables.

In the following analyses, the overall $\chi^2$ is calculated, and  use to determine the allowed parameter region at the $95\%$ C.L. For the one-, and two-parameter fits, the corresponding $\Delta\chi^2=\chi^2-\chi_{\rm{min}}^2$ values at the 95\% C.L. are 3.84, and 5.99 respectively. 

\subsection{Flavour constraints}
The charged Higgs $H^{\pm}$ boson couples to up and down type fermions, and thus observations from flavor physics put strong bounds on its mass and couplings ~\cite{Amhis:2016xyh}.
Among various flavor observations, measurements related to $B$ physics provide the most stringent limits on $\tan\beta$ and $m_{H^{\pm}}$.
For example, $m_{H^{\pm}}<580$~GeV in the Type-II 2HDM has been excluded by 
the measurement of $BR(B\to X_s\gamma)$~\cite{Arbey:2017gmh}.
$\Delta M_{B_s}$ and $BR(B_s \to \mu^+ \mu^-)$ further exclude $m_{H^{\pm}}<1$~TeV in the Type-II 2HDM when $\tan\beta < 0.7$.
The region with $\tan\beta < 1$ and $m_{H^{\pm}}<1$~TeV in the Type-I 2HDM has been excluded by $B\to X_s\gamma$~\cite{Arbey:2017gmh}. In our study, we take these constraints into account.  

\section{Study results}

Based on the diverse constraints above, in this section we will discuss their effects on the SFOEWPT in Type-I and Type-II 2HDMs.

\subsection{The Phase Transition of 2HDM}

To get a better understanding of the electroweak phase transition in 2HDMs, we will first discuss it in the context of some approximate or limiting cases, focusing on the relationship between the phase transition and the Higgs vacuum uplifting. Then we will consider several benchmark cases, varying one or two parameters to dig into the effects of constraints as well as features of the Higgs potential. Our general results will follow these specific cases.


\subsubsection{High Temperature Expansion}

Due to the complicated form of the thermal effective potential~\autoref{eq:TEP} and its intricate thermal evolution history, it is difficult to tell whether a specific point can successfully trigger a SFOEWPT in the early universe through a simple formula or argument. 
To simplify the analysis of the phase transition, people generally use a high temperature expansion, limited to the leading terms of the thermal correction functions $J_B$ and $J_F$. 
Then the thermal effective potential can be simplified to a polynomial function of the Higgs field value:
\begin{eqnarray}
V(\phi_h,T) \approx (D T^2 - \mu^2) \phi_h^2 - E T \phi_h^3 + \frac{\tilde{\lambda}}{4} \phi_h^4 
\label{eq:simple_V}
\end{eqnarray}
Here $\phi_h \equiv \cos\beta \phi_1 + \sin\beta \phi_2 $ is the scalar field that breaks the $SU(2)\times U(1)$ symmetry at zero temperature. 
Due to the simple form of Eq.~\ref{eq:simple_V}, we can use the minimization condition and $V(0,T_c) = V(v_c,T_c)$ to directly calculate the wash-out parameter:
\begin{eqnarray}
\xi_c \equiv \frac{v_c}{T_c} \approx \frac{2E}{\tilde{\lambda}}
\end{eqnarray}
At tree-level, the coefficients $\mu^2$ and $\tilde{\lambda}$ in ~\autoref{eq:simple_V} are:
\begin{eqnarray}
\label{coef}
\mu^2 &=& \frac{1}{4}m^2_h \ , \ \tilde{\lambda} = \frac{m^2_h}{2 v^2}   
\end{eqnarray}   
The coefficients $D$ and $E$ are induced from the leading thermal corrections:
\begin{eqnarray}
\frac{T^4}{2\pi^2}J_B(\frac{m^2(\phi_h)}{T^2}) &\approx& - \frac{\pi^2 T^4}{90} + \frac{1}{24}T^2 m^2(\phi_h) - \frac{1}{12\pi} T (m^2(\phi_h))^{3/2} + \cdots \\
\frac{T^4}{2\pi^2}J_F(\frac{m^2(\phi_h)}{T^2}) &\approx&  + \frac{7}{8} \frac{\pi^2 T^4}{90} - \frac{1}{48}T^2 m^2(\phi_h) + \cdots
\end{eqnarray} 
Here $m^2(\phi_h)$ is the mass square of a massive particle with $v^2$ in it being replaced by $\phi_h^2$ (For example, $m_W^2(\phi_h)=\frac{m^2_W}{v^2}\phi^2_h$). 
Considering the most massive particles in the 2HDM, $D$ and $E$ can be expressed as:
\begin{eqnarray}
D &=& \frac{1}{24} \left[6\frac{m_W^2}{v^2}+3\frac{m_Z^2}{v^2} +\frac{m_h^2}{v^2}+ 6\frac{m_t^2}{v^2} + \frac{m_H^2 - M^2}{v^2} + \frac{m_A^2 - M^2}{v^2} + 2\frac{m_{H^{\pm}}^2 - M^2}{v^2}  \right] \\\nonumber
E &=& \frac{1}{12\pi} \left[  6\frac{m_W^3}{v^3}+3\frac{m_Z^3}{v^3} +\frac{m_h^3}{v^3}  \right] + E_{(H/A/H^{\pm})}
\label{eq:PT_DE}
\end{eqnarray}

In the expression for $E$, the term $E_{(H/A/H^{\pm})}$ denotes the contributions from the non-SM Higgs bosons. 
We cannot explicitly write out the expression for $E_{(H/A/H^{\pm})}$ because, as we said in Section 2.1, the mass of the $H/A/H^{\pm}$ bosons come from two sources. 
Schematically, the $\phi_h$-dependent non-SM Higgs squared masses can be expressed as:
\begin{eqnarray}
m^2_\alpha (\phi_h) = M^2 + \lambda_\alpha \phi_h^2
\end{eqnarray}
Here $M^2 = \frac{m^2_{12}}{\sin\beta \cos\beta}$ is the scale at which the $\mathbb{Z}_2$ symmetry is broken.
$\alpha$ can be $A$, $H$, or $H^{\pm}$, and $\lambda_\alpha$ is a linear combination of the $\lambda_i (i=1,2,3,4,5)$ parameters.
In the alignment limit $\cos(\beta-\alpha)=0$, the expressions for $\lambda_\alpha$ are:
\begin{eqnarray}
\lambda_H &=&  ( \lambda_1 + \lambda_2 - 2\lambda_{345} ) ( \sin^2\beta \cos^2\beta )    \ , \\
\lambda_A &=&  -\lambda_5    \ , \\
\lambda_{H^{\pm}} &=&  - \frac{1}{2} (\lambda_4 + \lambda_5)  
\end{eqnarray}

So the non-SM Higgs bosons provide a term in $V(\phi_h,T)$ which is not exactly proportional to $\phi_h^3$:
\begin{eqnarray}
- \frac{1}{12\pi} T ( m^2_\alpha (\phi_h) )^{3/2} = - \frac{1}{12\pi} T (  M^2 + \lambda_\alpha \phi_h^2  )^{3/2}
\end{eqnarray}
We can simplify the above expression in two limiting cases:
\begin{eqnarray}
- \frac{1}{12\pi} T (  M^2 + \lambda_\alpha \phi_h^2  )^{3/2} \approx \left\{
             \begin{array}{lr}
             -\frac{T}{12\pi} \lambda_{\alpha}^{3/2} \phi_h^3  , & M^2 \ll \lambda_\alpha \phi_h^2  \\
             -\frac{T}{12\pi} M^3 \left( 1 + \frac{3}{2} \frac{\lambda_\alpha \phi_h^2}{M^2} \right) , & M^2 \gg \lambda_\alpha \phi_h^2
             \end{array}
\right.
\end{eqnarray}
And so in these two limiting cases:
\begin{eqnarray}
E_{(\alpha)} \approx \left\{
             \begin{array}{lr}
            \frac{1}{12\pi} \lambda_{\alpha}^{3/2} = \frac{1}{12\pi} \frac{m_\alpha^3}{v^3}  , & M^2 \ll \lambda_\alpha \phi_h^2  \\
             0 , & M^2 \gg \lambda_\alpha \phi_h^2
             \end{array}
\right.
\label{eq:expression_E}
\end{eqnarray}
The above expression needs to be multiplied by 2 if $\alpha$ is $H^{\pm}$. 

Although expression~\ref{eq:expression_E} is obtained in a limiting case, it helps us to understand which of the input parameters are particularly relevant for a SPOEWPT. 
When the non-SM Higgs masses are dominated by $M^2$, the spectrum tends to be degenerate, and the phase transition strength tends to be reduced as the non-SM Higgs boson masses increase. 
When the non-SM Higgs masses are dominated by $\lambda_\alpha v^2$, the spectrum tends to be split, and the phase transition strength tends to be increased as the non-SM Higgs boson masses increase.

\subsubsection{Higgs Vacuum Uplifting}

Another method that can help us to understand which parameters are important for SFOEWPT, is to calculate the depth of the zero temperature Higgs potential~\cite{Dorsch:2017nza}. 
For a shallow Higgs potential, it is easier to develop an energy barrier between the symmetric phase and the broken phase than for a deep Higgs potential, when the temperature is high.  
Thus generally speaking, there is an inverse relation between the phase transition strength and the depth of the vacuum energy. 
We follow the notation at Ref.~\cite{Dorsch:2017nza} and define the SM vacuum energy density as $\mathcal{F}_0^\sm$. The value of $\mathcal{F}_0^\sm$ is about $-1.25\times 10^8$GeV$^4$.
The vacuum energy density of the 2HDM is denoted by $\mathcal{F}_0$. 
We can further define a dimensionless parameter:
\begin{eqnarray}
\Delta\mathcal{F}_0/|\mathcal{F}_0^\sm| \equiv \frac{\mathcal{F}_0-\mathcal{F}_0^\sm}{|\mathcal{F}_0^\sm|}
\end{eqnarray}
$\Delta\mathcal{F}_0/|\mathcal{F}_0^\sm|>0$ means that the 2HDM vacuum energy is uplifted from the SM value, whilst $\Delta\mathcal{F}_0/|\mathcal{F}_0^\sm|$ cannot exceed 1, otherwise the zero temperature vacuum will be unstable. 
The numerical results in~\cite{Dorsch:2017nza} show a positive correlation between $\xi_c$ and the parameter $\Delta\mathcal{F}_0/|\mathcal{F}_0^\sm|$. However, we find that the relationship is only valid for $m_H\leq 500 \gev$, the range Ref~\cite{Dorsch:2017nza} explored, and the parameters may become negatively-correlated for large $m_H$. To illustrate this,
here we refine their analysis by considering a benchmark case:
\begin{eqnarray}
\label{eq:BM1}
\tan\beta = 3.0&,&\ \cos(\beta-\alpha) = 0, \ m_H \in (200, 1000)\gev \ , \\\nonumber 
& & \lambvs=0,\ \Delta m=m_{A/H^{\pm}} - m_H = 200\gev .
\end{eqnarray}
All parameters are fixed except $m_H$, and $\lambda v^2=m_H^2 - \frac{m^2_{12}}{\sin\beta\cos\beta} = 0$ (to meet the theoretical constraints, as in the right panel of~\autoref{fig:theory}). 
\begin{figure}[ht]
  \centering
  \includegraphics[width=0.47 \linewidth]{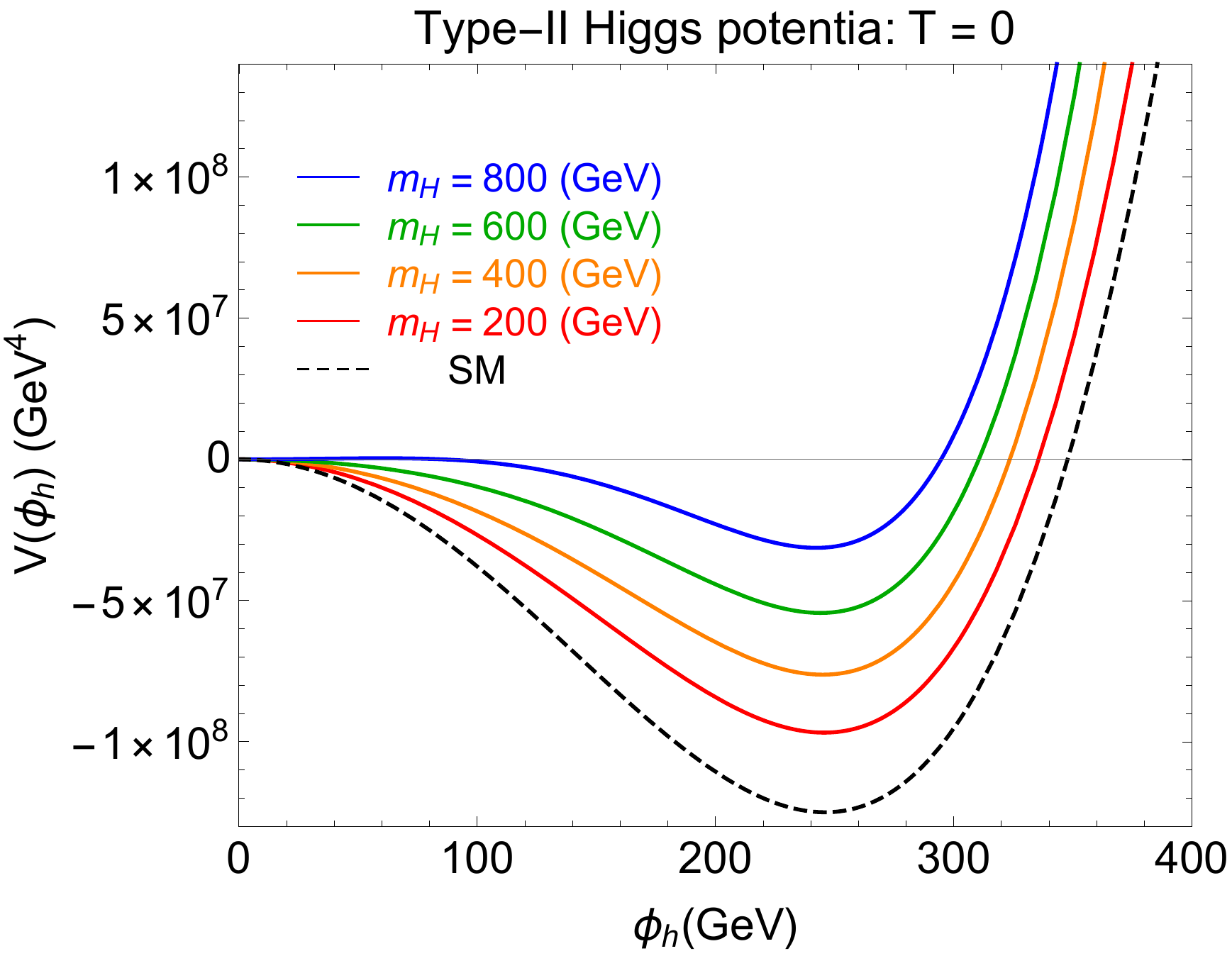}
  \includegraphics[width=0.47 \linewidth]{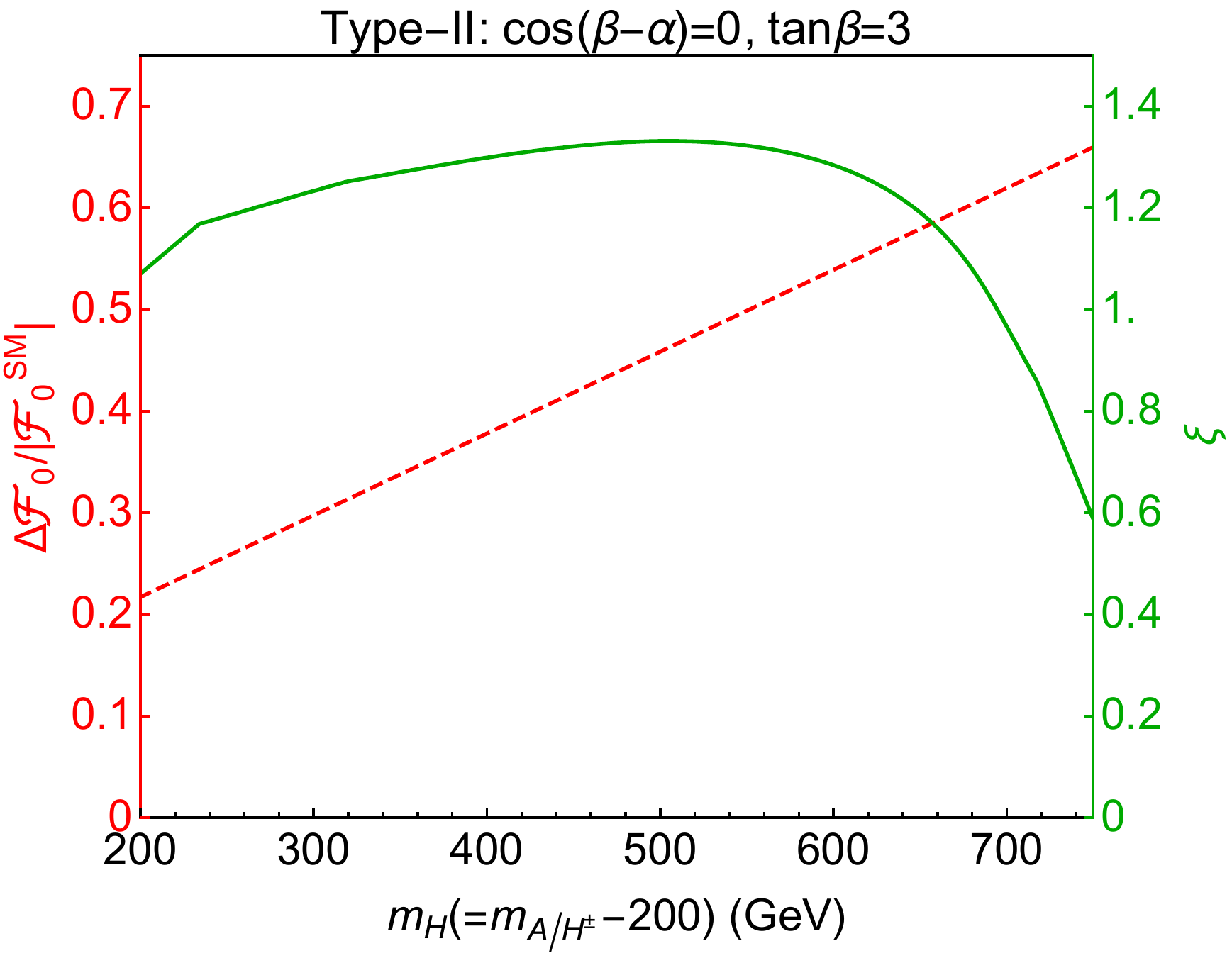} 
  \caption{Zero temperature Higgs potential and its connection with the electroweak phase transition.  (left): The zero temperature Higgs potential along the $\phi_h \equiv \cos\beta \phi_1 + \sin\beta \phi_2 $ direction with different $m_H$. (right): Vacuum energy uplifting $\Delta\mathcal{F}_0/|\mathcal{F}_0^\sm|$ and wash-out parameter $\xi_c$ as functions of $m_H$. }
  \label{fig:uplift1}
\end{figure}

The one-loop level Higgs vacuum uplifting in the alignment limit $\cos(\beta-\alpha) = 0$ has been given in Ref~\cite{Dorsch:2017nza}:
\begin{eqnarray}
\label{eq:df0}
\Delta \mathcal{F}_{0}&=& \frac{1}{64 \pi^{2}} \left[\left(m_{h}^{2}-2 M^{2}\right)^{2}\left(\frac{3}{2}+\frac{1}{2} \log \left[\frac{4 m_{A} m_{H} m_{H^{\pm}}^{2}}{\left(m_{h}^{2}-2 M^{2}\right)^{2}}\right]\right)\right.\nonumber \\
& &\left.+\frac{1}{2}\left(m_{A}^{4}+m_{H}^{4}+2 m_{H^{\pm}}^{4}\right)+\left(m_{h}^{2}-2 M^{2}\right)\left(m_{A}^{2}+m_{H}^{2}+2 m_{H^{\pm}}^{2}\right)\right]
\end{eqnarray}

\begin{figure}[ht]
  \centering
  \includegraphics[width=0.47 \linewidth]{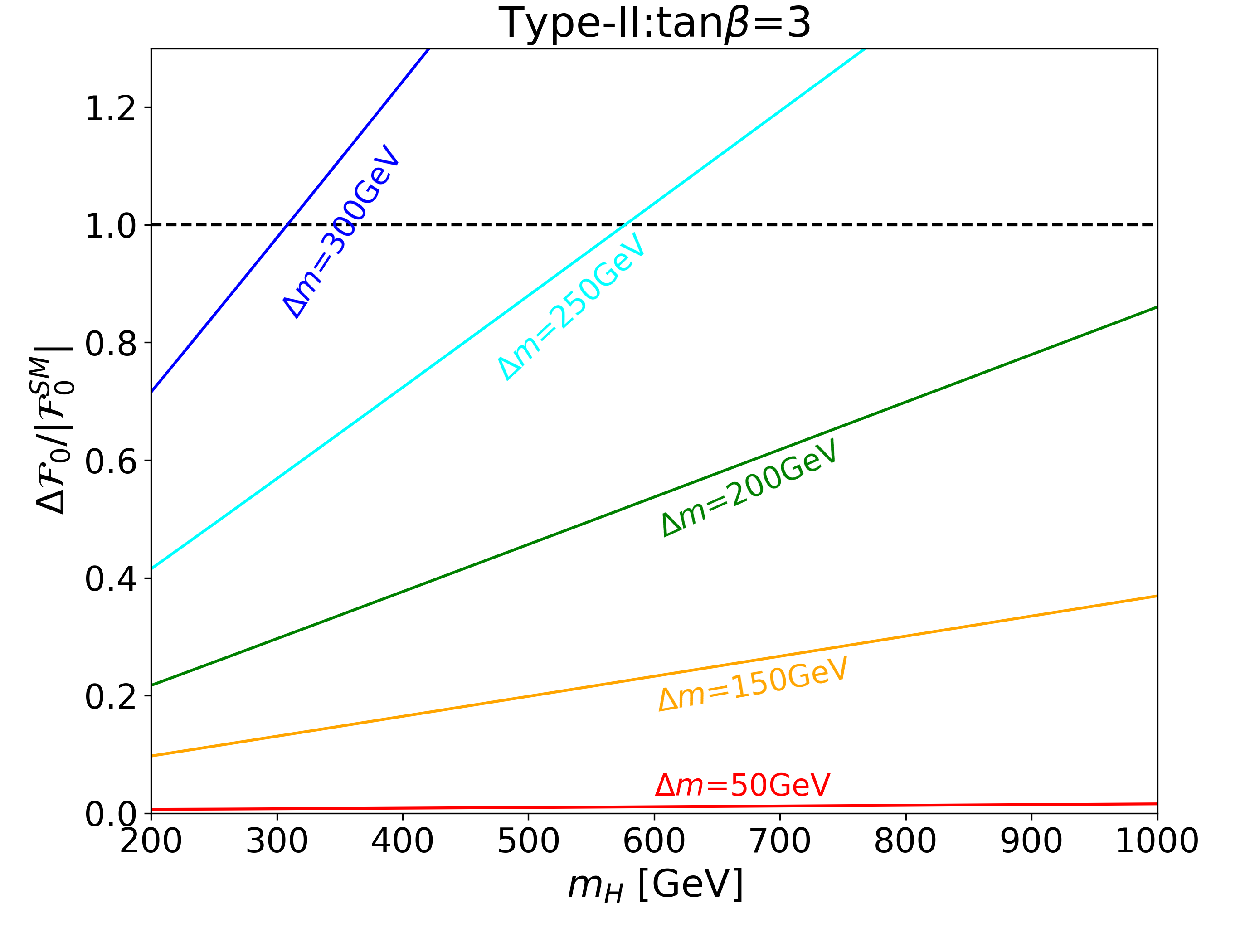}
  \includegraphics[width=0.47 \linewidth]{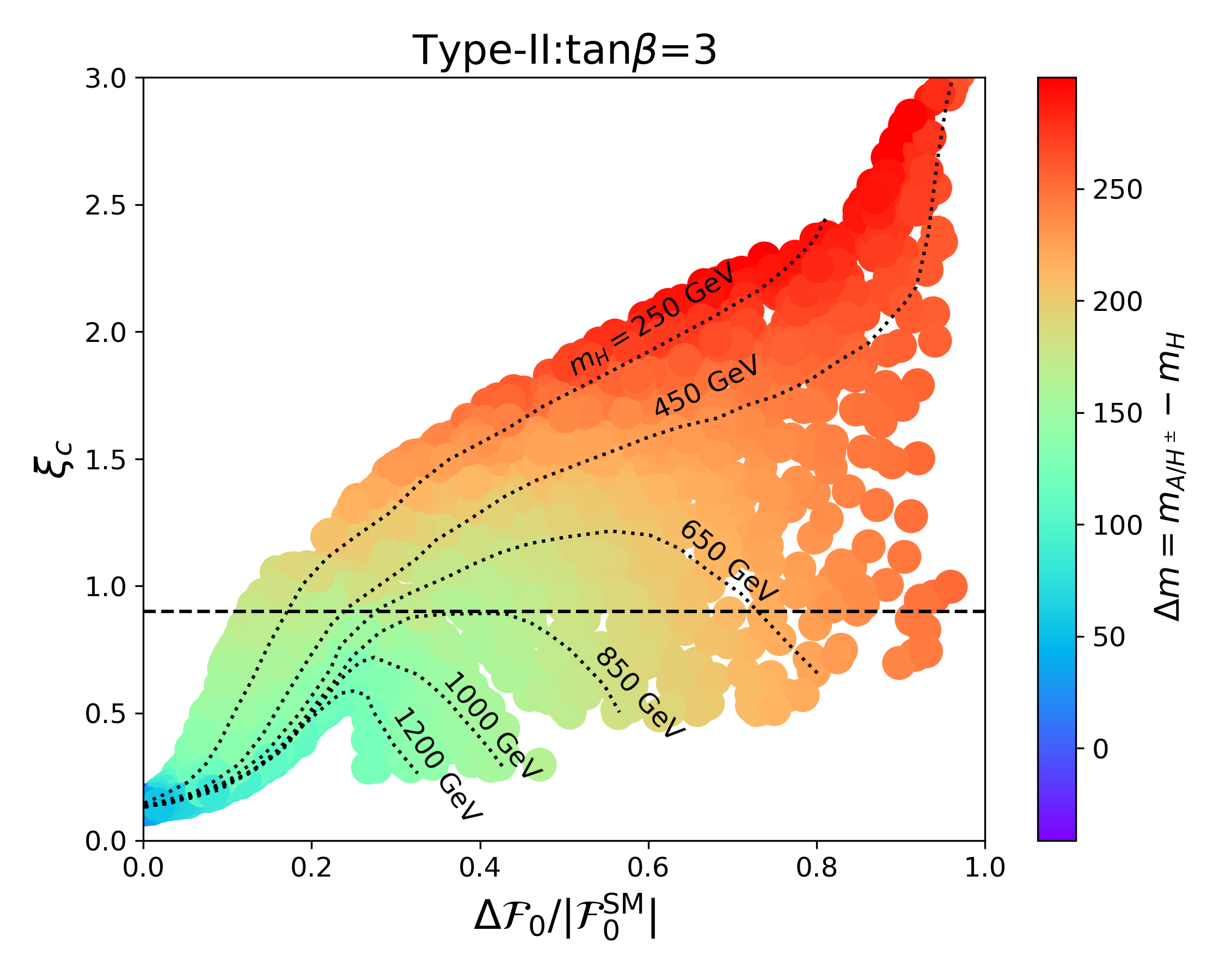}
  \caption{ (left): $\Delta\mathcal{F}_0/|\mathcal{F}_0^\sm|$ as functions of $m_H$, with different mass splitting. (right): Scan result projected in the $\Delta\mathcal{F}_0/|\mathcal{F}_0^\sm|$ - $\xi_c$ plane. Points with different mass splitting are tagged by different colors, and the dotted lines mark the CP-even Higgs mass $m_H$. }
  \label{fig:uplift2}
\end{figure}

To illustrate the idea underlying Higgs vacuum uplifting, here we display the whole shape of the zero temperature Higgs potential. 
In the left panel of \autoref{fig:uplift1}, we present the zero temperature Higgs potential along the $\phi_h \equiv \cos\beta \phi_1 + \sin\beta \phi_2 $ direction, with $m_H = 200$, 400, 600, 800 GeV represented by red, orange, green and blue lines respectively . The SM Higgs potential is also shown with black dashed line for comparison. 
It is clear that as $m_H$ increases, the height of the minimum point of the Higgs potential continues to rise, and the shape of the Higgs potential becomes shallower. For large $m_H$, $\mathcal{F}_0> 0 $, generating an unstable vacuum. 
Thus for a stable vacuum $m_H$ cannot be too large.

To find the relationship between $\xi_c$ and $\Delta\mathcal{F}_0/|\mathcal{F}_0^\sm|$, in the right panel of~\autoref{fig:uplift1} we present both $\Delta\mathcal{F}_0/|\mathcal{F}_0^\sm|$ and  $\xi_c$ as functions of $m_H$. 
In the plot, the left y axis is for $\Delta\mathcal{F}_0/|\mathcal{F}_0^\sm|$ with the red dashed line representing the relationship with $m_H$.
While  $\xi_c$, the right y axis, is shown by the solid green line. Here $\xi_c$ is calculated numerically from the package $\texttt{BSMPT}$.

%
As the red dashed line, it is clear that there is a linear relationship between $\Delta\mathcal{F}_0/|\mathcal{F}_0^\sm|$ and $m_H$, similar as the left panel. But as the green line shows, $\xi_c$ is not monotonically dependent on $m_H$ and gets the maximum value around $m_H = 500$ GeV. 
This result can be understood by our high temperature expansion analysis.
Generally as $m_H$, equal to $M^2$ in our scenario to meet theoretical constraints, becomes too large, the non-SM Higgs mass is dominated by $M^2$ and $E_{(H/A/H^\pm)}$ get smaller as \autoref{eq:expression_E}. Thus the phase transition strength becomes weaker as $m_H$ increases from \autoref{eq:PT_DE} and \autoref{eq:simple_V} . 

Since $\Delta\mathcal{F}_0/|\mathcal{F}_0^\sm|$ always gets larger when $m_H$ grows, while $\xi_c$ gets larger at first ($m_H<500\gev$ here), and then gets smaller, we can conclude $\Delta\mathcal{F}_0/|\mathcal{F}_0^\sm|$ is not monotonically correlated with $\xi_c$. This conclusion is different to the previous study~\cite{Dorsch:2013wja}. 

%

In order to get a more robust relationship between $\Delta\mathcal{F}_0/|\mathcal{F}_0^\sm|$ and $\xi_c$, as well as exploring the mass splitting effects $\Delta m=m_{A/H^\pm}-m_H$, we extend the benchmark case by including different mass splittings between the non-SM Higgs bosons:
\begin{eqnarray}
\label{eq:BM2}
\tan\beta&&= 3.0,\ \cos(\beta-\alpha) = 0, \ m_H \in (200, 1000)\gev \ , \\\nonumber 
& & \lambvs=0,\ \Delta m=m_{A/H^{\pm}} - m_H \in (0, 300)\gev .
\end{eqnarray}
The reason for us to consider different mass splittings is that the mass splitting between different non-SM Higgs bosons is roughly proportional to the size of the couplings $\lambda_i$.   
Generally speaking, the greater the couplings $\lambda_i$, the easier it is for the non-SM Higgs bosons to change the shape of the Higgs thermal potential from~\autoref{eq:expression_E} and \autoref{eq:pt_E}. 
%
However, as can be seen from \autoref{eq:df0}, a large mass splitting tends to be more limited by vacuum stability considerations, since too large mass splitting and vacuum uplifting $\Delta\mathcal{F}_0$ can result to $ \Delta\mathcal{F}_0/|\mathcal{F}_0^\sm| > 1$.
In the left panel of~\autoref{fig:uplift2}, we present $\Delta\mathcal{F}_0/|\mathcal{F}_0^\sm|$ as a function of $m_H$ under different mass splittings $\Delta m=m_{A/H^\pm}-m_H=50$ (red), 150 (orange), 200 (green), 250 (cyan), and 300 (blue) GeV. It is clear that the curves with the largest mass splittings quickly reach the unstable limit $\Delta\mathcal{F}_0=|\mathcal{F}_0^\sm|$ as $m_H$ increases. 
In the right panel of~\autoref{fig:uplift2}, we present our scan results in the plane of $\Delta\mathcal{F}_0/|\mathcal{F}_0^\sm|$ - $\xi_c$. Points with different mass splittings are tagged by different colors, with $m_H$ indicated by black dotted lines. 

To understand our scan results, we need to invoke the analysis we performed in the last subsection.
In our scenario, we have the following relationships between different parameters:

\begin{eqnarray}
 \lambda_{A/H^{\pm}} v^2 &=& (\Delta m)^2 + 2 m_H \Delta m 
 \label{eq:pt_E}
\end{eqnarray}
with $\Delta m=m_{A/H^{\pm}}-m_H, m_H^2=M^2$.
Thus, following the discussion we presented in the last subsection, if the value of $\Delta m$ is fixed and $m_H$ is not too large, the phase transition strength will increase as $m_H$ and $\Delta\mathcal{F}_0/|\mathcal{F}_0^\sm|$ increase.
But if $m_H$ becomes too large and dominates $m_{A/H^{\pm}}$, the phase transition strength will decrease as $m_H$ increases, until the vacuum becomes unstable, i.e. $\Delta\mathcal{F}_0 = |\mathcal{F}_0^\sm|$. 
In the right panel of \autoref{fig:uplift2}, we therefore observe that $\xi_c$ first rises as $\Delta\mathcal{F}_0 = |\mathcal{F}_0^\sm|$ increases (equivalent to $m_H$ increasing), and then $\xi_c$ decreases as $\Delta\mathcal{F}_0 = |\mathcal{F}_0^\sm|$ (and $m_H$) continues to increase. 

For the right panel of \autoref{fig:uplift2}, depending on the mass splitting and the phase transition features, we can divide the parameter space into three regions:
\begin{enumerate}
\item The small mass splitting region, with mass splitting $\Delta m < $ 160GeV. In this case, the Higgs vacuum energy cannot be uplifted too high, which means that these points are safe from vacuum stability bounds, and $m_H$ can vary from 200GeV to 1TeV. 
Due to the small value of $\Delta m $, however,  $\lambda_{A/H^{\pm}} v^2$ is too small to satisfy the requirement of a SFOEWPT. 
\item The medium mass splitting region, with mass splitting $\Delta m \in [\text{160GeV},\text{230GeV}]$. 
In this case, most of the parameter space is still safe from the vacuum stability constraint. When $m_H$ is not too large, it helps to enhance the phase transition strength. 
As $m_H$ grows to dominate the mass expression of $m_{A/H^{\pm}}$, $\xi_c$ rapidly decreases. 
We also observe that $\xi_c$ rises firstly and then falls as $\Delta\mathcal{F}_0/|\mathcal{F}_0^\sm|$ increases. 
The middle region of $\Delta\mathcal{F}_0/|\mathcal{F}_0^\sm|$ is favored by the existence of a SFOEWPT. 
\item The large mass splitting region, with mass splitting $\Delta m > $ 230GeV. In this region $\Delta\mathcal{F}_0/|\mathcal{F}_0^\sm|$ starts from a value that is larger than 0.4, and quickly touches the vacuum stability bound $\Delta\mathcal{F}_0/|\mathcal{F}_0^\sm|=1$ as $m_H$ increases. 
This means that, before $m_H$ has increased to be able to dominate $m_{A/H^{\pm}}$, the vacuum is already unstable. We therefore observe that $\xi_c$ increases nearly monotonically as $\Delta\mathcal{F}_0/|\mathcal{F}_0^\sm|$ increasing. 
\end{enumerate}

Through the above discussion, it is clear that the upper limits on the non-SM Higgs boson masses in the 2HDM come from vacuum stability (when the mass splitting is large), or the SFOEWPT requirement (when the mass splitting is medium-large). 
Without the SFOEWPT requirement, the non-SM Higgs bosons can be arbitrarily heavy without violating vacuum stability, provided the mass splitting between them is small enough.

On the other hand, the black dotted lines in the right panel of \autoref{fig:uplift2} clearly show the relationship between $\xi_c$ and $m_H$. 
We found that $\xi_c$ is a monotonically increasing function of $\Delta\mathcal{F}_0/|\mathcal{F}_0^\sm|$ when $m_H<500\gev$ 
, such as $m_H=250, 450 \gev$ in the right panel of \autoref{fig:uplift2}. But with larger mass, the phase transition strength $\xi_c$ gets smaller, and when $m_H > 850$ GeV, $\xi_c$  can no longer reach 0.9.
To avoid the unstable vacuum, larger $m_{H}$ needs smaller $\Delta m$ as in~\autoref{fig:uplift1}. Therefore, a too large $m_{H}$ will result in a too small $\lambda_{A/H^{\pm}} v^2$, which could not generate a SFOEWPT. 
In \autoref{tab:DF0_xi_mH} we present the range of $\Delta m$ and $\Delta\mathcal{F}_0/|\mathcal{F}_0^\sm|$ for different values of $m_H$. 
This clearly shows that the SFOEWPT-satisfied region keeps shrinking as $m_H$ gets larger and larger.


%
\begin{table}[h]
\begin{center}
\begin{tabular}{|c|c|c|c|c|c|}
\hline
$m_H (\gev$) & 250 & 450 & 650 & 700 & 850 \\\hline
$\Delta m (\gev$) & (170, 280)& (160, 280)& (150, 230)& (155, 210) & (160, 165)\\ \hline
$\Delta\mathcal{F}_0/|\mathcal{F}_0^\sm| $& (0.18, 0.83) 
& (0.25, 0.95) & (0.28, 0.73) & (0.3, 0.7) & (0.38,0.42)
\\ \hline
\end{tabular}
\end{center}
\caption{The parameter space for which a SFOEWPT is obtained, with $\tanb=3, \Delta m=m_{A/H^\pm}-m_H$ in the Type-II 2HDM, similar to \autoref{fig:uplift2}.}
\label{tab:DF0_xi_mH}
\end{table}

\subsection{Case1: alignment limit with  fixed  mass  splitting}

\begin{figure}[t]
\label{fig:case1} 
\centering
\includegraphics[width=0.63\textwidth]{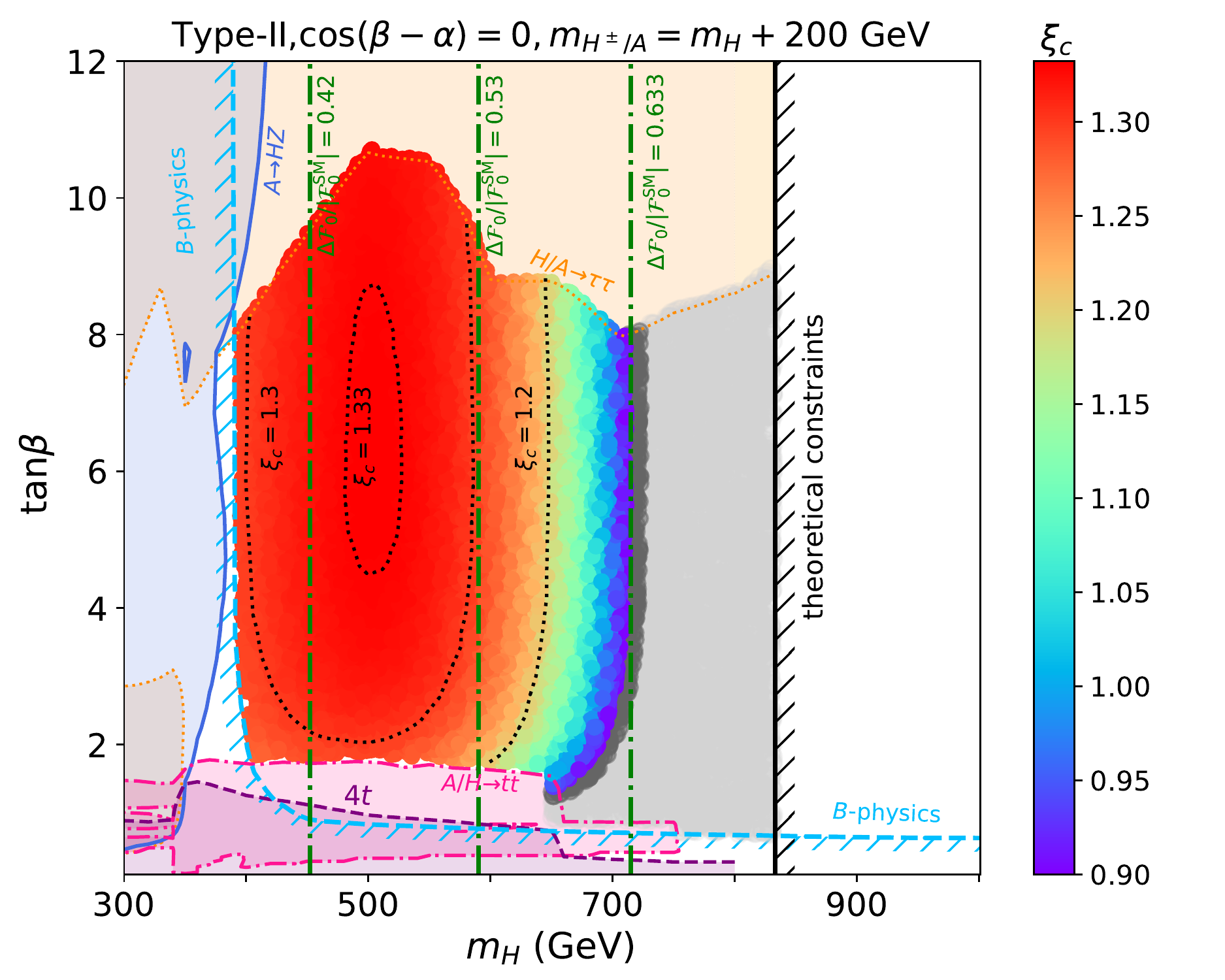}
\caption{Electroweak phase transition and other constraints analyzed in the plane of $m_H-\tanb$ for the Type-II 2HDD. Here we fix $\Delta m=m_{A/H^\pm} - m_H=200 \gev$ and assume the tree-level alignment limit $\cosba=0$, which is same as the right panel of~\autoref{fig:LHC}. The central colored region is allowed after all constraints. The null results of $H/A \to \tau\tau$ searches (orange region with dotted line boundary) provide an upper bound on $\tanb$ for a given $m_H$, the $A\to HZ$ results (blue region) constrain the small mass region, and the $A/H\to tt$ (red region with  dash-dotted line boundary) and $4t$ (purple region with dashed line boundary) channels constrain the small $\tanb$ region. The hatched cyan dashed line show the constraints from $B$ physics observables, and the hatched black line indicates the theoretical constraints. The gradient-filled regions show the parameter space that can generate a SFOEWPT, with the black dashed lines indicating the phase transition strength $\xi_c$. The black region means there is an electroweak phase transition with $\xi_c<0.9$. The grey regions are allowed by various constraints,but there is no first order phase transition. We also show $\Delta \mathcal{F}_0/|\mathcal{F}_0^\sm|$ by green dash-dotted lines.
}
\end{figure}

Following the previous approximate analysis of the electroweak phase transition, we now investigate a series of benchmark cases, starting with,
\begin{eqnarray}
\label{eq:case1}
&&\tan\beta \in (0.2, 50), \ m_H \in (200, 1000) \gev \ , \nonumber \\ 
 \cos(\beta-\alpha)&&= 0, \lambvs=0,\ \Delta m=m_{A/H^{\pm}} - m_H = 200\gev .
\end{eqnarray}
Here we take $\lambvs=0$ to allow for the largest range of $\tanb$ as shown in the second and third panel of~\autoref{fig:theory}. To explore the dependence on $m_H$, we fix the mass splitting $\Delta m=200 \gev$, and assume the tree-level alignment limit $\cosba=0$. The parameter space is the same as the right panel of~\autoref{fig:LHC}, where there are important constraints from direct non-SM Higgs boson searches at LHC Run-II including $H/A \to \tau\tau$ (orange region with dotted line boundary, providing an upper boound), $A\to HZ$ (blue region, constraining the small mass region), and $A/H\to tt$ (red region with  dash-dotted line boundary) and $4t$ (purple region with dashed line boundary), which constrain the small $\tanb$ region. 
For the Type-II 2HDM, there are important constraints on the mass of the charged Higgs boson from $B$ physics, which are represented by the hatched cyan dashed line. $B$ physics observables also give effective constraints at small $\tanb$.  The hatched black line indicates the theoretical constraints, as discussed in~\autoref{fig:theory}, requiring $m_H<835 \gev$ for $\cosba=0$.

After these theoretical and experimental constraints, the allowed parameter region is approximately located around $m_H \in (380, 830)\gev$, and $\tanb\in(1,10)$.
The colored region $m_H \in (380, 700)\gev$ shows the parameter space which can generate a SFOEWPT, with dashed lines indicating the phase transition strength $\xi_c$. We can see that, generally, the strength $\xi_c$ gets its maximal value around $m_H=500\gev$, which is discussed in the right panel of \autoref{fig:uplift1}. The green dash-dotted lines show $\Delta \mathcal{F}_0/|\mathcal{F}_0^\sm| = 0.42, 0.53, 0.63$, which grows with larger $m_H$ and is independent of $\tanb$. We can therefore again conclude that the SFOEWPT strength is not monotonically dependent on $m_H$ or $\Delta \mathcal{F}_0/|\mathcal{F}_0^\sm|$. 

Finally there is a black band region round $m_H=700\gev$, which means that the phase transition strength $\xi_c<0.9$.  Beside the black band region, there is a grey region which is allowed by various constraints, but $\xi_c$ in this region has no value. This is because the phase transition in this region is not first order, and thus we can not find the critical temperature and calculate $\xi_c$. 
We have also checked that Higgs and Z-pole precision measurements give no constraints in this case since $\cosba=0$ and $\Delta m=200 \gev$.

\begin{figure}[b]
\label{fig:case2} 
\centering
\includegraphics[width=0.63\textwidth]{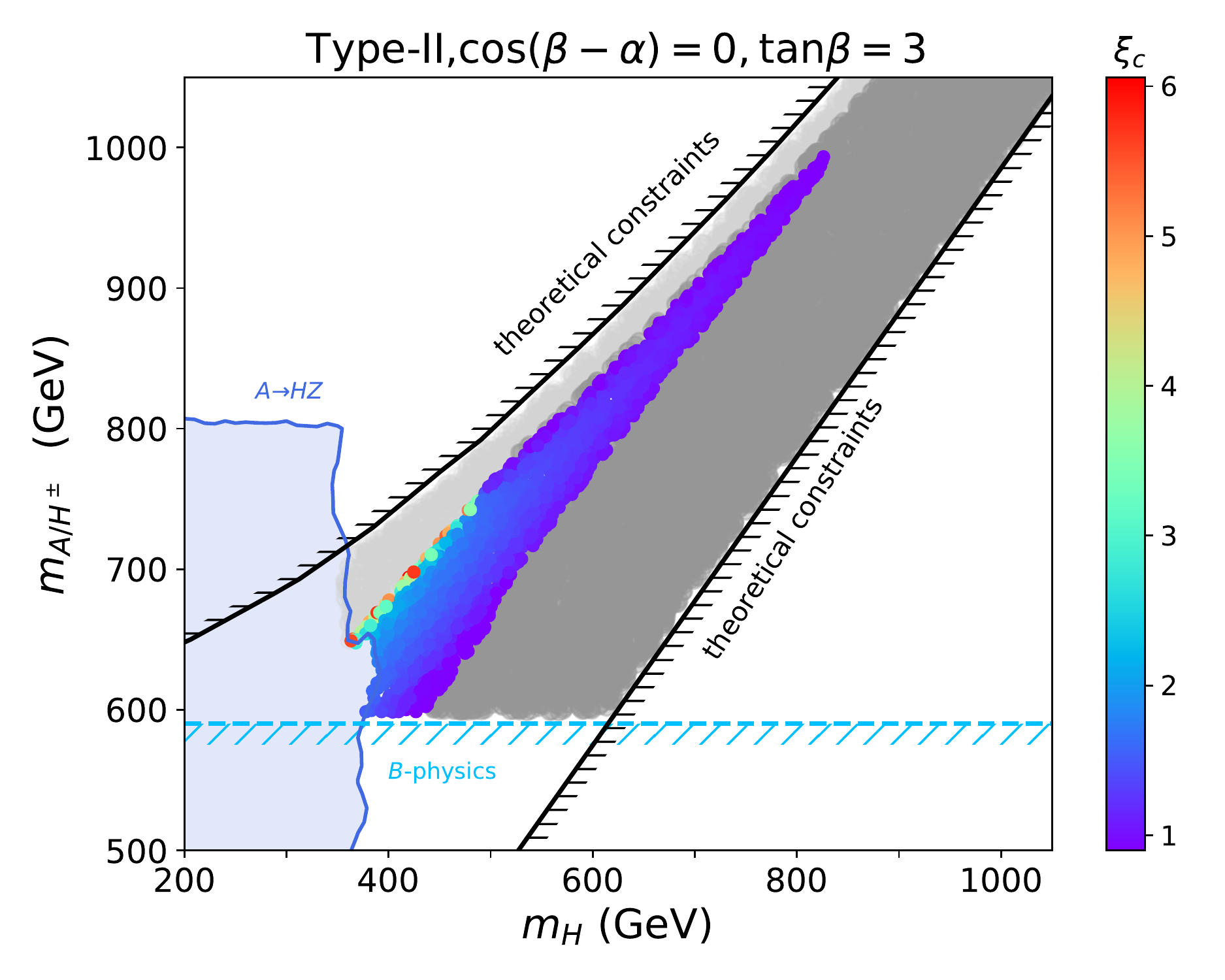}
\caption{Electroweak phase transition and other constraints analyzed in the plane of $m_H-m_{A/H^\pm}$, with $\cosba=0$ and $\tanb=3$, for the Type-II 2HDM. Of the various heavy Higgs search channels, only $A\to HZ$ gives visible constraints, shown by the blue region. Again the hatched cyan dashed line shows the $B$-physics constraints, and the hatched black lines are for theoretical constraints. The allowed regions are divided into three parts, the colorful region with $\xi_c>0.9$, the light grey region (mostly above the colorful region) with $\xi_c<0.9$, and the dark grey region in which a SFOEWPT cannot occur.}
\end{figure}
\subsection{Case2: alignment limit with  $m_A=m_{H^\pm}$}
Based on the results in~\autoref{fig:uplift2}, here we show our second benchmark  case, the alignment limit with fixed $\tanb$,
\begin{eqnarray}
\label{eq:case2}
&& m_{A/H^{\pm}} \in (500, 1200) \gev, \ m_H \in (200, 1000) \gev \ , \nonumber \\ 
 &&\cos(\beta-\alpha)= 0, \lambvs=0,\ \tan\beta=3 .
\end{eqnarray}

Again here $\lambvs=0$ is set to avoid the constraints on the parameter $\tanb$. In \autoref{fig:case2}, we show the constraints arising from the requirement of a SFOEWPT and other observables in the $m_H-m_{A/H^\pm}$ plane of the Type-II 2HDM. For the various heavy Higgs search channels, only $A\to HZ$ gives a visible constraint (shown by the blue region), which can exclude the region with $m_H<350 \gev, m_{A/H^\pm}<800 \gev$. $B$-physics constraints, shown by the hatched cyan dashed line, exclude $m_{H^\pm} <580\gev$. Since here we have $m_A=m_{H^\pm}$ and $\cosba=0$, the Higgs and Z-pole precision constraints are satisfied automatically. On the other hand, the theoretical constraints, indicated by hatched black lines, give a strong limit on the mass splitting range, roughly $\Delta m = m_{A/H^\pm}-m_H \in (-50, 200) \gev$.

The allowed regions are divided into three parts, the colorful region with $\xi_c>0.9$, the light grey region which is mostly above the colorful region with $\Delta m = m_{A/H^\pm}-m_H \approx200 \gev$.) with $\xi_c<0.9$, and the dark grey region in which a phase transition cannot occur. From the colored region, we find that, both a too large or too small $\Delta m$ will not allow for a SFOEWPT.
As discussed  in~\autoref{fig:uplift2}, for a too small $\Delta m$, the Higgs vacuum energy cannot be uplifted high enough to generate a phase transition, while too large a value of $\Delta m$ will result in an unstable potential $\mathcal{F}_0=|\mathcal{F}_0^\sm|$, where the the potential at second EW minimal is higher than the it at the origin. This is also responsible for the upper limit on $m_H$, as the analysis around \autoref{eq:pt_E} shows, since too small a value of $m_H\Delta m$ cannot generate a proper barrier for a SFOEWPT. 

\begin{figure}[t]
\label{fig:case3} 
\centering
\includegraphics[width=0.63\textwidth]{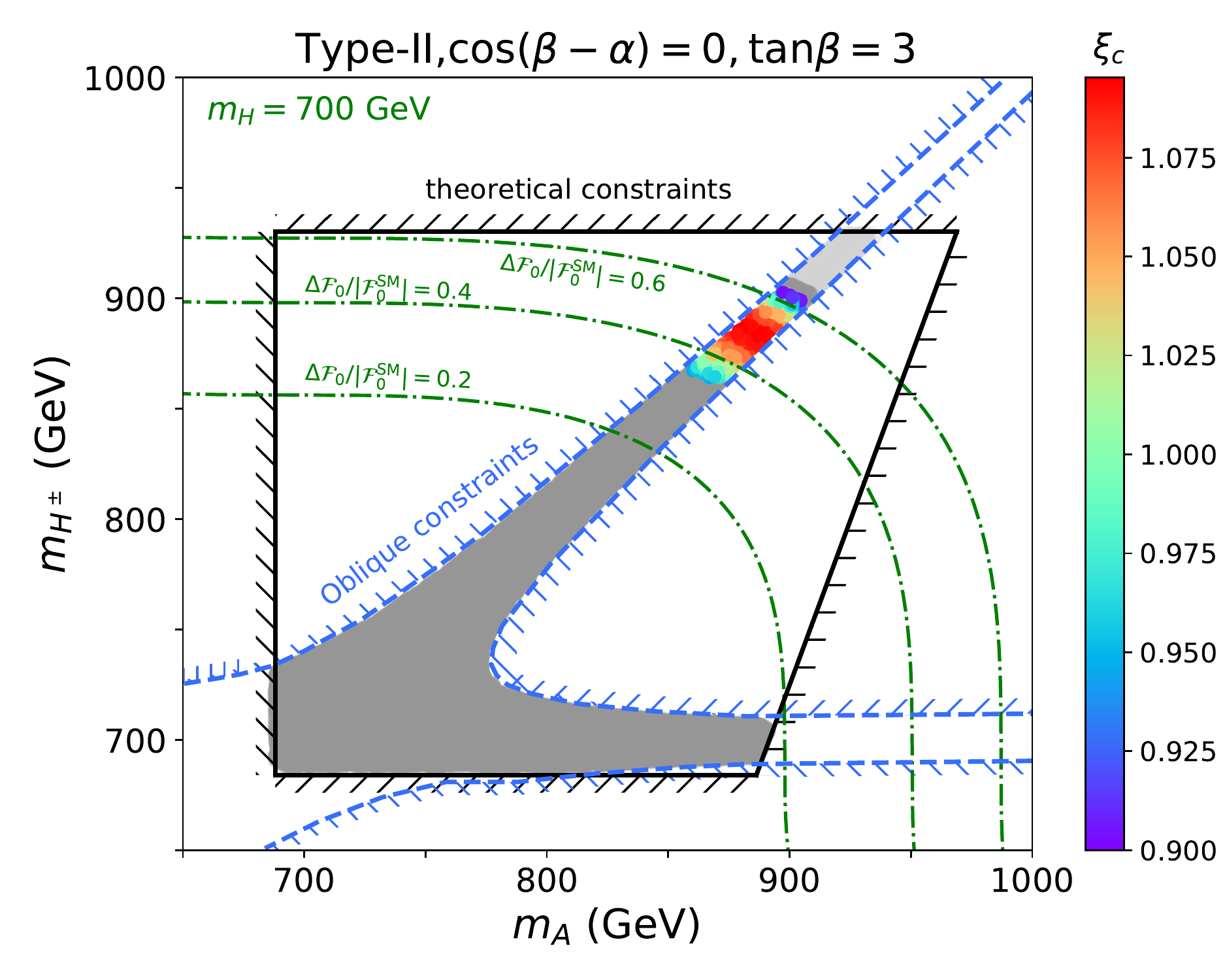}
\caption{Electroweak phase transition and other constraints analyzed in the plane of $m_A-m_{H^\pm}$, with $\cosba=0$ and $\tanb=3$, in the Type-II 2HDM. For the various new physics search channels, only the oblique constraints make an important contribution, represented by the hatched blue dashed lines. Theoretical constraints are shown by the hatched black lines. The allowed regions are divided into three parts, the colorful region with $\xi_c>0.9$, the light grey region (mostly above the colorful region) with $\xi_c<0.9$, and the dark grey region where a first order phase transition does not occur.  We also show green dash-dotted lines for $\Delta \mathcal{F}_0/|\mathcal{F}_0^\sm|$.}
\end{figure}
\subsection{Case3: alignment limit with  $m_H$=700 GeV}
\label{sec:case3}
In our previous case studies, we always had the simple assumption of $m_A=m_{H^\pm}$ to satisfy the oblique constraints from Z-pole measurements, and also to simplify the parameter space. Here to study the general mass splitting region, we take another benchmark case, 
\begin{align*}
\label{eq:case3}
m_{A} \in (500&, 1200) \gev,\ m_{H^{\pm}} \in (500, 1200) \gev , \nonumber \\ 
 m_H=700\gev&,\ \cos(\beta-\alpha)= 0,  \lambvs=0,\ \tan\beta=3 .
\end{align*}

$\lambvs=0$ is once more set to avoid the constraints on the parameter $\tanb$, and we take $m_H=700\gev$ as an example. In~\autoref{fig:case3}, we show the electroweak phase transition and other constraints in the plane of $m_A-m_{H^\pm}$ in the Type-II 2HDM. The theoretical constraints are now particularly important, as the region with hatched black lines acts as a boundary on the allowed parameter space. The lower limits on both $m_A$ and $m_{H^\pm}$ are approximately 670 GeV, while the upper limits are 970 and 930 GeV respectively. This is because, once there is a large mass splitting, $\lambda_{1-5}$ will be enlarged~\cite{Chen:2018shg}.
For the various new physics search channels, only the oblique constraints make an effect here. As the hatched blue dashed lines show, the allowed regions are around either $m_A=m_{H^\pm}$ or $m_{H^\pm}=m_H=700 \gev$. 

The allowed regions are divided into three parts, the colorful region with $\xi_c>0.9$ allowing a SFOEWPT, the light grey region (mostly above the colorful region) with $\xi_c<0.9$, and the dark grey region without a first order phase transition.  To understand the features here, we also have green dash-dotted lines for $\Delta \mathcal{F}_0/|\mathcal{F}_0^\sm|$, which gets large when  $m_A, m_{H^\pm}$ increases. We also note that, to get a proper vacuum energy uplifting, at least one of  $m_A$ or $ m_{H^\pm}$ should be large. For instance, $\mathcal{F}_0/|\mathcal{F}_0^\sm|=0.4$ requires $ m_{H^\pm}=900\gev$ when $m_A=m_H$, 
or $m_A=950\gev$ when $m_A=m_{H^\pm}$, or $m_A\approx m_{H^\pm} \approx  870\gev$.
The region with $\xi_c>0.9$ is located at $\mathcal{F}_0/|\mathcal{F}_0^\sm| \in (0.37,0.63)$. The large mass limit comes from $\mathcal{F}_0/|\mathcal{F}_0^\sm| \to 1$, where the vacuum is not stable, while the small mass limit comes from ~\autoref{eq:pt_E}, where there is only limited vacuum uplifting and a barrier to generating a SFOEWPT. 

\subsection{General results}
During the last section, we presented three benchmark cases to discuss the effects of the heavy Higgs masses on the existence of a SFOEWPT in the alignment limit, as well as the influence of a variety of theoretical and current experimental constraints up to the one-loop level. 

In this section, we present a more general study of Type-I and Type-II 2HDMs. At the same time, we will explore the impact of future results from Higgs factories, presented in ~\autoref{sec:cons_indrect}, taking the CEPC precision measurements as an example.

Our parameter scan regions for both Type-I and Type-II are :
\begin{eqnarray}
& & |\alpha| < \frac{\pi}{2}, \tan\beta \in (0.2, 50), \  m_A \in (10, 1500)\text{ GeV} \ , \  m_{H^{\pm}} \in(10, 1500)\text{ GeV} \ , \nonumber \\
& &   m^2_{12} \in(0,  1500^2) \text{ GeV}^2, \ m_h=125.1 \gev, m_H \in (130, 1500) \gev . 
\end{eqnarray}
We perform a random parameter scan in the above parameter region, with the total number of samples exceeding 1 billion, for both Type-I and Type-II models. 

\begin{figure}[t]
  \centering
  \includegraphics[width=0.48\linewidth]{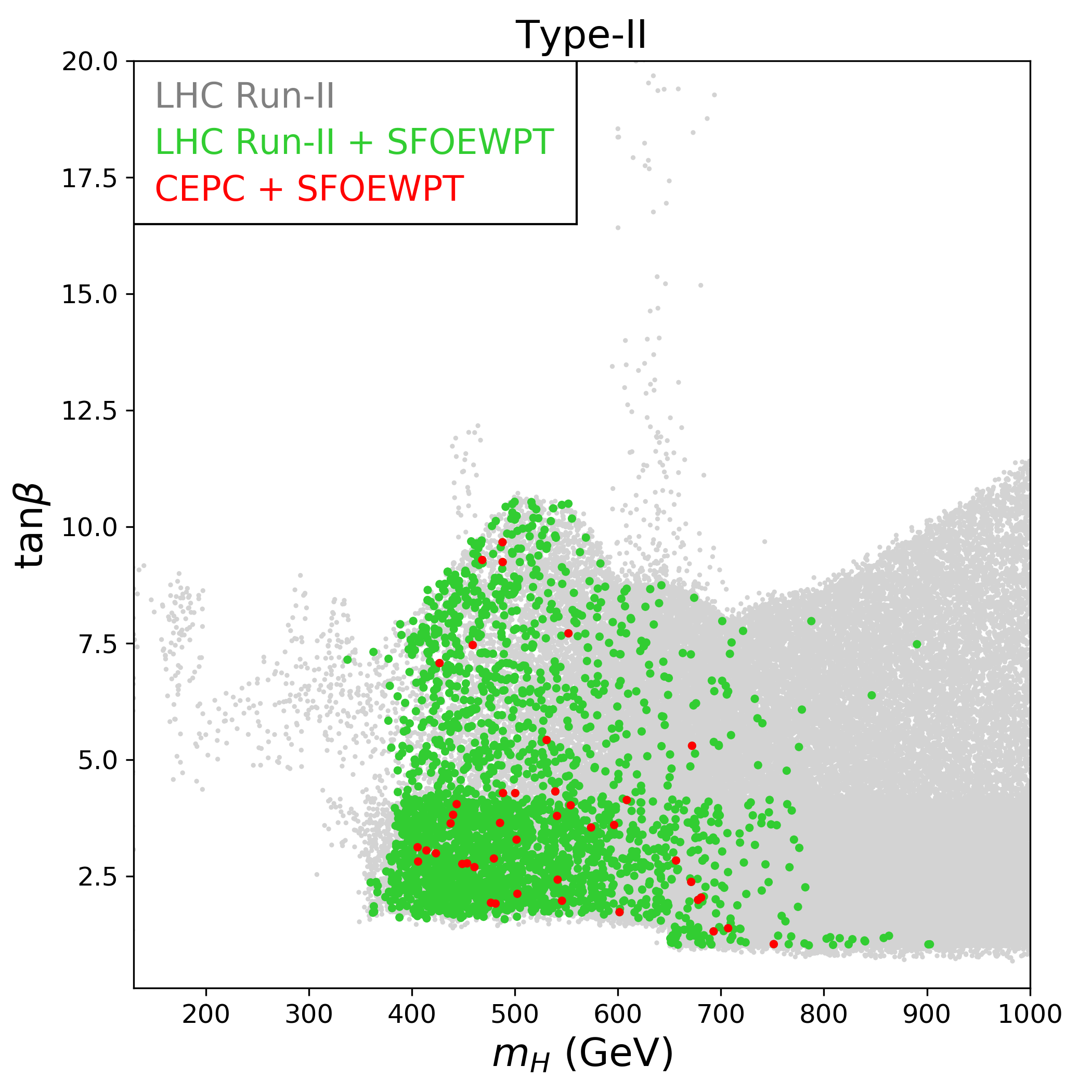}
   \includegraphics[width=0.48\linewidth]{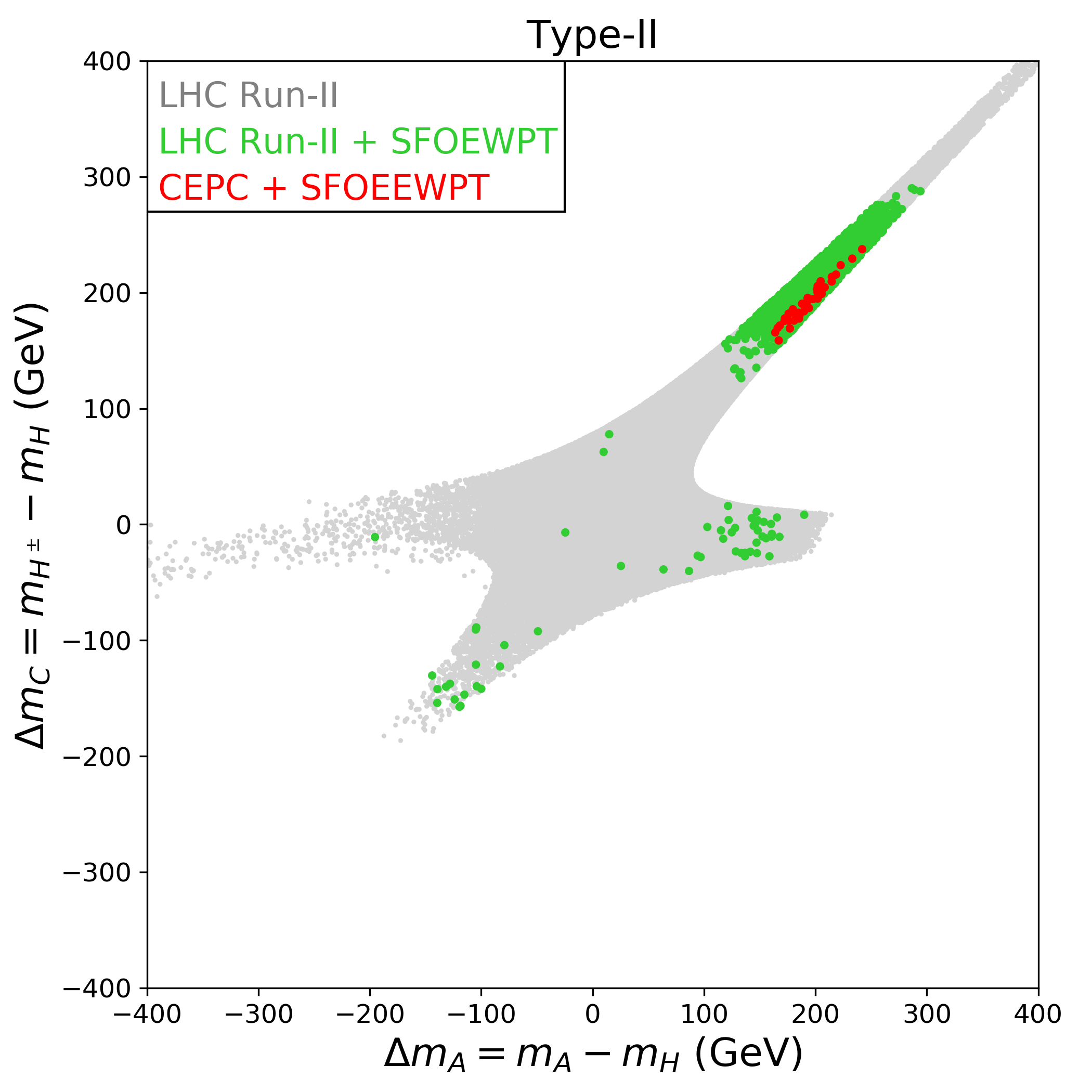}
  \caption{The allowed parameter space in the plane of $m_H - \tanb$ (left), $\Delta m_A-\Delta m_C$ (right). The grey points survive all theoretical and current experimental constraints. The green ones are able to provide a SFOEWPT, while the red ones are allowed by future precision measurements from CEPC.}
  \label{fig:t_ma_tanb}
\end{figure}

In~\autoref{fig:t_ma_tanb} we show the scan results for the Type-II 2HDM. The grey scatter points are the regions allowed by $B$ physics, theoretical constraints, heavy Higgs direct searches and SM Higgs precision measurements at the current LHC Run-II, and constraints from EW oblique operators. The green points are a subset of the grey ones, which can generate a SFOEWPT, and the red points are further required to meet the constraints from future Higgs precision measurements at CEPC.  Compared to {\bf Case 1} (\autoref{fig:case1}), which assumed the alignment limit and set $m_{H^\pm}=m_A$, here we could divide the whole allowed region into 4 classes, 

\begin{itemize}
    \item {\bf Class A:} Regions with $m_H<350$ GeV. Here the region has $m_{H^\pm}\approx m_A > m_H$, and the mass splitting is about (300,500) GeV to meet the constraint $m_{H^\pm}>580$ GeV. Generally $\lambvs \approx 0$ to allow for such a large mass splitting and $\tanb$ is within the region selected by the theoretical constraints shown in~\autoref{fig:theory}. This region can also be divided into two subgroups based on $\rm sign(\kappa_b)$. When $\rm sign(\kappa_b)=+$, $m_H<200~\gev, \tanb \in (5,10)$ can escape the constraints from the $H\to \tau\tau$ channel as in the right panel of \autoref{fig:LHC}. At the same time, the large mass splitting $m_A-m_H>450$ GeV weakens the constraint from the $A\to HZ$ channel~\cite{Kling:2020hmi}. Another subgroup is $\rm sign(\kappa_b)=-$, the so-called wrong-sign Yukawa coupling region with $\cosba\approx2/\tanb$. Here $m_H$ can reach 350 GeV, $\cosba \in(0.2, 0.4)$, and LHC direct searches require $\tanb<10$~\cite{Su:2019ibd}. Because of the large mass splitting in this region, $\Delta \mathcal{F}_0/ |\mathcal{F}_0^\sm|$ is too large to produce a stable vacuum.
    \item {\bf Class B:} Regions with $5<\tanb<12$ for $m_H\approx 450$ GeV. This region is also a wrong-sign Yukawa coupling region with $\rm sign(\kappa_b)=-$. Generally $\lambvs \approx 0$ to meet theoretical constraints, and  $m_{H^\pm}\approx m_A = m_H+140 $ GeV with $\tanb<12$ to meet constraints from the $A\to HZ$ and $A/H\to \tau\tau$ channels (see \autoref{fig:LHC}). $m_{A/H^\pm}-m_H>140$ GeV to meet $B$ physics constraints, while a larger mass splitting is not allowed by theoretical constraints even though $\lambvs \approx 0$. As the right panel of~\autoref{fig:uplift2} shows, because $m_H=450 \gev$ and $\Delta m=140 \gev$, $\Delta \mathcal{F}_0/ |\mathcal{F}_0^\sm| <0.2$ is too small, thus the vacuum uplifting is too small, and there is no SFOEWPT here.
    %
    \item {\bf Class C:} Regions with $5<\tanb<45$ for $600< m_H< 700$ GeV. Here $m_{H^\pm}=m_H$ with $m_A< m_h=125$ GeV. Again it is a wrong-sign Yukawa coupling region with $\lambvs \approx 0$. The lower limit of $m_H$ comes from $B$ physics and EW oblique constraints, and the upper limit comes from theoretical constraints $m_H-m_A<650$ GeV. In the region, $\Delta \mathcal{F}_0/ |\mathcal{F}_0^\sm| <0$, thus there is no chance to generate a SFOEWPT. 
    \item {\bf Class D:} The main allowed region with $m_H>350$ GeV. The region is similar to the white allowed region in the right panel of \autoref{fig:LHC}. Compared to {\bf Case 1} with $m_{H^\pm}=m_A=m_H=200$ GeV in the alignment limit, here the allowed grey region by current LHC Run-II has no upper limit on $m_H$ anymore from theoretical constraints when all parameters are free. When $m_H<900$ GeV, $1<\tanb<10$ is required to satisfy the constraints from $H/A \to \tau\tau$, top searches and $B$ physics. When $m_H>900$ GeV, $\tanb$ can take a larger value as the constraining power of the $H/A \to \tau\tau$ channel gets weaker. In this region, there are a number of points with $\xi_c>0.9$, as shown by green points. We can see the green parameter space has an upper limit of about 900 GeV. For points that also satisfy CEPC constraints as the red points, the parameter space has an upper limit of about 800 GeV. 
\end{itemize}
\begin{figure}[ht]
  \centering
  \includegraphics[width=0.48\linewidth]{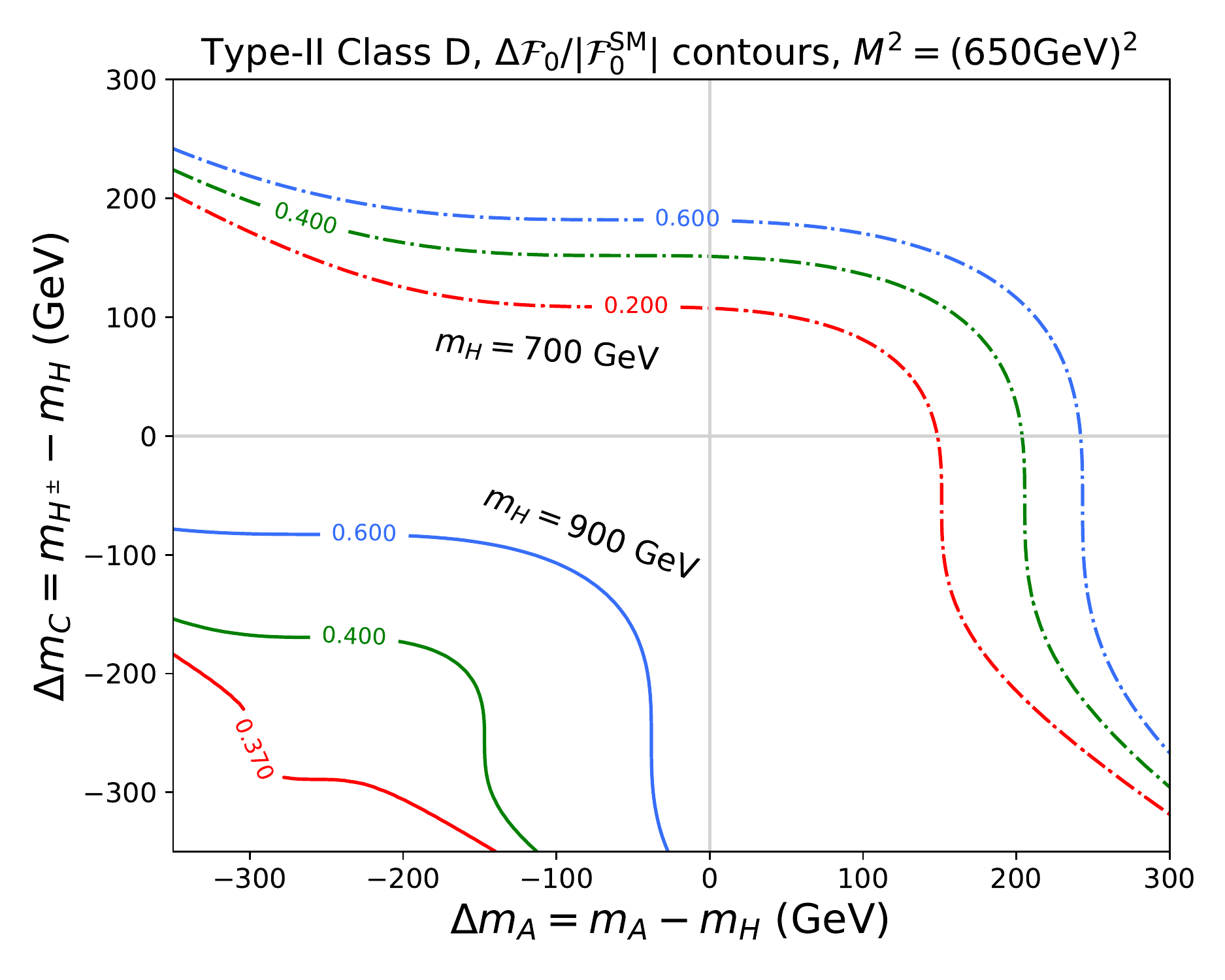}
  \caption{$\Delta \mathcal{F}_0|\mathcal{F}_0^\sm|$ contours in the plane of $\Delta m_A- m_C$ to explore {\bf Class D}. Here $M^2=\frac{m_{12}^2}{\sin\beta\cos\beta} = (650\gev)^2$, $m_H=$ 900 GeV (solid lines), 700 GeV (dashed lines). }
  \label{fig:t2_dmac_ana}
\end{figure}

The right panel of \autoref{fig:t_ma_tanb}  shows the scan results in the plane of $\Delta m_A= m_A-m_H$ and $\Delta m_C= m_A-m_{H^\pm}$, allowing us to analyze the {\bf Class D} parameter space. Here the general structure is $\Delta m_C \approx \Delta m_A$ or $\Delta m_C =0$ because of Z-pole oblique constraints. 

For the green points from {\bf Class D} that satisfy LHC Run-II constraints whilst producing a SFOEWPT, there are mainly three regions. For {\bf Class D1}, $\Delta m_C \approx \Delta m_A \in (100,350) \gev$. The region has $m_H\in (350,600)
\gev$, $\lambvs \approx 0$, $\tanb \in (1,5)$, and the features are similar to {\bf Case 3} results in \autoref{sec:case3}. For {\bf Class D2}, $\Delta m_C \approx \Delta m_A \in (-200,-50) \gev$. The region has  $m_H\in (750,900)
\gev$, $M^2=\frac{m_{12}^2}{\sin\beta\cos\beta} \approx (650\gev)^2$, $\tanb \approx 1$. In the left panel of \autoref{fig:t2_dmac_ana}, we show the $\Delta \mathcal{F}_0|\mathcal{F}_0^\sm|$ contours in the plane of $\Delta m_A- m_C$. We can see that when $m_H=900\gev$ and $M=650\gev$, $\Delta m_C \approx \Delta m_A \in (-200,-50) \gev$, which  results in $\Delta \mathcal{F}_0|\mathcal{F}_0^\sm| \in (0.37,0.6)$. This is one of the essential conditions for a SFOEWPT. In this region, $\lambvs \in (500, 600)\gev$, thus theoretical constraints impose $\tanb\approx 1$ as shown in \autoref{fig:theory}. 
For {\bf Class D3}, $\Delta m_C \approx 0, \Delta m_A \in (50,200) \gev$. The region has  $m_H\in (650,750)
\gev$, $M^2=\frac{m_{12}^2}{\sin\beta\cos\beta} \approx (650\gev)^2$, $\tanb \approx 1$,$\lambvs \in (450, 550)\gev$. Similarly in the left panel of \autoref{fig:t2_dmac_ana}, we show the $\Delta \mathcal{F}_0|\mathcal{F}_0^\sm|$ contours for $m_H=700\gev$  with dash-dotted lines.
{\bf Class D2} and {\bf Class D3}, which are allowed by current LHC indirect Higgs precision measurements and direct heavy Higgs searches, will be excluded by Higgs precision observables at the CEPC. This is because the large $\lambvs$ in the two regions will lead to large one-loop level corrections to the SM-like Higgs couplings~\cite{Gu:2017ckc,Chen:2018shg,Chen:2019pkq}, and large mass splittings around $\tanb=1$ are not allowed by precise measurements of the Higgs couplings.

\begin{figure}[ht]
  \centering
  \includegraphics[width=0.48\linewidth]{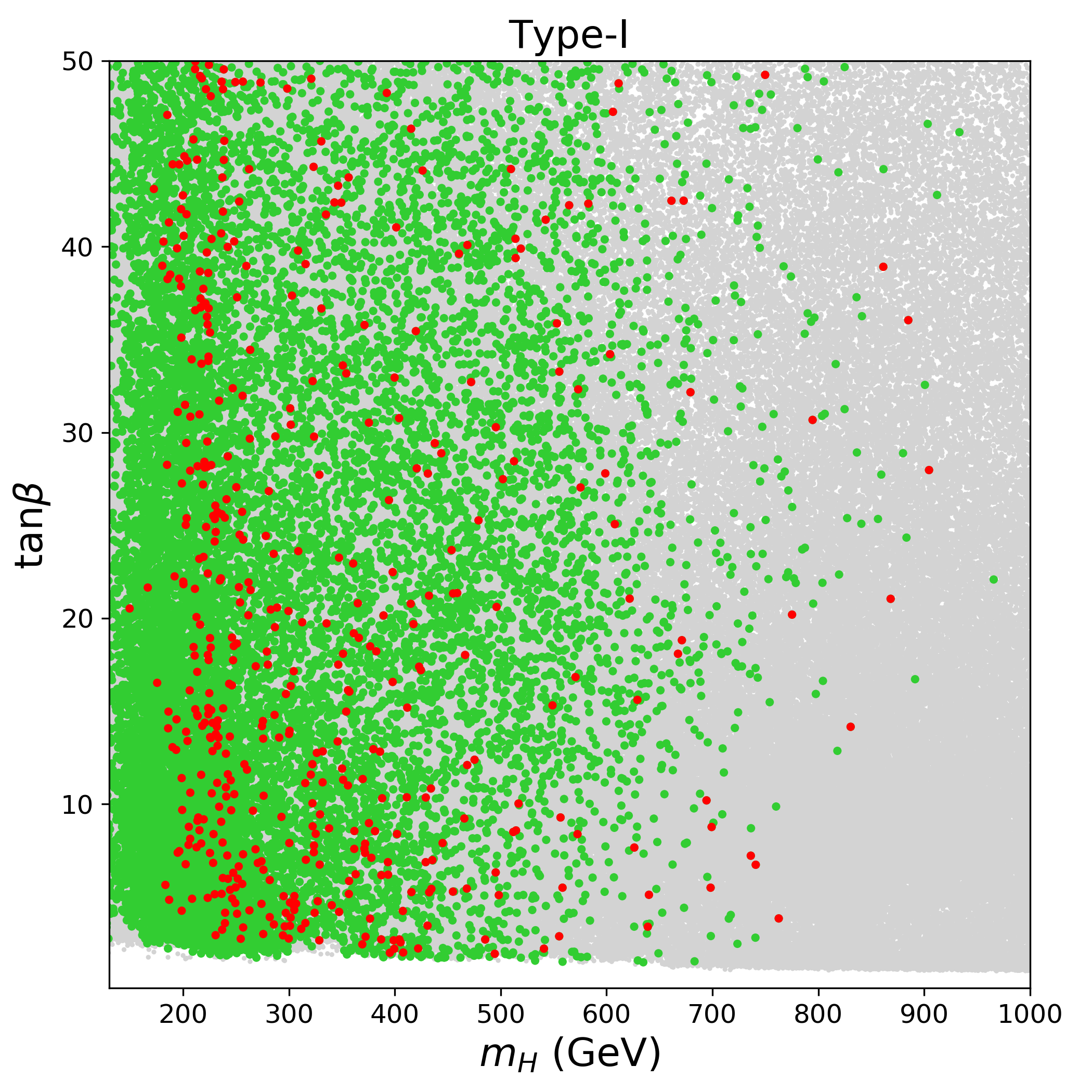}
   \includegraphics[width=0.48\linewidth]{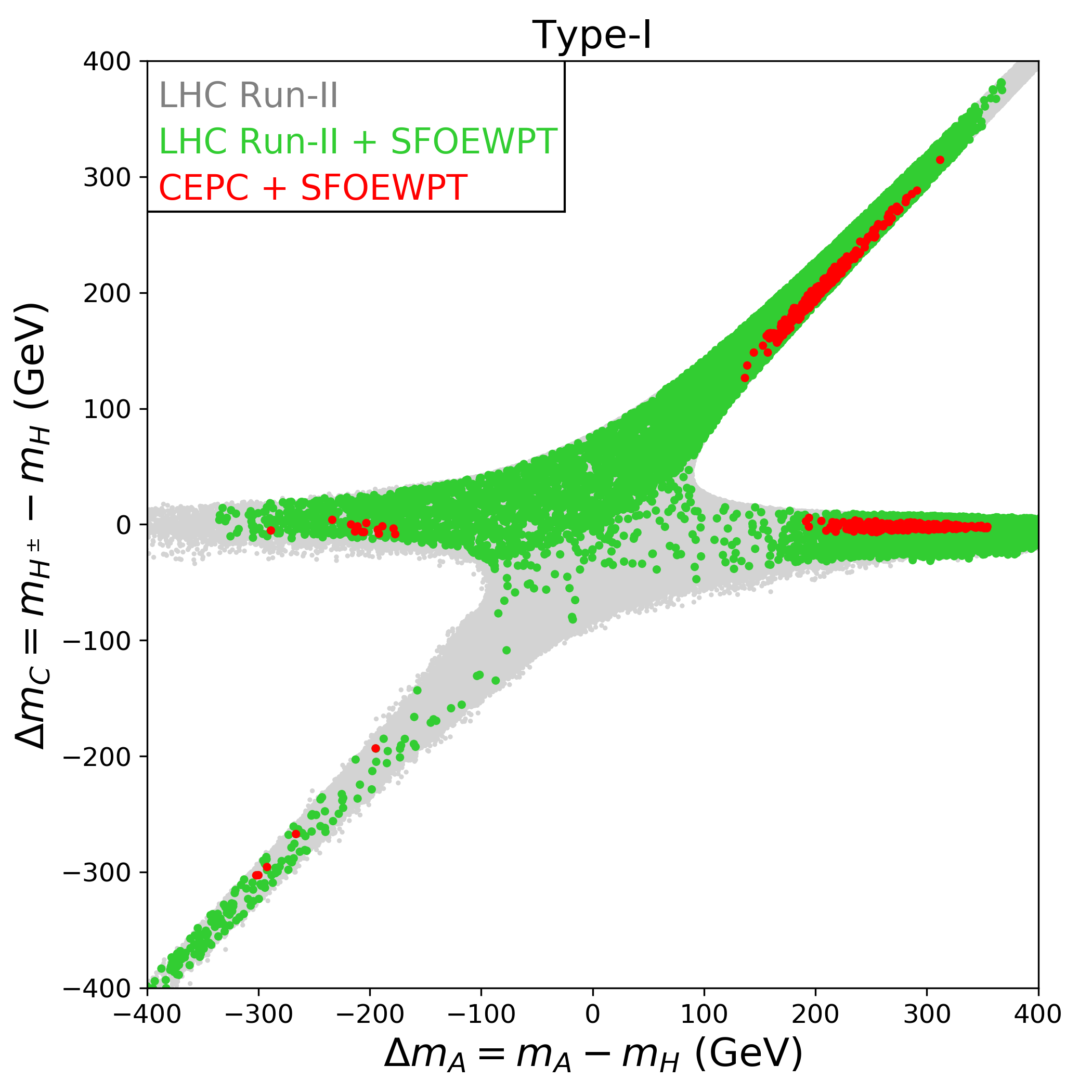}
  \caption{Allowed parameter space in the plane of $m_H - \tanb$ (left), $\Delta m_A-\Delta m_C$ (right). Same as~\autoref{fig:t_ma_tanb}, but for the Type-I model.}
  \label{fig:t1_ma_tanb}
\end{figure}

We show our general scan results for the Type-I 2HDM in  \autoref{fig:t1_ma_tanb}. The allowed grey, green and red points here cover a larger area than for the Type-II model, which mainly comes from heavy Higgs direct search constraints on the large $\tanb$ region. As the benchmark case shown in~\autoref{fig:LHC} shows, there is no constraint on $\tanb>2$ when $m_H> 2m_t$ in the Type-I 2HDM because all $Hf\bar f$ couplings are reduced as  $\tanb$ increases. 

At the same time, there is also a larger range for $\cosba$ at $\tanb>2$ compared to the Type-II 2HDM~\cite{Su:2019ibd}. Thus terms involving $\cosba$ will also become important, and from Ref.~\cite{Dorsch:2017nza} we can get,
\begin{align}
    \Delta \mathcal{F}_0|_{\rm general} = &\Delta \mathcal{F}_0|_{\cosba=0}+ \frac{1}{128 \pi^{2}}\cosba \sin(\beta-\alpha)(\tanb-\frac{1}{\tanb}) \nonumber \\
&(m_H^2-m_h^2)(2m_{H^\pm}^2+2m_A^2+5m_H^2-6M^2) + \mathcal{O}(\cos^2(\beta-\alpha)),
\end{align}
here $M^2=\frac{m_{12}^2}{\sin\beta\cos\beta}$. Because of this additional term, once there is sizable $\tanb,\cosba$, the allowed $\Delta m_A, \Delta m_C$ to generate the proper $\Delta \mathcal{F}_0/ |\mathcal{F}_0^\sm| $ range will be a little different to that in the Type-II case. In other words, the allowed parameter space in the Type-I model is larger than that in the Type-II model.

Generally speaking, compared to the Type-II 2HDM, the upper limit of $m_H$ allowed by a SFOEWPT in the Type-I model can still reach to 900 GeV. In the Type-II model, such points have $\Delta m_{A, C} <0$,  and are excluded by Higgs and $Z$-pole precision measurements. But larger $\tanb$ values allow larger mass splittings between the heavy Higgs bosons~\cite{Chen:2018shg,Chen:2019pkq}, and thus in the Type-I model $m_H\to 900\gev$ still satisfy these precision measurements. Similarly the regions with $\Delta m_{C}\approx 0,\Delta m_{A} <0$ or $\Delta m_{C}\approx 0,\Delta m_{A} >0$ which are not allowed in the Type-II model can still generate a SFOEWPT in the Type-I model.





\section{Conclusion}

In this work, we have revisited the existence of a  strong  first  order electroweak phase transition (SFOEWPT)  in the Type-I and Type-II 2HDMs. 
Using both numerical and analytical analysis methods, we pointed out that $\Delta\mathcal{F}_0/|\mathcal{F}_0^\sm|$ is not monotonically related to $\xi_c$ as shown in \autoref{fig:uplift1} and \autoref{fig:uplift2}. This conclusion is different to that of a previous study~\cite{Dorsch:2013wja}.

We also found, SFOEWPT suggests the non-SM Higgs bosons, $H/A/H^{\pm}$, have upper limits on their mass as our benchmark {\bf Case 1}~\autoref{fig:case1} and general scan results~\autoref{fig:t_ma_tanb} and \autoref{fig:t1_ma_tanb}. This limits comes from the combined requirements of vacuum stability at zero temperature and $\lambda_{H/A/H^{\pm}} v^2$ corrections term at high temperature.  

By combining current bounds from LHC direct and indirect Higgs searches, current electroweak precision measurements, flavour physics, and anticipated precision measurements at the future CEPC $Z$ and Higgs factory, we have shown that the requirement of a SFOEWPT puts strong constraints on the mass spectrum of $H/A/H^{\pm}$:
\begin{eqnarray}
\text{For the type-I}& &\text{2HDM:} \\\nonumber
& & \text{200 GeV} \lsim m_H,m_A,m_{H^{\pm}} \lsim \text{1 TeV} \ , |m_A - m_H| \in (\text{150GeV},\text{350GeV})\\\nonumber
& &|m_{H^{\pm}} - m_H| \in (\text{150GeV},\text{350GeV})\\
\text{For the type-II}& &\text{2HDM:} \\\nonumber
& & \text{400 GeV} \lsim m_H,m_A,m_{H^{\pm}} \lsim \text{1 TeV} \ , m_A - m_H \in (\text{150GeV},\text{250GeV})\\\nonumber
& &m_{H^{\pm}} - m_H \in (\text{150GeV},\text{250GeV})
\end{eqnarray}
In Type-II 2HDM, parameter space
{\bf Class D2} ($m_A=m_{H^\pm} <m_H$) and {\bf D3 } ($m_A<m_{H^\pm}=m_H$) are allowed by SM Higgs precision measurements and heavy Higgs searches at LHC Run-II, but can be excluded by Higgs precision observables at the CEPC because of the one-loop level corrections to the SM-like Higgs couplings. The allowed region has $m_A=m_{H^\pm} >m_H$ and small $\lambda  v^2=m_H^2-\frac{m_{12}^2}{\sin\beta\cos\beta}$. 
In Type-I 2HDM, because of allowed large $\tanb$ region from Higgs precision measurements, {\bf Class D2} and {\bf D3} are still allowed. 

Both Type-I and Type-II requires a sizable mass splitting between different heavy non-SM Higgs. And the suggested upper limits of $m_{A/H/H^\pm}$ is 900 GeV at current stage, and 800 GeV after including Higgs and Z-pole precisions at CEPC.
Such a constrained spectrum points out a clear direction for direct searches at the LHC and future colliders.

\appendix
\begin{acknowledgments}
 We thank Martin White for useful discussion and a careful reading of the manuscript. WS and AGW are supported by the Australian Research Council (ARC) Centre of Excellence for Dark Matter Particle Physics (CE200100008). M.Z. is supported by the National Natural Science Foundation of China (Grant No. 11947118). 
 \end{acknowledgments}
\clearpage

\bibliographystyle{utphys}
\bibliography{references}

\providecommand{\href}[2]{#2}\begingroup\raggedright\begin{thebibliography}{100}

\bibitem{Aad:2012tfa}
{\bf ATLAS} Collaboration, G.~Aad {\em et al.}, ``{Observation of a new
  particle in the search for the Standard Model Higgs boson with the ATLAS
  detector at the LHC},''
  \href{http://dx.doi.org/10.1016/j.physletb.2012.08.020}{{\em Phys. Lett.}
  {\bf B716} (2012)  1--29},
\href{http://arxiv.org/abs/1207.7214}{{\tt arXiv:1207.7214 [hep-ex]}}.

\bibitem{Chatrchyan:2012xdj}
{\bf CMS} Collaboration, S.~Chatrchyan {\em et al.}, ``{Observation of a new
  boson at a mass of 125 GeV with the CMS experiment at the LHC},''
  \href{http://dx.doi.org/10.1016/j.physletb.2012.08.021}{{\em Phys. Lett.}
  {\bf B716} (2012)  30--61},
\href{http://arxiv.org/abs/1207.7235}{{\tt arXiv:1207.7235 [hep-ex]}}.

\bibitem{Sakharov:1967dj}
A.~Sakharov, ``{Violation of CP Invariance, C asymmetry, and baryon asymmetry
  of the universe},''
  \href{http://dx.doi.org/10.1070/PU1991v034n05ABEH002497}{{\em Sov. Phys.
  Usp.} {\bf 34} (1991) no.~5, 392--393}.

\bibitem{Kuzmin:1985mm}
V.~Kuzmin, V.~Rubakov, and M.~Shaposhnikov, ``{On the Anomalous Electroweak
  Baryon Number Nonconservation in the Early Universe},''
  \href{http://dx.doi.org/10.1016/0370-2693(85)91028-7}{{\em Phys. Lett. B}
  {\bf 155} (1985)  36}.

\bibitem{Shaposhnikov:1986jp}
M.~Shaposhnikov, ``{Possible Appearance of the Baryon Asymmetry of the Universe
  in an Electroweak Theory},'' {\em JETP Lett.} {\bf 44} (1986)  465--468.

\bibitem{Shaposhnikov:1987tw}
M.~Shaposhnikov, ``{Baryon Asymmetry of the Universe in Standard Electroweak
  Theory},'' \href{http://dx.doi.org/10.1016/0550-3213(87)90127-1}{{\em Nucl.
  Phys. B} {\bf 287} (1987)  757--775}.

\bibitem{Manton:1983nd}
N.~Manton, ``{Topology in the Weinberg-Salam Theory},''
  \href{http://dx.doi.org/10.1103/PhysRevD.28.2019}{{\em Phys. Rev. D} {\bf 28}
  (1983)  2019}.

\bibitem{Klinkhamer:1984di}
F.~R. Klinkhamer and N.~Manton, ``{A Saddle Point Solution in the
  Weinberg-Salam Theory},''
  \href{http://dx.doi.org/10.1103/PhysRevD.30.2212}{{\em Phys. Rev. D} {\bf 30}
  (1984)  2212}.

\bibitem{Huet:1994jb}
P.~Huet and E.~Sather, ``{Electroweak baryogenesis and standard model CP
  violation},'' \href{http://dx.doi.org/10.1103/PhysRevD.51.379}{{\em Phys.
  Rev. D} {\bf 51} (1995)  379--394},
  \href{http://arxiv.org/abs/hep-ph/9404302}{{\tt arXiv:hep-ph/9404302}}.

\bibitem{Kajantie:1996mn}
K.~Kajantie, M.~Laine, K.~Rummukainen, and M.~E. Shaposhnikov, ``{Is there a
  hot electroweak phase transition at m(H) larger or equal to m(W)?},''
  \href{http://dx.doi.org/10.1103/PhysRevLett.77.2887}{{\em Phys. Rev. Lett.}
  {\bf 77} (1996)  2887--2890}, \href{http://arxiv.org/abs/hep-ph/9605288}{{\tt
  arXiv:hep-ph/9605288}}.

\bibitem{Csikor:1998eu}
F.~Csikor, Z.~Fodor, and J.~Heitger, ``{Endpoint of the hot electroweak phase
  transition},'' \href{http://dx.doi.org/10.1103/PhysRevLett.82.21}{{\em Phys.
  Rev. Lett.} {\bf 82} (1999)  21--24},
  \href{http://arxiv.org/abs/hep-ph/9809291}{{\tt arXiv:hep-ph/9809291}}.

\bibitem{Trodden:1998ym}
M.~Trodden, ``{Electroweak baryogenesis},''
  \href{http://dx.doi.org/10.1103/RevModPhys.71.1463}{{\em Rev. Mod. Phys.}
  {\bf 71} (1999)  1463--1500}, \href{http://arxiv.org/abs/hep-ph/9803479}{{\tt
  arXiv:hep-ph/9803479}}.

\bibitem{Konstandin:2013caa}
T.~Konstandin, ``{Quantum Transport and Electroweak Baryogenesis},''
  \href{http://dx.doi.org/10.3367/UFNe.0183.201308a.0785}{{\em Phys. Usp.} {\bf
  56} (2013)  747--771}, \href{http://arxiv.org/abs/1302.6713}{{\tt
  arXiv:1302.6713 [hep-ph]}}.

\bibitem{Carena:2018vpt}
M.~Carena, Z.~Liu, and M.~Riembau, ``{Probing the electroweak phase transition
  via enhanced di-Higgs boson production},''
  \href{http://dx.doi.org/10.1103/PhysRevD.97.095032}{{\em Phys. Rev. D} {\bf
  97} (2018) no.~9, 095032}, \href{http://arxiv.org/abs/1801.00794}{{\tt
  arXiv:1801.00794 [hep-ph]}}.

\bibitem{Cline:2012hg}
J.~M. Cline and K.~Kainulainen, ``{Electroweak baryogenesis and dark matter
  from a singlet Higgs},''
  \href{http://dx.doi.org/10.1088/1475-7516/2013/01/012}{{\em JCAP} {\bf 01}
  (2013)  012}, \href{http://arxiv.org/abs/1210.4196}{{\tt arXiv:1210.4196
  [hep-ph]}}.

\bibitem{Cline:2017qpe}
J.~M. Cline, K.~Kainulainen, and D.~Tucker-Smith, ``{Electroweak baryogenesis
  from a dark sector},''
  \href{http://dx.doi.org/10.1103/PhysRevD.95.115006}{{\em Phys. Rev. D} {\bf
  95} (2017) no.~11, 115006}, \href{http://arxiv.org/abs/1702.08909}{{\tt
  arXiv:1702.08909 [hep-ph]}}.

\bibitem{Carena:2018cjh}
M.~Carena, M.~Quirós, and Y.~Zhang, ``{Electroweak Baryogenesis from
  Dark-Sector CP Violation},''
  \href{http://dx.doi.org/10.1103/PhysRevLett.122.201802}{{\em Phys. Rev.
  Lett.} {\bf 122} (2019) no.~20, 201802},
  \href{http://arxiv.org/abs/1811.09719}{{\tt arXiv:1811.09719 [hep-ph]}}.

\bibitem{Cline:2009sn}
J.~M. Cline, G.~Laporte, H.~Yamashita, and S.~Kraml, ``{Electroweak Phase
  Transition and LHC Signatures in the Singlet Majoron Model},''
  \href{http://dx.doi.org/10.1088/1126-6708/2009/07/040}{{\em JHEP} {\bf 07}
  (2009)  040}, \href{http://arxiv.org/abs/0905.2559}{{\tt arXiv:0905.2559
  [hep-ph]}}.

\bibitem{Profumo:2014opa}
S.~Profumo, M.~J. Ramsey-Musolf, C.~L. Wainwright, and P.~Winslow,
  ``{Singlet-catalyzed electroweak phase transitions and precision Higgs boson
  studies},'' \href{http://dx.doi.org/10.1103/PhysRevD.91.035018}{{\em Phys.
  Rev. D} {\bf 91} (2015) no.~3, 035018},
  \href{http://arxiv.org/abs/1407.5342}{{\tt arXiv:1407.5342 [hep-ph]}}.

\bibitem{Curtin:2014jma}
D.~Curtin, P.~Meade, and C.-T. Yu, ``{Testing Electroweak Baryogenesis with
  Future Colliders},'' \href{http://dx.doi.org/10.1007/JHEP11(2014)127}{{\em
  JHEP} {\bf 11} (2014)  127}, \href{http://arxiv.org/abs/1409.0005}{{\tt
  arXiv:1409.0005 [hep-ph]}}.

\bibitem{Huang:2015bta}
F.~P. Huang and C.~S. Li, ``{Electroweak baryogenesis in the framework of the
  effective field theory},''
  \href{http://dx.doi.org/10.1103/PhysRevD.92.075014}{{\em Phys. Rev. D} {\bf
  92} (2015) no.~7, 075014}, \href{http://arxiv.org/abs/1507.08168}{{\tt
  arXiv:1507.08168 [hep-ph]}}.

\bibitem{Kotwal:2016tex}
A.~V. Kotwal, M.~J. Ramsey-Musolf, J.~M. No, and P.~Winslow,
  ``{Singlet-catalyzed electroweak phase transitions in the 100 TeV
  frontier},'' \href{http://dx.doi.org/10.1103/PhysRevD.94.035022}{{\em Phys.
  Rev. D} {\bf 94} (2016) no.~3, 035022},
  \href{http://arxiv.org/abs/1605.06123}{{\tt arXiv:1605.06123 [hep-ph]}}.

\bibitem{Vaskonen:2016yiu}
V.~Vaskonen, ``{Electroweak baryogenesis and gravitational waves from a real
  scalar singlet},'' \href{http://dx.doi.org/10.1103/PhysRevD.95.123515}{{\em
  Phys. Rev. D} {\bf 95} (2017) no.~12, 123515},
  \href{http://arxiv.org/abs/1611.02073}{{\tt arXiv:1611.02073 [hep-ph]}}.

\bibitem{Beniwal:2017eik}
A.~Beniwal, M.~Lewicki, J.~D. Wells, M.~White, and A.~G. Williams,
  ``{Gravitational wave, collider and dark matter signals from a scalar singlet
  electroweak baryogenesis},''
  \href{http://dx.doi.org/10.1007/JHEP08(2017)108}{{\em JHEP} {\bf 08} (2017)
  108}, \href{http://arxiv.org/abs/1702.06124}{{\tt arXiv:1702.06124
  [hep-ph]}}.

\bibitem{Kurup:2017dzf}
G.~Kurup and M.~Perelstein, ``{Dynamics of Electroweak Phase Transition In
  Singlet-Scalar Extension of the Standard Model},''
  \href{http://dx.doi.org/10.1103/PhysRevD.96.015036}{{\em Phys. Rev. D} {\bf
  96} (2017) no.~1, 015036}, \href{http://arxiv.org/abs/1704.03381}{{\tt
  arXiv:1704.03381 [hep-ph]}}.

\bibitem{Chiang:2017nmu}
C.-W. Chiang, M.~J. Ramsey-Musolf, and E.~Senaha, ``{Standard Model with a
  Complex Scalar Singlet: Cosmological Implications and Theoretical
  Considerations},'' \href{http://dx.doi.org/10.1103/PhysRevD.97.015005}{{\em
  Phys. Rev. D} {\bf 97} (2018) no.~1, 015005},
  \href{http://arxiv.org/abs/1707.09960}{{\tt arXiv:1707.09960 [hep-ph]}}.

\bibitem{Alves:2018jsw}
A.~Alves, T.~Ghosh, H.-K. Guo, K.~Sinha, and D.~Vagie, ``{Collider and
  Gravitational Wave Complementarity in Exploring the Singlet Extension of the
  Standard Model},'' \href{http://dx.doi.org/10.1007/JHEP04(2019)052}{{\em
  JHEP} {\bf 04} (2019)  052}, \href{http://arxiv.org/abs/1812.09333}{{\tt
  arXiv:1812.09333 [hep-ph]}}.

\bibitem{Li:2019tfd}
H.-L. Li, M.~Ramsey-Musolf, and S.~Willocq, ``{Probing a scalar
  singlet-catalyzed electroweak phase transition with resonant di-Higgs boson
  production in the $4b$ channel},''
  \href{http://dx.doi.org/10.1103/PhysRevD.100.075035}{{\em Phys. Rev. D} {\bf
  100} (2019) no.~7, 075035}, \href{http://arxiv.org/abs/1906.05289}{{\tt
  arXiv:1906.05289 [hep-ph]}}.

\bibitem{Bell:2019mbn}
N.~F. Bell, M.~J. Dolan, L.~S. Friedrich, M.~J. Ramsey-Musolf, and R.~R.
  Volkas, ``{Electroweak Baryogenesis with Vector-like Leptons and Scalar
  Singlets},'' \href{http://dx.doi.org/10.1007/JHEP09(2019)012}{{\em JHEP} {\bf
  19} (2020)  012}, \href{http://arxiv.org/abs/1903.11255}{{\tt
  arXiv:1903.11255 [hep-ph]}}.

\bibitem{Grzadkowski:2018nbc}
B.~Grzadkowski and D.~Huang, ``{Spontaneous $CP$-Violating Electroweak
  Baryogenesis and Dark Matter from a Complex Singlet Scalar},''
  \href{http://dx.doi.org/10.1007/JHEP08(2018)135}{{\em JHEP} {\bf 08} (2018)
  135}, \href{http://arxiv.org/abs/1807.06987}{{\tt arXiv:1807.06987
  [hep-ph]}}.

\bibitem{Huang:2018aja}
F.~P. Huang, Z.~Qian, and M.~Zhang, ``{Exploring dynamical CP violation induced
  baryogenesis by gravitational waves and colliders},''
  \href{http://dx.doi.org/10.1103/PhysRevD.98.015014}{{\em Phys. Rev. D} {\bf
  98} (2018) no.~1, 015014}, \href{http://arxiv.org/abs/1804.06813}{{\tt
  arXiv:1804.06813 [hep-ph]}}.

\bibitem{Bochkarev:1990fx}
A.~Bochkarev, S.~Kuzmin, and M.~Shaposhnikov, ``{Electroweak baryogenesis and
  the Higgs boson mass problem},''
  \href{http://dx.doi.org/10.1016/0370-2693(90)90069-I}{{\em Phys. Lett. B}
  {\bf 244} (1990)  275--278}.

\bibitem{McLerran:1990zh}
L.~D. McLerran, M.~E. Shaposhnikov, N.~Turok, and M.~B. Voloshin, ``{Why the
  baryon asymmetry of the universe is approximately 10**-10},''
  \href{http://dx.doi.org/10.1016/0370-2693(91)91794-V}{{\em Phys. Lett. B}
  {\bf 256} (1991)  451--456}.

\bibitem{Bochkarev:1990gb}
A.~Bochkarev, S.~Kuzmin, and M.~Shaposhnikov, ``{On the Model Dependence of the
  Cosmological Upper Bound on the Higgs Boson and Top Quark Masses},''
  \href{http://dx.doi.org/10.1103/PhysRevD.43.369}{{\em Phys. Rev. D} {\bf 43}
  (1991)  369--374}.

\bibitem{Turok:1990zg}
N.~Turok and J.~Zadrozny, ``{Electroweak baryogenesis in the two doublet
  model},'' \href{http://dx.doi.org/10.1016/0550-3213(91)90356-3}{{\em Nucl.
  Phys. B} {\bf 358} (1991)  471--493}.

\bibitem{Cohen:1991iu}
A.~G. Cohen, D.~Kaplan, and A.~Nelson, ``{Spontaneous baryogenesis at the weak
  phase transition},''
  \href{http://dx.doi.org/10.1016/0370-2693(91)91711-4}{{\em Phys. Lett. B}
  {\bf 263} (1991)  86--92}.

\bibitem{Turok:1991uc}
N.~Turok and J.~Zadrozny, ``{Phase transitions in the two doublet model},''
  \href{http://dx.doi.org/10.1016/0550-3213(92)90284-I}{{\em Nucl. Phys. B}
  {\bf 369} (1992)  729--742}.

\bibitem{Nelson:1991ab}
A.~Nelson, D.~Kaplan, and A.~G. Cohen, ``{Why there is something rather than
  nothing: Matter from weak interactions},''
  \href{http://dx.doi.org/10.1016/0550-3213(92)90440-M}{{\em Nucl. Phys. B}
  {\bf 373} (1992)  453--478}.

\bibitem{Funakubo:1993jg}
K.~Funakubo, A.~Kakuto, and K.~Takenaga, ``{The Effective potential of
  electroweak theory with two massless Higgs doublets at finite temperature},''
  \href{http://dx.doi.org/10.1143/PTP.91.341}{{\em Prog. Theor. Phys.} {\bf 91}
  (1994)  341--352}, \href{http://arxiv.org/abs/hep-ph/9310267}{{\tt
  arXiv:hep-ph/9310267}}.

\bibitem{Davies:1994id}
A.~Davies, C.~froggatt, G.~Jenkins, and R.~Moorhouse, ``{Baryogenesis
  constraints on two Higgs doublet models},''
  \href{http://dx.doi.org/10.1016/0370-2693(94)90559-2}{{\em Phys. Lett. B}
  {\bf 336} (1994)  464--470}.

\bibitem{Funakubo:1995kw}
K.~Funakubo, A.~Kakuto, S.~Otsuki, K.~Takenaga, and F.~Toyoda, ``{CP violating
  profile of the electroweak bubble wall},''
  \href{http://dx.doi.org/10.1143/PTP.94.845}{{\em Prog. Theor. Phys.} {\bf 94}
  (1995)  845--860}, \href{http://arxiv.org/abs/hep-ph/9507452}{{\tt
  arXiv:hep-ph/9507452}}.

\bibitem{Funakubo:1996iw}
K.~Funakubo, A.~Kakuto, S.~Otsuki, and F.~Toyoda, ``{Explicit CP breaking and
  electroweak baryogenesis},'' \href{http://dx.doi.org/10.1143/PTP.96.771}{{\em
  Prog. Theor. Phys.} {\bf 96} (1996)  771--780},
  \href{http://arxiv.org/abs/hep-ph/9606282}{{\tt arXiv:hep-ph/9606282}}.

\bibitem{Cline:1995dg}
J.~M. Cline, K.~Kainulainen, and A.~P. Vischer, ``{Dynamics of two Higgs
  doublet CP violation and baryogenesis at the electroweak phase transition},''
  \href{http://dx.doi.org/10.1103/PhysRevD.54.2451}{{\em Phys. Rev. D} {\bf 54}
  (1996)  2451--2472}, \href{http://arxiv.org/abs/hep-ph/9506284}{{\tt
  arXiv:hep-ph/9506284}}.

\bibitem{Fuyuto:2015jha}
K.~Fuyuto and E.~Senaha, ``{Sphaleron and critical bubble in the scale
  invariant two Higgs doublet model},''
  \href{http://dx.doi.org/10.1016/j.physletb.2015.05.061}{{\em Phys. Lett. B}
  {\bf 747} (2015)  152--157}, \href{http://arxiv.org/abs/1504.04291}{{\tt
  arXiv:1504.04291 [hep-ph]}}.

\bibitem{Chiang:2016vgf}
C.-W. Chiang, K.~Fuyuto, and E.~Senaha, ``{Electroweak Baryogenesis with Lepton
  Flavor Violation},''
  \href{http://dx.doi.org/10.1016/j.physletb.2016.09.052}{{\em Phys. Lett. B}
  {\bf 762} (2016)  315--320}, \href{http://arxiv.org/abs/1607.07316}{{\tt
  arXiv:1607.07316 [hep-ph]}}.

\bibitem{Dorsch:2013wja}
G.~Dorsch, S.~Huber, and J.~No, ``{A strong electroweak phase transition in the
  2HDM after LHC8},'' \href{http://dx.doi.org/10.1007/JHEP10(2013)029}{{\em
  JHEP} {\bf 10} (2013)  029}, \href{http://arxiv.org/abs/1305.6610}{{\tt
  arXiv:1305.6610 [hep-ph]}}.

\bibitem{Dorsch:2014qja}
G.~Dorsch, S.~Huber, K.~Mimasu, and J.~No, ``{Echoes of the Electroweak Phase
  Transition: Discovering a second Higgs doublet through $A_0 \rightarrow
  ZH_0$},'' \href{http://dx.doi.org/10.1103/PhysRevLett.113.211802}{{\em Phys.
  Rev. Lett.} {\bf 113} (2014) no.~21, 211802},
  \href{http://arxiv.org/abs/1405.5537}{{\tt arXiv:1405.5537 [hep-ph]}}.

\bibitem{Cline:1996mga}
J.~M. Cline and P.-A. Lemieux, ``{Electroweak phase transition in two Higgs
  doublet models},'' \href{http://dx.doi.org/10.1103/PhysRevD.55.3873}{{\em
  Phys. Rev. D} {\bf 55} (1997)  3873--3881},
  \href{http://arxiv.org/abs/hep-ph/9609240}{{\tt arXiv:hep-ph/9609240}}.

\bibitem{Fromme:2006cm}
L.~Fromme, S.~J. Huber, and M.~Seniuch, ``{Baryogenesis in the two-Higgs
  doublet model},'' \href{http://dx.doi.org/10.1088/1126-6708/2006/11/038}{{\em
  JHEP} {\bf 11} (2006)  038}, \href{http://arxiv.org/abs/hep-ph/0605242}{{\tt
  arXiv:hep-ph/0605242}}.

\bibitem{Cline:2011mm}
J.~M. Cline, K.~Kainulainen, and M.~Trott, ``{Electroweak Baryogenesis in Two
  Higgs Doublet Models and B meson anomalies},''
  \href{http://dx.doi.org/10.1007/JHEP11(2011)089}{{\em JHEP} {\bf 11} (2011)
  089}, \href{http://arxiv.org/abs/1107.3559}{{\tt arXiv:1107.3559 [hep-ph]}}.

\bibitem{Dorsch:2016nrg}
G.~Dorsch, S.~Huber, T.~Konstandin, and J.~No, ``{A Second Higgs Doublet in the
  Early Universe: Baryogenesis and Gravitational Waves},''
  \href{http://dx.doi.org/10.1088/1475-7516/2017/05/052}{{\em JCAP} {\bf 05}
  (2017)  052}, \href{http://arxiv.org/abs/1611.05874}{{\tt arXiv:1611.05874
  [hep-ph]}}.

\bibitem{Basler:2016obg}
P.~Basler, M.~Krause, M.~Muhlleitner, J.~Wittbrodt, and A.~Wlotzka, ``{Strong
  First Order Electroweak Phase Transition in the CP-Conserving 2HDM
  Revisited},'' \href{http://dx.doi.org/10.1007/JHEP02(2017)121}{{\em JHEP}
  {\bf 02} (2017)  121}, \href{http://arxiv.org/abs/1612.04086}{{\tt
  arXiv:1612.04086 [hep-ph]}}.

\bibitem{Haarr:2016qzq}
A.~Haarr, A.~Kvellestad, and T.~C. Petersen, ``{Disfavouring Electroweak
  Baryogenesis and a hidden Higgs in a CP-violating Two-Higgs-Doublet Model},''
  \href{http://arxiv.org/abs/1611.05757}{{\tt arXiv:1611.05757 [hep-ph]}}.

\bibitem{Fuyuto:2017ewj}
K.~Fuyuto, W.-S. Hou, and E.~Senaha, ``{Electroweak baryogenesis driven by
  extra top Yukawa couplings},''
  \href{http://dx.doi.org/10.1016/j.physletb.2017.11.073}{{\em Phys. Lett. B}
  {\bf 776} (2018)  402--406}, \href{http://arxiv.org/abs/1705.05034}{{\tt
  arXiv:1705.05034 [hep-ph]}}.

\bibitem{Dorsch:2017nza}
G.~Dorsch, S.~Huber, K.~Mimasu, and J.~No, ``{The Higgs Vacuum Uplifted:
  Revisiting the Electroweak Phase Transition with a Second Higgs Doublet},''
  \href{http://dx.doi.org/10.1007/JHEP12(2017)086}{{\em JHEP} {\bf 12} (2017)
  086}, \href{http://arxiv.org/abs/1705.09186}{{\tt arXiv:1705.09186
  [hep-ph]}}.

\bibitem{Cherchiglia:2017gko}
A.~Cherchiglia and C.~Nishi, ``{One-loop considerations for coexisting vacua in
  the CP conserving 2HDM},''
  \href{http://dx.doi.org/10.1007/JHEP11(2017)106}{{\em JHEP} {\bf 11} (2017)
  106}, \href{http://arxiv.org/abs/1707.04595}{{\tt arXiv:1707.04595
  [hep-ph]}}.

\bibitem{Basler:2017uxn}
P.~Basler, M.~Mühlleitner, and J.~Wittbrodt, ``{The CP-Violating 2HDM in Light
  of a Strong First Order Electroweak Phase Transition and Implications for
  Higgs Pair Production},''
  \href{http://dx.doi.org/10.1007/JHEP03(2018)061}{{\em JHEP} {\bf 03} (2018)
  061}, \href{http://arxiv.org/abs/1711.04097}{{\tt arXiv:1711.04097
  [hep-ph]}}.

\bibitem{Andersen:2017ika}
J.~O. Andersen, T.~Gorda, A.~Helset, L.~Niemi, T.~V.~I. Tenkanen, A.~Tranberg,
  A.~Vuorinen, and D.~J. Weir, ``{Nonperturbative Analysis of the Electroweak
  Phase Transition in the Two Higgs Doublet Model},''
  \href{http://dx.doi.org/10.1103/PhysRevLett.121.191802}{{\em Phys. Rev.
  Lett.} {\bf 121} (2018) no.~19, 191802},
  \href{http://arxiv.org/abs/1711.09849}{{\tt arXiv:1711.09849 [hep-ph]}}.

\bibitem{Bernon:2017jgv}
J.~Bernon, L.~Bian, and Y.~Jiang, ``{A new insight into the phase transition in
  the early Universe with two Higgs doublets},''
  \href{http://dx.doi.org/10.1007/JHEP05(2018)151}{{\em JHEP} {\bf 05} (2018)
  151}, \href{http://arxiv.org/abs/1712.08430}{{\tt arXiv:1712.08430
  [hep-ph]}}.

\bibitem{Gorda:2018hvi}
T.~Gorda, A.~Helset, L.~Niemi, T.~V. Tenkanen, and D.~J. Weir,
  ``{Three-dimensional effective theories for the two Higgs doublet model at
  high temperature},'' \href{http://dx.doi.org/10.1007/JHEP02(2019)081}{{\em
  JHEP} {\bf 02} (2019)  081}, \href{http://arxiv.org/abs/1802.05056}{{\tt
  arXiv:1802.05056 [hep-ph]}}.

\bibitem{Basler:2018cwe}
P.~Basler and M.~Mühlleitner, ``{BSMPT (Beyond the Standard Model Phase
  Transitions): A tool for the electroweak phase transition in extended Higgs
  sectors},'' \href{http://dx.doi.org/10.1016/j.cpc.2018.11.006}{{\em Comput.
  Phys. Commun.} {\bf 237} (2019)  62--85},
  \href{http://arxiv.org/abs/1803.02846}{{\tt arXiv:1803.02846 [hep-ph]}}.

\bibitem{Wang:2018hnw}
L.~Wang, J.~M. Yang, M.~Zhang, and Y.~Zhang, ``{Revisiting lepton-specific 2HDM
  in light of muon g-2 anomaly},''
  \href{http://dx.doi.org/10.1016/j.physletb.2018.11.045}{{\em Phys. Lett. B}
  {\bf 788} (2019)  519--529}, \href{http://arxiv.org/abs/1809.05857}{{\tt
  arXiv:1809.05857 [hep-ph]}}.

\bibitem{Kainulainen:2019kyp}
K.~Kainulainen, V.~Keus, L.~Niemi, K.~Rummukainen, T.~V. Tenkanen, and
  V.~Vaskonen, ``{On the validity of perturbative studies of the electroweak
  phase transition in the Two Higgs Doublet model},''
  \href{http://dx.doi.org/10.1007/JHEP06(2019)075}{{\em JHEP} {\bf 06} (2019)
  075}, \href{http://arxiv.org/abs/1904.01329}{{\tt arXiv:1904.01329
  [hep-ph]}}.

\bibitem{Wang:2019pet}
X.~Wang, F.~P. Huang, and X.~Zhang, ``{Gravitational wave and collider signals
  in complex two-Higgs doublet model with dynamical CP-violation at finite
  temperature},'' \href{http://dx.doi.org/10.1103/PhysRevD.101.015015}{{\em
  Phys. Rev. D} {\bf 101} (2020) no.~1, 015015},
  \href{http://arxiv.org/abs/1909.02978}{{\tt arXiv:1909.02978 [hep-ph]}}.

\bibitem{Borah:2012pu}
D.~Borah and J.~M. Cline, ``{Inert Doublet Dark Matter with Strong Electroweak
  Phase Transition},'' \href{http://dx.doi.org/10.1103/PhysRevD.86.055001}{{\em
  Phys. Rev. D} {\bf 86} (2012)  055001},
  \href{http://arxiv.org/abs/1204.4722}{{\tt arXiv:1204.4722 [hep-ph]}}.

\bibitem{Cline:2013bln}
J.~M. Cline and K.~Kainulainen, ``{Improved Electroweak Phase Transition with
  Subdominant Inert Doublet Dark Matter},''
  \href{http://dx.doi.org/10.1103/PhysRevD.87.071701}{{\em Phys. Rev. D} {\bf
  87} (2013) no.~7, 071701}, \href{http://arxiv.org/abs/1302.2614}{{\tt
  arXiv:1302.2614 [hep-ph]}}.

\bibitem{Fuyuto:2015ida}
K.~Fuyuto, J.~Hisano, and E.~Senaha, ``{Toward verification of electroweak
  baryogenesis by electric dipole moments},''
  \href{http://dx.doi.org/10.1016/j.physletb.2016.02.053}{{\em Phys. Lett. B}
  {\bf 755} (2016)  491--497}, \href{http://arxiv.org/abs/1510.04485}{{\tt
  arXiv:1510.04485 [hep-ph]}}.

\bibitem{Modak:2018csw}
T.~Modak and E.~Senaha, ``{Electroweak baryogenesis via bottom transport},''
  \href{http://dx.doi.org/10.1103/PhysRevD.99.115022}{{\em Phys. Rev. D} {\bf
  99} (2019) no.~11, 115022}, \href{http://arxiv.org/abs/1811.08088}{{\tt
  arXiv:1811.08088 [hep-ph]}}.

\bibitem{Chao:2015uoa}
W.~Chao and M.~J. Ramsey-Musolf, ``{Catalysis of Electroweak Baryogenesis via
  Fermionic Higgs Portal Dark Matter},''
  \href{http://arxiv.org/abs/1503.00028}{{\tt arXiv:1503.00028 [hep-ph]}}.

\bibitem{Inoue:2015pza}
S.~Inoue, G.~Ovanesyan, and M.~J. Ramsey-Musolf, ``{Two-Step Electroweak
  Baryogenesis},'' \href{http://dx.doi.org/10.1103/PhysRevD.93.015013}{{\em
  Phys. Rev. D} {\bf 93} (2016)  015013},
  \href{http://arxiv.org/abs/1508.05404}{{\tt arXiv:1508.05404 [hep-ph]}}.

\bibitem{Niemi:2018asa}
L.~Niemi, H.~H. Patel, M.~J. Ramsey-Musolf, T.~V. Tenkanen, and D.~J. Weir,
  ``{Electroweak phase transition in the real triplet extension of the SM:
  Dimensional reduction},''
  \href{http://dx.doi.org/10.1103/PhysRevD.100.035002}{{\em Phys. Rev. D} {\bf
  100} (2019) no.~3, 035002}, \href{http://arxiv.org/abs/1802.10500}{{\tt
  arXiv:1802.10500 [hep-ph]}}.

\bibitem{Chala:2018opy}
M.~Chala, M.~Ramos, and M.~Spannowsky, ``{Gravitational wave and collider
  probes of a triplet Higgs sector with a low cutoff},''
  \href{http://dx.doi.org/10.1140/epjc/s10052-019-6655-1}{{\em Eur. Phys. J. C}
  {\bf 79} (2019) no.~2, 156}, \href{http://arxiv.org/abs/1812.01901}{{\tt
  arXiv:1812.01901 [hep-ph]}}.

\bibitem{Zhou:2018zli}
R.~Zhou, W.~Cheng, X.~Deng, L.~Bian, and Y.~Wu, ``{Electroweak phase transition
  and Higgs phenomenology in the Georgi-Machacek model},''
  \href{http://dx.doi.org/10.1007/JHEP01(2019)216}{{\em JHEP} {\bf 01} (2019)
  216}, \href{http://arxiv.org/abs/1812.06217}{{\tt arXiv:1812.06217
  [hep-ph]}}.

\bibitem{Ellis:2019flb}
S.~A. Ellis, S.~Ipek, and G.~White, ``{Electroweak Baryogenesis from
  Temperature-Varying Couplings},''
  \href{http://dx.doi.org/10.1007/JHEP08(2019)002}{{\em JHEP} {\bf 08} (2019)
  002}, \href{http://arxiv.org/abs/1905.11994}{{\tt arXiv:1905.11994
  [hep-ph]}}.

\bibitem{Cao:2017oez}
Q.-H. Cao, F.~P. Huang, K.-P. Xie, and X.~Zhang, ``{Testing the electroweak
  phase transition in scalar extension models at lepton colliders},''
  \href{http://dx.doi.org/10.1088/1674-1137/42/2/023103}{{\em Chin. Phys. C}
  {\bf 42} (2018) no.~2, 023103}, \href{http://arxiv.org/abs/1708.04737}{{\tt
  arXiv:1708.04737 [hep-ph]}}.

\bibitem{Huang:2015izx}
F.~P. Huang, P.-H. Gu, P.-F. Yin, Z.-H. Yu, and X.~Zhang, ``{Testing the
  electroweak phase transition and electroweak baryogenesis at the LHC and a
  circular electron-positron collider},''
  \href{http://dx.doi.org/10.1103/PhysRevD.93.103515}{{\em Phys. Rev. D} {\bf
  93} (2016) no.~10, 103515}, \href{http://arxiv.org/abs/1511.03969}{{\tt
  arXiv:1511.03969 [hep-ph]}}.

\bibitem{Huang:2016odd}
F.~P. Huang, Y.~Wan, D.-G. Wang, Y.-F. Cai, and X.~Zhang, ``{Hearing the echoes
  of electroweak baryogenesis with gravitational wave detectors},''
  \href{http://dx.doi.org/10.1103/PhysRevD.94.041702}{{\em Phys. Rev. D} {\bf
  94} (2016) no.~4, 041702}, \href{http://arxiv.org/abs/1601.01640}{{\tt
  arXiv:1601.01640 [hep-ph]}}.

\bibitem{Balazs:2016yvi}
C.~Balazs, G.~White, and J.~Yue, ``{Effective field theory, electric dipole
  moments and electroweak baryogenesis},''
  \href{http://dx.doi.org/10.1007/JHEP03(2017)030}{{\em JHEP} {\bf 03} (2017)
  030}, \href{http://arxiv.org/abs/1612.01270}{{\tt arXiv:1612.01270
  [hep-ph]}}.

\bibitem{deVries:2017ncy}
J.~de~Vries, M.~Postma, J.~van~de Vis, and G.~White, ``{Electroweak
  Baryogenesis and the Standard Model Effective Field Theory},''
  \href{http://dx.doi.org/10.1007/JHEP01(2018)089}{{\em JHEP} {\bf 01} (2018)
  089}, \href{http://arxiv.org/abs/1710.04061}{{\tt arXiv:1710.04061
  [hep-ph]}}.

\bibitem{Cline:2008hr}
J.~M. Cline, M.~Jarvinen, and F.~Sannino, ``{The Electroweak Phase Transition
  in Nearly Conformal Technicolor},''
  \href{http://dx.doi.org/10.1103/PhysRevD.78.075027}{{\em Phys. Rev. D} {\bf
  78} (2008)  075027}, \href{http://arxiv.org/abs/0808.1512}{{\tt
  arXiv:0808.1512 [hep-ph]}}.

\bibitem{Bian:2019kmg}
L.~Bian, Y.~Wu, and K.-P. Xie, ``{Electroweak phase transition with composite
  Higgs models: calculability, gravitational waves and collider searches},''
  \href{http://dx.doi.org/10.1007/JHEP12(2019)028}{{\em JHEP} {\bf 12} (2019)
  028}, \href{http://arxiv.org/abs/1909.02014}{{\tt arXiv:1909.02014
  [hep-ph]}}.

\bibitem{Xie:2020bkl}
K.-P. Xie, Y.~Wu, and L.~Bian, ``{Electroweak baryogenesis and gravitational
  waves in a composite Higgs model with high dimensional fermion
  representations},'' \href{http://arxiv.org/abs/2005.13552}{{\tt
  arXiv:2005.13552 [hep-ph]}}.

\bibitem{Cline:1997vk}
J.~M. Cline, M.~Joyce, and K.~Kainulainen, ``{Supersymmetric electroweak
  baryogenesis in the WKB approximation},''
  \href{http://dx.doi.org/10.1016/S0370-2693(97)01361-0}{{\em Phys. Lett. B}
  {\bf 417} (1998)  79--86}, \href{http://arxiv.org/abs/hep-ph/9708393}{{\tt
  arXiv:hep-ph/9708393}}. [Erratum: Phys.Lett.B 448, 321--321 (1999)].

\bibitem{Menon:2004wv}
A.~Menon, D.~Morrissey, and C.~Wagner, ``{Electroweak baryogenesis and dark
  matter in the nMSSM},''
  \href{http://dx.doi.org/10.1103/PhysRevD.70.035005}{{\em Phys. Rev. D} {\bf
  70} (2004)  035005}, \href{http://arxiv.org/abs/hep-ph/0404184}{{\tt
  arXiv:hep-ph/0404184}}.

\bibitem{Carena:2011jy}
M.~Carena, N.~R. Shah, and C.~E. Wagner, ``{Light Dark Matter and the
  Electroweak Phase Transition in the NMSSM},''
  \href{http://dx.doi.org/10.1103/PhysRevD.85.036003}{{\em Phys. Rev. D} {\bf
  85} (2012)  036003}, \href{http://arxiv.org/abs/1110.4378}{{\tt
  arXiv:1110.4378 [hep-ph]}}.

\bibitem{Bi:2015qva}
X.-J. Bi, L.~Bian, W.~Huang, J.~Shu, and P.-F. Yin, ``{Interpretation of the
  Galactic Center excess and electroweak phase transition in the NMSSM},''
  \href{http://dx.doi.org/10.1103/PhysRevD.92.023507}{{\em Phys. Rev. D} {\bf
  92} (2015)  023507}, \href{http://arxiv.org/abs/1503.03749}{{\tt
  arXiv:1503.03749 [hep-ph]}}.

\bibitem{Demidov:2016wcv}
S.~Demidov, D.~Gorbunov, and D.~Kirpichnikov, ``{Split NMSSM with electroweak
  baryogenesis},'' \href{http://dx.doi.org/10.1007/JHEP11(2016)148}{{\em JHEP}
  {\bf 11} (2016)  148}, \href{http://arxiv.org/abs/1608.01985}{{\tt
  arXiv:1608.01985 [hep-ph]}}. [Erratum: JHEP 08, 080 (2017)].

\bibitem{Huang:2014ifa}
W.~Huang, Z.~Kang, J.~Shu, P.~Wu, and J.~M. Yang, ``{New insights in the
  electroweak phase transition in the NMSSM},''
  \href{http://dx.doi.org/10.1103/PhysRevD.91.025006}{{\em Phys. Rev. D} {\bf
  91} (2015) no.~2, 025006}, \href{http://arxiv.org/abs/1405.1152}{{\tt
  arXiv:1405.1152 [hep-ph]}}.

\bibitem{Cheung:2012pg}
K.~Cheung, T.-J. Hou, J.~S. Lee, and E.~Senaha, ``{Singlino-driven Electroweak
  Baryogenesis in the Next-to-MSSM},''
  \href{http://dx.doi.org/10.1016/j.physletb.2012.02.070}{{\em Phys. Lett. B}
  {\bf 710} (2012)  188--191}, \href{http://arxiv.org/abs/1201.3781}{{\tt
  arXiv:1201.3781 [hep-ph]}}.

\bibitem{Balazs:2013cia}
C.~Balázs, A.~Mazumdar, E.~Pukartas, and G.~White, ``{Baryogenesis, dark
  matter and inflation in the Next-to-Minimal Supersymmetric Standard Model},''
  \href{http://dx.doi.org/10.1007/JHEP01(2014)073}{{\em JHEP} {\bf 01} (2014)
  073}, \href{http://arxiv.org/abs/1309.5091}{{\tt arXiv:1309.5091 [hep-ph]}}.

\bibitem{Huber:2006wf}
S.~J. Huber, T.~Konstandin, T.~Prokopec, and M.~G. Schmidt, ``{Electroweak
  Phase Transition and Baryogenesis in the nMSSM},''
  \href{http://dx.doi.org/10.1016/j.nuclphysb.2006.09.003}{{\em Nucl. Phys. B}
  {\bf 757} (2006)  172--196}, \href{http://arxiv.org/abs/hep-ph/0606298}{{\tt
  arXiv:hep-ph/0606298}}.

\bibitem{Bian:2017wfv}
L.~Bian, H.-K. Guo, and J.~Shu, ``{Gravitational Waves, baryon asymmetry of the
  universe and electric dipole moment in the CP-violating NMSSM},''
  \href{http://dx.doi.org/10.1088/1674-1137/42/9/093106}{{\em Chin. Phys. C}
  {\bf 42} (2018) no.~9, 093106}, \href{http://arxiv.org/abs/1704.02488}{{\tt
  arXiv:1704.02488 [hep-ph]}}. [Erratum: Chin.Phys.C 43, 129101 (2019)].

\bibitem{Kozaczuk:2014kva}
J.~Kozaczuk, S.~Profumo, L.~S. Haskins, and C.~L. Wainwright, ``{Cosmological
  Phase Transitions and their Properties in the NMSSM},''
  \href{http://dx.doi.org/10.1007/JHEP01(2015)144}{{\em JHEP} {\bf 01} (2015)
  144}, \href{http://arxiv.org/abs/1407.4134}{{\tt arXiv:1407.4134 [hep-ph]}}.

\bibitem{Katz:2015uja}
A.~Katz, M.~Perelstein, M.~J. Ramsey-Musolf, and P.~Winslow, ``{Stop-Catalyzed
  Baryogenesis Beyond the MSSM},''
  \href{http://dx.doi.org/10.1103/PhysRevD.92.095019}{{\em Phys. Rev. D} {\bf
  92} (2015) no.~9, 095019}, \href{http://arxiv.org/abs/1509.02934}{{\tt
  arXiv:1509.02934 [hep-ph]}}.

\bibitem{Akula:2017yfr}
S.~Akula, C.~Balázs, L.~Dunn, and G.~White, ``{Electroweak baryogenesis in the
  $ {\mathbb{Z}}_3 $ -invariant NMSSM},''
  \href{http://dx.doi.org/10.1007/JHEP11(2017)051}{{\em JHEP} {\bf 11} (2017)
  051}, \href{http://arxiv.org/abs/1706.09898}{{\tt arXiv:1706.09898
  [hep-ph]}}.

\bibitem{Lee:2004we}
C.~Lee, V.~Cirigliano, and M.~J. Ramsey-Musolf, ``{Resonant relaxation in
  electroweak baryogenesis},''
  \href{http://dx.doi.org/10.1103/PhysRevD.71.075010}{{\em Phys. Rev. D} {\bf
  71} (2005)  075010}, \href{http://arxiv.org/abs/hep-ph/0412354}{{\tt
  arXiv:hep-ph/0412354}}.

\bibitem{Balazs:2004ae}
C.~Balazs, M.~Carena, A.~Menon, D.~Morrissey, and C.~Wagner, ``{The
  Supersymmetric origin of matter},''
  \href{http://dx.doi.org/10.1103/PhysRevD.71.075002}{{\em Phys. Rev. D} {\bf
  71} (2005)  075002}, \href{http://arxiv.org/abs/hep-ph/0412264}{{\tt
  arXiv:hep-ph/0412264}}.

\bibitem{Liebler:2015ddv}
S.~Liebler, S.~Profumo, and T.~Stefaniak, ``{Light Stop Mass Limits from Higgs
  Rate Measurements in the MSSM: Is MSSM Electroweak Baryogenesis Still Alive
  After All?},'' \href{http://dx.doi.org/10.1007/JHEP04(2016)143}{{\em JHEP}
  {\bf 04} (2016)  143}, \href{http://arxiv.org/abs/1512.09172}{{\tt
  arXiv:1512.09172 [hep-ph]}}.

\bibitem{Kobakhidze:2015xlz}
A.~Kobakhidze, L.~Wu, and J.~Yue, ``{Electroweak Baryogenesis with Anomalous
  Higgs Couplings},'' \href{http://dx.doi.org/10.1007/JHEP04(2016)011}{{\em
  JHEP} {\bf 04} (2016)  011}, \href{http://arxiv.org/abs/1512.08922}{{\tt
  arXiv:1512.08922 [hep-ph]}}.

\bibitem{Ramsey-Musolf:2017tgh}
M.~J. Ramsey-Musolf, P.~Winslow, and G.~White, ``{Color Breaking
  Baryogenesis},'' \href{http://dx.doi.org/10.1103/PhysRevD.97.123509}{{\em
  Phys. Rev. D} {\bf 97} (2018) no.~12, 123509},
  \href{http://arxiv.org/abs/1708.07511}{{\tt arXiv:1708.07511 [hep-ph]}}.

\bibitem{YaserAyazi:2019caf}
S.~Yaser~Ayazi and A.~Mohamadnejad, ``{Conformal vector dark matter and
  strongly first-order electroweak phase transition},''
  \href{http://dx.doi.org/10.1007/JHEP03(2019)181}{{\em JHEP} {\bf 03} (2019)
  181}, \href{http://arxiv.org/abs/1901.04168}{{\tt arXiv:1901.04168
  [hep-ph]}}.

\bibitem{Mohamadnejad:2019vzg}
A.~Mohamadnejad, ``{Gravitational waves from scale-invariant vector dark matter
  model: Probing below the neutrino-floor},''
  \href{http://dx.doi.org/10.1140/epjc/s10052-020-7756-6}{{\em Eur. Phys. J. C}
  {\bf 80} (2020) no.~3, 197}, \href{http://arxiv.org/abs/1907.08899}{{\tt
  arXiv:1907.08899 [hep-ph]}}.

\bibitem{Lee:1973iz}
T.~Lee, ``{A Theory of Spontaneous T Violation},''
  \href{http://dx.doi.org/10.1103/PhysRevD.8.1226}{{\em Phys. Rev. D} {\bf 8}
  (1973)  1226--1239}.

\bibitem{Branco:2011iw}
G.~C. Branco, P.~M. Ferreira, L.~Lavoura, M.~N. Rebelo, M.~Sher, and J.~P.
  Silva, ``{Theory and phenomenology of two-Higgs-doublet models},''
  \href{http://dx.doi.org/10.1016/j.physrep.2012.02.002}{{\em Phys. Rept.} {\bf
  516} (2012)  1--102},
\href{http://arxiv.org/abs/1106.0034}{{\tt arXiv:1106.0034 [hep-ph]}}.

\bibitem{Fukugita:1986hr}
M.~Fukugita and T.~Yanagida, ``{Baryogenesis Without Grand Unification},''
  \href{http://dx.doi.org/10.1016/0370-2693(86)91126-3}{{\em Phys. Lett. B}
  {\bf 174} (1986)  45--47}.

\bibitem{Affleck:1984fy}
I.~Affleck and M.~Dine, ``{A New Mechanism for Baryogenesis},''
  \href{http://dx.doi.org/10.1016/0550-3213(85)90021-5}{{\em Nucl. Phys. B}
  {\bf 249} (1985)  361--380}.

\bibitem{Abada:2019ono}
{\bf FCC} Collaboration, A.~Abada {\em et al.}, ``{HE-LHC: The High-Energy
  Large Hadron Collider}: {Future Circular Collider Conceptual Design Report
  Volume 4},'' \href{http://dx.doi.org/10.1140/epjst/e2019-900088-6}{{\em Eur.
  Phys. J. ST} {\bf 228} (2019) no.~5, 1109--1382}.

\bibitem{CEPCStudyGroup:2018ghi}
{\bf CEPC Study Group} Collaboration, ``{CEPC Conceptual Design Report: Volume
  2 - Physics \& Detector},''
\href{http://arxiv.org/abs/1811.10545}{{\tt arXiv:1811.10545 [hep-ex]}}.

\bibitem{CEPCPhysics-DetectorStudyGroup:2019wir}
{\bf CEPC Physics-Detector Study Group} Collaboration, ``{The CEPC input for
  the European Strategy for Particle Physics - Physics and Detector},''
  \href{http://arxiv.org/abs/1901.03170}{{\tt arXiv:1901.03170 [hep-ex]}}.

\bibitem{Li:2020hao}
S.~Li, H.~Song, and S.~Su, ``{Probing Exotic Charged Higgs Decays in the
  Type-II 2HDM through Top Rich Signal at a Future 100 TeV pp Collider},''
  \href{http://arxiv.org/abs/2005.00576}{{\tt arXiv:2005.00576 [hep-ph]}}.

\bibitem{Kling:2018xud}
F.~Kling, H.~Li, A.~Pyarelal, H.~Song, and S.~Su, ``{Exotic Higgs Decays in
  Type-II 2HDMs at the LHC and Future 100 TeV Hadron Colliders},''
  \href{http://dx.doi.org/10.1007/JHEP06(2019)031}{{\em JHEP} {\bf 06} (2019)
  031},
\href{http://arxiv.org/abs/1812.01633}{{\tt arXiv:1812.01633 [hep-ph]}}.

\bibitem{Chen:2017dwb}
C.-R. Chen, J.~Hajer, T.~Liu, I.~Low, and H.~Zhang, ``{Testing naturalness at
  100 TeV},'' \href{http://dx.doi.org/10.1007/JHEP09(2017)129}{{\em JHEP} {\bf
  09} (2017)  129}, \href{http://arxiv.org/abs/1705.07743}{{\tt
  arXiv:1705.07743 [hep-ph]}}.

\bibitem{Craig:2016ygr}
N.~Craig, J.~Hajer, Y.-Y. Li, T.~Liu, and H.~Zhang, ``{Heavy Higgs bosons at
  low $\tan \beta$: from the LHC to 100 TeV},''
  \href{http://dx.doi.org/10.1007/JHEP01(2017)018}{{\em JHEP} {\bf 01} (2017)
  018}, \href{http://arxiv.org/abs/1605.08744}{{\tt arXiv:1605.08744
  [hep-ph]}}.

\bibitem{Bambade:2019fyw}
P.~Bambade {\em et al.}, ``{The International Linear Collider: A Global
  Project},'' \href{http://arxiv.org/abs/1903.01629}{{\tt arXiv:1903.01629
  [hep-ex]}}.

\bibitem{Abada:2019lih}
{\bf FCC} Collaboration, A.~Abada {\em et al.}, ``{FCC Physics Opportunities}:
  {Future Circular Collider Conceptual Design Report Volume 1},''
  \href{http://dx.doi.org/10.1140/epjc/s10052-019-6904-3}{{\em Eur. Phys. J. C}
  {\bf 79} (2019) no.~6, 474}.

\bibitem{Abada:2019zxq}
{\bf FCC} Collaboration, A.~Abada {\em et al.}, ``{FCC-ee: The Lepton
  Collider}: {Future Circular Collider Conceptual Design Report Volume 2},''
  \href{http://dx.doi.org/10.1140/epjst/e2019-900045-4}{{\em Eur. Phys. J. ST}
  {\bf 228} (2019) no.~2, 261--623}.

\bibitem{Lacey:2283082}
{\bf ATLAS Collaboration} Collaboration, J.~Lacey, ``{Higgs results from the
  combination of ATLAS and CMS},''. \url{https://cds.cern.ch/record/2283082}.

\bibitem{Gu:2017ckc}
J.~Gu, H.~Li, Z.~Liu, S.~Su, and W.~Su, ``{Learning from Higgs Physics at
  Future Higgs Factories},''
  \href{http://dx.doi.org/10.1007/JHEP12(2017)153}{{\em JHEP} {\bf 12} (2017)
  153},
\href{http://arxiv.org/abs/1709.06103}{{\tt arXiv:1709.06103 [hep-ph]}}.

\bibitem{Laine:2016hma}
M.~Laine and A.~Vuorinen,
  \href{http://dx.doi.org/10.1007/978-3-319-31933-9}{{\em {Basics of Thermal
  Field Theory}}}, vol.~925.
\newblock Springer, 2016.
\newblock \href{http://arxiv.org/abs/1701.01554}{{\tt arXiv:1701.01554
  [hep-ph]}}.

\bibitem{Coleman:1973jx}
S.~R. Coleman and E.~J. Weinberg, ``{Radiative Corrections as the Origin of
  Spontaneous Symmetry Breaking},''
  \href{http://dx.doi.org/10.1103/PhysRevD.7.1888}{{\em Phys. Rev. D} {\bf 7}
  (1973)  1888--1910}.

\bibitem{Quiros:1999jp}
M.~Quiros, ``{Finite temperature field theory and phase transitions},'' in {\em
  {ICTP Summer School in High-Energy Physics and Cosmology}}, pp.~187--259.
\newblock 1, 1999.
\newblock \href{http://arxiv.org/abs/hep-ph/9901312}{{\tt
  arXiv:hep-ph/9901312}}.

\bibitem{Arnold:1992rz}
P.~B. Arnold and O.~Espinosa, ``{The Effective potential and first order phase
  transitions: Beyond leading-order},''
  \href{http://dx.doi.org/10.1103/PhysRevD.47.3546}{{\em Phys. Rev. D} {\bf 47}
  (1993)  3546}, \href{http://arxiv.org/abs/hep-ph/9212235}{{\tt
  arXiv:hep-ph/9212235}}. [Erratum: Phys.Rev.D 50, 6662 (1994)].

\bibitem{Moore:1998swa}
G.~D. Moore, ``{Measuring the broken phase sphaleron rate nonperturbatively},''
  \href{http://dx.doi.org/10.1103/PhysRevD.59.014503}{{\em Phys. Rev. D} {\bf
  59} (1999)  014503}, \href{http://arxiv.org/abs/hep-ph/9805264}{{\tt
  arXiv:hep-ph/9805264}}.

\bibitem{Nielsen:1975fs}
N.~Nielsen, ``{On the Gauge Dependence of Spontaneous Symmetry Breaking in
  Gauge Theories},'' \href{http://dx.doi.org/10.1016/0550-3213(75)90301-6}{{\em
  Nucl. Phys. B} {\bf 101} (1975)  173--188}.

\bibitem{DiLuzio:2014bua}
L.~Di~Luzio and L.~Mihaila, ``{On the gauge dependence of the Standard Model
  vacuum instability scale},''
  \href{http://dx.doi.org/10.1007/JHEP06(2014)079}{{\em JHEP} {\bf 06} (2014)
  079}, \href{http://arxiv.org/abs/1404.7450}{{\tt arXiv:1404.7450 [hep-ph]}}.

\bibitem{Patel:2011th}
H.~H. Patel and M.~J. Ramsey-Musolf, ``{Baryon Washout, Electroweak Phase
  Transition, and Perturbation Theory},''
  \href{http://dx.doi.org/10.1007/JHEP07(2011)029}{{\em JHEP} {\bf 07} (2011)
  029}, \href{http://arxiv.org/abs/1101.4665}{{\tt arXiv:1101.4665 [hep-ph]}}.

\bibitem{Laine:2017hdk}
M.~Laine, M.~Meyer, and G.~Nardini, ``{Thermal phase transition with full
  2-loop effective potential},''
  \href{http://dx.doi.org/10.1016/j.nuclphysb.2017.04.023}{{\em Nucl. Phys. B}
  {\bf 920} (2017)  565--600}, \href{http://arxiv.org/abs/1702.07479}{{\tt
  arXiv:1702.07479 [hep-ph]}}.

\bibitem{Dine:1991ck}
M.~Dine, P.~Huet, and J.~Singleton, Robert~L., ``{Baryogenesis at the
  electroweak scale},''
  \href{http://dx.doi.org/10.1016/0550-3213(92)90113-P}{{\em Nucl. Phys. B}
  {\bf 375} (1992)  625--648}.

\bibitem{Wainwright:2011kj}
C.~L. Wainwright, ``{CosmoTransitions: Computing Cosmological Phase Transition
  Temperatures and Bubble Profiles with Multiple Fields},''
  \href{http://dx.doi.org/10.1016/j.cpc.2012.04.004}{{\em Comput. Phys.
  Commun.} {\bf 183} (2012)  2006--2013},
  \href{http://arxiv.org/abs/1109.4189}{{\tt arXiv:1109.4189 [hep-ph]}}.

\bibitem{Athron:2020sbe}
P.~Athron, C.~Balázs, A.~Fowlie, and Y.~Zhang, ``{PhaseTracer: tracing
  cosmological phases and calculating transition properties},''
  \href{http://dx.doi.org/10.1140/epjc/s10052-020-8035-2}{{\em Eur. Phys. J. C}
  {\bf 80} (2020) no.~6, 567}, \href{http://arxiv.org/abs/2003.02859}{{\tt
  arXiv:2003.02859 [hep-ph]}}.

\bibitem{Han:2015yys}
T.~Han, S.~K. Kang, and J.~Sayre, ``{Muon $g-2$ in the aligned two Higgs
  doublet model},'' \href{http://dx.doi.org/10.1007/JHEP02(2016)097}{{\em JHEP}
  {\bf 02} (2016)  097}, \href{http://arxiv.org/abs/1511.05162}{{\tt
  arXiv:1511.05162 [hep-ph]}}.

\bibitem{Kling:2020hmi}
F.~Kling, S.~Su, and W.~Su, ``{2HDM Neutral Scalars under the LHC},''
  \href{http://dx.doi.org/10.1007/JHEP06(2020)163}{{\em JHEP} {\bf 06} (2020)
  163}, \href{http://arxiv.org/abs/2004.04172}{{\tt arXiv:2004.04172
  [hep-ph]}}.

\bibitem{Deshpande:1977rw}
N.~G. Deshpande and E.~Ma, ``{Pattern of Symmetry Breaking with Two Higgs
  Doublets},'' \href{http://dx.doi.org/10.1103/PhysRevD.18.2574}{{\em Phys.
  Rev. D} {\bf 18} (1978)  2574}.

\bibitem{Sher:1988mj}
M.~Sher, ``{Electroweak Higgs Potentials and Vacuum Stability},''
  \href{http://dx.doi.org/10.1016/0370-1573(89)90061-6}{{\em Phys. Rept.} {\bf
  179} (1989)  273--418}.

\bibitem{Nie:1998yn}
S.~Nie and M.~Sher, ``{Vacuum stability bounds in the two Higgs doublet
  model},'' \href{http://dx.doi.org/10.1016/S0370-2693(99)00019-2}{{\em Phys.
  Lett. B} {\bf 449} (1999)  89--92},
  \href{http://arxiv.org/abs/hep-ph/9811234}{{\tt arXiv:hep-ph/9811234}}.

\bibitem{Kanemura:1999xf}
S.~Kanemura, T.~Kasai, and Y.~Okada, ``{Mass bounds of the lightest CP even
  Higgs boson in the two Higgs doublet model},''
  \href{http://dx.doi.org/10.1016/S0370-2693(99)01351-9}{{\em Phys. Lett. B}
  {\bf 471} (1999)  182--190}, \href{http://arxiv.org/abs/hep-ph/9903289}{{\tt
  arXiv:hep-ph/9903289}}.

\bibitem{Huffel:1980sk}
H.~Huffel and G.~Pocsik, ``{Unitarity Bounds on Higgs Boson Masses in the
  {Weinberg-Salam} Model With Two Higgs Doublets},''
  \href{http://dx.doi.org/10.1007/BF01429824}{{\em Z. Phys. C} {\bf 8} (1981)
  13}.

\bibitem{Maalampi:1991fb}
J.~Maalampi, J.~Sirkka, and I.~Vilja, ``{Tree level unitarity and triviality
  bounds for two Higgs models},''
  \href{http://dx.doi.org/10.1016/0370-2693(91)90068-2}{{\em Phys. Lett. B}
  {\bf 265} (1991)  371--376}.

\bibitem{Kanemura:1993hm}
S.~Kanemura, T.~Kubota, and E.~Takasugi, ``{Lee-Quigg-Thacker bounds for Higgs
  boson masses in a two doublet model},''
  \href{http://dx.doi.org/10.1016/0370-2693(93)91205-2}{{\em Phys. Lett. B}
  {\bf 313} (1993)  155--160}, \href{http://arxiv.org/abs/hep-ph/9303263}{{\tt
  arXiv:hep-ph/9303263}}.

\bibitem{Akeroyd:2000wc}
A.~G. Akeroyd, A.~Arhrib, and E.-M. Naimi, ``{Note on tree level unitarity in
  the general two Higgs doublet model},''
  \href{http://dx.doi.org/10.1016/S0370-2693(00)00962-X}{{\em Phys. Lett. B}
  {\bf 490} (2000)  119--124}, \href{http://arxiv.org/abs/hep-ph/0006035}{{\tt
  arXiv:hep-ph/0006035}}.

\bibitem{Ginzburg:2005dt}
I.~Ginzburg and I.~Ivanov, ``{Tree-level unitarity constraints in the most
  general 2HDM},'' \href{http://dx.doi.org/10.1103/PhysRevD.72.115010}{{\em
  Phys. Rev. D} {\bf 72} (2005)  115010},
  \href{http://arxiv.org/abs/hep-ph/0508020}{{\tt arXiv:hep-ph/0508020}}.

\bibitem{Su:2019ibd}
W.~Su, ``{Probing loop effects in wrong-sign Yukawa region of 2HDM},''
\href{http://arxiv.org/abs/1910.06269}{{\tt arXiv:1910.06269 [hep-ph]}}.

\bibitem{Sirunyan:2019tkw}
{\bf CMS} Collaboration, A.~M. Sirunyan {\em et al.}, ``{Search for MSSM Higgs
  bosons decaying to $\mu^+\mu^-$ in proton-proton collisions at $\sqrt{s}=$ 13
  TeVSearch for MSSM Higgs bosons decaying to $\mu^+\mu^-$ in proton-proton
  collisions at s=13TeV},''
  \href{http://dx.doi.org/10.1016/j.physletb.2019.134992}{{\em Phys. Lett.}
  {\bf B798} (2019)  134992},
\href{http://arxiv.org/abs/1907.03152}{{\tt arXiv:1907.03152 [hep-ex]}}.

\bibitem{Aaboud:2019sgt}
{\bf ATLAS} Collaboration, M.~Aaboud {\em et al.}, ``{Search for scalar
  resonances decaying into $\mu^{+}\mu^{-}$ in events with and without
  $b$-tagged jets produced in proton-proton collisions at $\sqrt{s}=13$ TeV
  with the ATLAS detector},''
  \href{http://dx.doi.org/10.1007/JHEP07(2019)117}{{\em JHEP} {\bf 07} (2019)
  117},
\href{http://arxiv.org/abs/1901.08144}{{\tt arXiv:1901.08144 [hep-ex]}}.

\bibitem{Sirunyan:2018taj}
{\bf CMS} Collaboration, A.~M. Sirunyan {\em et al.}, ``{Search for beyond the
  standard model Higgs bosons decaying into a $\mathrm{b\overline{b}}$ pair in
  pp collisions at $\sqrt{s} =$ 13 TeV},''
  \href{http://dx.doi.org/10.1007/JHEP08(2018)113}{{\em JHEP} {\bf 08} (2018)
  113},
\href{http://arxiv.org/abs/1805.12191}{{\tt arXiv:1805.12191 [hep-ex]}}.

\bibitem{Aad:2019zwb}
{\bf ATLAS} Collaboration, G.~Aad {\em et al.}, ``{Search for heavy neutral
  Higgs bosons produced in association with $b$-quarks and decaying to
  $b$-quarks at $\sqrt{s}=13$ TeV with the ATLAS detector},''
\href{http://arxiv.org/abs/1907.02749}{{\tt arXiv:1907.02749 [hep-ex]}}.

\bibitem{Sirunyan:2018zut}
{\bf CMS} Collaboration, A.~M. Sirunyan {\em et al.}, ``{Search for additional
  neutral MSSM Higgs bosons in the $\tau\tau$ final state in proton-proton
  collisions at $\sqrt{s}=$ 13 TeV},''
  \href{http://dx.doi.org/10.1007/JHEP09(2018)007}{{\em JHEP} {\bf 09} (2018)
  007},
\href{http://arxiv.org/abs/1803.06553}{{\tt arXiv:1803.06553 [hep-ex]}}.

\bibitem{CMS:2019hvr}
{\bf CMS} Collaboration, A.~M. Sirunyan {\em et al.}, ``{Search for a low-mass
  $\tau^+\tau^-$ resonance in association with a bottom quark in proton-proton
  collisions at $\sqrt{s}=$ 13 TeV},''
  \href{http://dx.doi.org/10.1007/JHEP05(2019)210}{{\em JHEP} {\bf 05} (2019)
  210},
\href{http://arxiv.org/abs/1903.10228}{{\tt arXiv:1903.10228 [hep-ex]}}.

\bibitem{Aad:2020zxo}
{\bf ATLAS} Collaboration, G.~Aad {\em et al.}, ``{Search for heavy Higgs
  bosons decaying into two tau leptons with the ATLAS detector using $pp$
  collisions at $\sqrt{s}=13$ TeV},''
\href{http://arxiv.org/abs/2002.12223}{{\tt arXiv:2002.12223 [hep-ex]}}.

\bibitem{Sirunyan:2018aui}
{\bf CMS} Collaboration, A.~M. Sirunyan {\em et al.}, ``{Search for a standard
  model-like Higgs boson in the mass range between 70 and 110 GeV in the
  diphoton final state in proton-proton collisions at $\sqrt{s}=$ 8 and 13
  TeV},'' \href{http://dx.doi.org/10.1016/j.physletb.2019.03.064}{{\em Phys.
  Lett.} {\bf B793} (2019)  320--347},
\href{http://arxiv.org/abs/1811.08459}{{\tt arXiv:1811.08459 [hep-ex]}}.

\bibitem{Sirunyan:2018wnk}
{\bf CMS} Collaboration, A.~M. Sirunyan {\em et al.}, ``{Search for physics
  beyond the standard model in high-mass diphoton events from proton-proton
  collisions at $\sqrt{s} =$ 13 TeV},''
  \href{http://dx.doi.org/10.1103/PhysRevD.98.092001}{{\em Phys. Rev.} {\bf
  D98} (2018) no.~9, 092001},
\href{http://arxiv.org/abs/1809.00327}{{\tt arXiv:1809.00327 [hep-ex]}}.

\bibitem{Aad:2014ioa}
{\bf ATLAS} Collaboration, G.~Aad {\em et al.}, ``{Search for Scalar Diphoton
  Resonances in the Mass Range $65-600$ GeV with the ATLAS Detector in $pp$
  Collision Data at $\sqrt{s}$ = 8 $TeV$},''
  \href{http://dx.doi.org/10.1103/PhysRevLett.113.171801}{{\em Phys. Rev.
  Lett.} {\bf 113} (2014) no.~17, 171801},
\href{http://arxiv.org/abs/1407.6583}{{\tt arXiv:1407.6583 [hep-ex]}}.

\bibitem{Aaboud:2017yyg}
{\bf ATLAS} Collaboration, M.~Aaboud {\em et al.}, ``{Search for new phenomena
  in high-mass diphoton final states using 37 fb$^{-1}$ of proton--proton
  collisions collected at $\sqrt{s}=13$ TeV with the ATLAS detector},''
  \href{http://dx.doi.org/10.1016/j.physletb.2017.10.039}{{\em Phys. Lett.}
  {\bf B775} (2017)  105--125},
\href{http://arxiv.org/abs/1707.04147}{{\tt arXiv:1707.04147 [hep-ex]}}.

\bibitem{ATLAS:2018xad}
{\bf ATLAS} Collaboration, T.~A. collaboration,
``{Search for resonances in the 65 to 110 GeV diphoton invariant mass range
  using 80 fb$^{-1}$ of $pp$ collisions collected at $\sqrt{s}=13$ TeV with the
  ATLAS detector},''.

\bibitem{Sirunyan:2019wph}
{\bf CMS} Collaboration, A.~M. Sirunyan {\em et al.}, ``{Search for heavy Higgs
  bosons decaying to a top quark pair in proton-proton collisions at $\sqrt{s}
  =$ 13 TeV},''
\href{http://arxiv.org/abs/1908.01115}{{\tt arXiv:1908.01115 [hep-ex]}}.

\bibitem{Sirunyan:2018qlb}
{\bf CMS} Collaboration, A.~M. Sirunyan {\em et al.}, ``{Search for a new
  scalar resonance decaying to a pair of Z bosons in proton-proton collisions
  at $\sqrt{s}=13 $ TeV},'' \href{http://dx.doi.org/10.1007/JHEP06(2018)127,
  10.1007/JHEP03(2019)128}{{\em JHEP} {\bf 06} (2018)  127},
  \href{http://arxiv.org/abs/1804.01939}{{\tt arXiv:1804.01939 [hep-ex]}}.
[Erratum: JHEP03,128(2019)].

\bibitem{Aaboud:2017rel}
{\bf ATLAS} Collaboration, M.~Aaboud {\em et al.}, ``{Search for heavy ZZ
  resonances in the $\ell ^+\ell ^-\ell ^+\ell ^-$ and $\ell ^+\ell ^-\nu
  \bar{\nu }$ final states using proton–proton collisions at $\sqrt{s}= 13$
  $\text {TeV}$ with the ATLAS detector},''
  \href{http://dx.doi.org/10.1140/epjc/s10052-018-5686-3}{{\em Eur. Phys. J.}
  {\bf C78} (2018) no.~4, 293},
\href{http://arxiv.org/abs/1712.06386}{{\tt arXiv:1712.06386 [hep-ex]}}.

\bibitem{Sirunyan:2019pqw}
{\bf CMS} Collaboration, A.~M. Sirunyan {\em et al.}, ``{Search for a heavy
  Higgs boson decaying to a pair of W bosons in proton-proton collisions at
  $\sqrt{s} =$ 13 TeV},''
\href{http://arxiv.org/abs/1912.01594}{{\tt arXiv:1912.01594 [hep-ex]}}.

\bibitem{Aaboud:2017gsl}
{\bf ATLAS} Collaboration, M.~Aaboud {\em et al.}, ``{Search for heavy
  resonances decaying into $WW$ in the $e\nu\mu\nu$ final state in $pp$
  collisions at $\sqrt{s}=13$ TeV with the ATLAS detector},''
  \href{http://dx.doi.org/10.1140/epjc/s10052-017-5491-4}{{\em Eur. Phys. J.}
  {\bf C78} (2018) no.~1, 24},
\href{http://arxiv.org/abs/1710.01123}{{\tt arXiv:1710.01123 [hep-ex]}}.

\bibitem{Khachatryan:2015lba}
{\bf CMS} Collaboration, V.~Khachatryan {\em et al.}, ``{Search for a
  pseudoscalar boson decaying into a Z boson and the 125 GeV Higgs boson in
  llbb final states},''
  \href{http://dx.doi.org/10.1016/j.physletb.2015.07.010}{{\em Phys. Lett.}
  {\bf B748} (2015)  221--243},
\href{http://arxiv.org/abs/1504.04710}{{\tt arXiv:1504.04710 [hep-ex]}}.

\bibitem{Sirunyan:2019xls}
{\bf CMS} Collaboration, A.~M. Sirunyan {\em et al.}, ``{Search for a heavy
  pseudoscalar boson decaying to a Z and a Higgs boson at $\sqrt{s} =$ 13
  TeV},'' \href{http://dx.doi.org/10.1140/epjc/s10052-019-7058-z}{{\em Eur.
  Phys. J.} {\bf C79} (2019) no.~7, 564},
\href{http://arxiv.org/abs/1903.00941}{{\tt arXiv:1903.00941 [hep-ex]}}.

\bibitem{Aad:2015wra}
{\bf ATLAS} Collaboration, G.~Aad {\em et al.}, ``{Search for a CP-odd Higgs
  boson decaying to Zh in pp collisions at $\sqrt{s} = 8$ TeV with the ATLAS
  detector},'' \href{http://dx.doi.org/10.1016/j.physletb.2015.03.054}{{\em
  Phys. Lett.} {\bf B744} (2015)  163--183},
\href{http://arxiv.org/abs/1502.04478}{{\tt arXiv:1502.04478 [hep-ex]}}.

\bibitem{Aaboud:2017cxo}
{\bf ATLAS} Collaboration, M.~Aaboud {\em et al.}, ``{Search for heavy
  resonances decaying into a $W$ or $Z$ boson and a Higgs boson in final states
  with leptons and $b$-jets in 36 fb$^{-1}$ of $\sqrt s = 13$ TeV $pp$
  collisions with the ATLAS detector},''
  \href{http://dx.doi.org/10.1007/JHEP11(2018)051,
  10.1007/JHEP03(2018)174}{{\em JHEP} {\bf 03} (2018)  174},
  \href{http://arxiv.org/abs/1712.06518}{{\tt arXiv:1712.06518 [hep-ex]}}.
[Erratum: JHEP11,051(2018)].

\bibitem{Khachatryan:2015tha}
{\bf CMS} Collaboration, V.~Khachatryan {\em et al.}, ``{Searches for a heavy
  scalar boson H decaying to a pair of 125 GeV Higgs bosons hh or for a heavy
  pseudoscalar boson A decaying to Zh, in the final states with $h \to \tau
  \tau$},'' \href{http://dx.doi.org/10.1016/j.physletb.2016.01.056}{{\em Phys.
  Lett.} {\bf B755} (2016)  217--244},
\href{http://arxiv.org/abs/1510.01181}{{\tt arXiv:1510.01181 [hep-ex]}}.

\bibitem{Sirunyan:2019xjg}
{\bf CMS} Collaboration, A.~M. Sirunyan {\em et al.}, ``{Search for a heavy
  pseudoscalar Higgs boson decaying into a 125 GeV Higgs boson and a Z boson in
  final states with two tau and two light leptons at $\sqrt{s}=$ 13 TeV},''
\href{http://arxiv.org/abs/1910.11634}{{\tt arXiv:1910.11634 [hep-ex]}}.

\bibitem{Sirunyan:2017tqo}
{\bf CMS} Collaboration, A.~M. Sirunyan {\em et al.}, ``{Search for Higgs boson
  pair production in the $bb\tau\tau$ final state in proton-proton collisions
  at $\sqrt{(}s)=8\text{ }\text{ }\mathrm{TeV}$},''
  \href{http://dx.doi.org/10.1103/PhysRevD.96.072004}{{\em Phys. Rev.} {\bf
  D96} (2017) no.~7, 072004},
\href{http://arxiv.org/abs/1707.00350}{{\tt arXiv:1707.00350 [hep-ex]}}.

\bibitem{Sirunyan:2018two}
{\bf CMS} Collaboration, A.~M. Sirunyan {\em et al.}, ``{Combination of
  searches for Higgs boson pair production in proton-proton collisions at
  $\sqrt{s} = $ 13 TeV},''
  \href{http://dx.doi.org/10.1103/PhysRevLett.122.121803}{{\em Phys. Rev.
  Lett.} {\bf 122} (2019) no.~12, 121803},
\href{http://arxiv.org/abs/1811.09689}{{\tt arXiv:1811.09689 [hep-ex]}}.

\bibitem{Aad:2015xja}
{\bf ATLAS} Collaboration, G.~Aad {\em et al.}, ``{Searches for Higgs boson
  pair production in the $hh\to bb\tau\tau, \gamma\gamma WW^*, \gamma\gamma bb,
  bbbb$ channels with the ATLAS detector},''
  \href{http://dx.doi.org/10.1103/PhysRevD.92.092004}{{\em Phys. Rev.} {\bf
  D92} (2015)  092004},
\href{http://arxiv.org/abs/1509.04670}{{\tt arXiv:1509.04670 [hep-ex]}}.

\bibitem{Aad:2019uzh}
{\bf ATLAS} Collaboration, G.~Aad {\em et al.}, ``{Combination of searches for
  Higgs boson pairs in $pp$ collisions at $\sqrt{s} = $13 TeV with the ATLAS
  detector},''
\href{http://arxiv.org/abs/1906.02025}{{\tt arXiv:1906.02025 [hep-ex]}}.

\bibitem{Aaboud:2018eoy}
{\bf ATLAS} Collaboration, M.~Aaboud {\em et al.}, ``{Search for a heavy Higgs
  boson decaying into a $Z$ boson and another heavy Higgs boson in the
  $\ell\ell bb$ final state in $pp$ collisions at $\sqrt{s}=13$ TeV with the
  ATLAS detector},''
  \href{http://dx.doi.org/10.1016/j.physletb.2018.07.006}{{\em Phys. Lett.}
  {\bf B783} (2018)  392--414},
\href{http://arxiv.org/abs/1804.01126}{{\tt arXiv:1804.01126 [hep-ex]}}.

\bibitem{Sirunyan:2019wrn}
{\bf CMS} Collaboration, A.~M. Sirunyan {\em et al.}, ``{Search for new neutral
  Higgs bosons through the H$\to$ ZA $\to \ell^{+}\ell^{-} \mathrm{b\bar{b}}$
  process in pp collisions at $\sqrt{s} =$ 13 TeV},''
\href{http://arxiv.org/abs/1911.03781}{{\tt arXiv:1911.03781 [hep-ex]}}.

\bibitem{Liebler:2016ceh}
S.~Liebler, S.~Patel, and G.~Weiglein, ``{Phenomenology of on-shell Higgs
  production in the MSSM with complex parameters},''
  \href{http://dx.doi.org/10.1140/epjc/s10052-017-4849-y}{{\em Eur. Phys. J.}
  {\bf C77} (2017) no.~5, 305},
\href{http://arxiv.org/abs/1611.09308}{{\tt arXiv:1611.09308 [hep-ph]}}.

\bibitem{Eriksson:2009ws}
D.~Eriksson, J.~Rathsman, and O.~Stal, ``{2HDMC: Two-Higgs-Doublet Model
  Calculator Physics and Manual},''
  \href{http://dx.doi.org/10.1016/j.cpc.2009.09.011}{{\em Comput. Phys.
  Commun.} {\bf 181} (2010)  189--205},
\href{http://arxiv.org/abs/0902.0851}{{\tt arXiv:0902.0851 [hep-ph]}}.

\bibitem{Su:2019dsf}
W.~Su, M.~White, A.~G. Williams, and Y.~Wu, ``{Exploring the low $\tan\beta$
  region of two Higgs doublet models at the LHC},''
  \href{http://arxiv.org/abs/1909.09035}{{\tt arXiv:1909.09035 [hep-ph]}}.

\bibitem{ALEPH:2005ab}
{\bf ALEPH, DELPHI, L3, OPAL, SLD, LEP Electroweak Working Group, SLD
  Electroweak Group, SLD Heavy Flavour Group} Collaboration, S.~Schael {\em et
  al.}, ``{Precision electroweak measurements on the $Z$ resonance},''
  \href{http://dx.doi.org/10.1016/j.physrep.2005.12.006}{{\em Phys. Rept.} {\bf
  427} (2006)  257--454},
\href{http://arxiv.org/abs/hep-ex/0509008}{{\tt arXiv:hep-ex/0509008
  [hep-ex]}}.

\bibitem{Haller:2018nnx}
J.~Haller, A.~Hoecker, R.~Kogler, K.~Mönig, T.~Peiffer, and J.~Stelzer,
  ``{Update of the global electroweak fit and constraints on two-Higgs-doublet
  models},'' \href{http://dx.doi.org/10.1140/epjc/s10052-018-6131-3}{{\em Eur.
  Phys. J.} {\bf C78} (2018) no.~8, 675},
\href{http://arxiv.org/abs/1803.01853}{{\tt arXiv:1803.01853 [hep-ph]}}.

\bibitem{Chen:2019pkq}
N.~Chen, T.~Han, S.~Li, S.~Su, W.~Su, and Y.~Wu, ``{Type-I 2HDM under the Higgs
  and Electroweak Precision Measurements},''
\href{http://arxiv.org/abs/1912.01431}{{\tt arXiv:1912.01431 [hep-ph]}}.

\bibitem{Chen:2018shg}
N.~Chen, T.~Han, S.~Su, W.~Su, and Y.~Wu, ``{Type-II 2HDM under the Precision
  Measurements at the $Z$-pole and a Higgs Factory},''
  \href{http://dx.doi.org/10.1007/JHEP03(2019)023}{{\em JHEP} {\bf 03} (2019)
  023},
\href{http://arxiv.org/abs/1808.02037}{{\tt arXiv:1808.02037 [hep-ph]}}.

\bibitem{Amhis:2016xyh}
{\bf HFLAV} Collaboration, Y.~Amhis {\em et al.}, ``{Averages of $b$-hadron,
  $c$-hadron, and $\tau$-lepton properties as of summer 2016},''
  \href{http://dx.doi.org/10.1140/epjc/s10052-017-5058-4}{{\em Eur. Phys. J.}
  {\bf C77} (2017) no.~12, 895},
\href{http://arxiv.org/abs/1612.07233}{{\tt arXiv:1612.07233 [hep-ex]}}.

\bibitem{Arbey:2017gmh}
A.~Arbey, F.~Mahmoudi, O.~Stal, and T.~Stefaniak, ``{Status of the Charged
  Higgs Boson in Two Higgs Doublet Models},''
  \href{http://dx.doi.org/10.1140/epjc/s10052-018-5651-1}{{\em Eur. Phys. J. C}
  {\bf 78} (2018) no.~3, 182}, \href{http://arxiv.org/abs/1706.07414}{{\tt
  arXiv:1706.07414 [hep-ph]}}.

\end{thebibliography}\endgroup

\end{document}